\documentclass[11pt]{article}
\usepackage{jheppub}
\usepackage{epsfig,comment}
\usepackage{graphicx}
\usepackage{url,hyperref}
\usepackage{amsmath,amssymb,bm}
\usepackage{slashed}
\usepackage{float}
\usepackage{orcidlink}
\usepackage{physics,braket,booktabs}
\usepackage{dsfont}
\usepackage{tikz}
\usetikzlibrary{quantikz2}
\usetikzlibrary{shapes.geometric}
\usetikzlibrary{arrows.meta,calc,positioning}
\usepackage{bbold}
\usepackage{amsthm}
\usepackage{yfonts}
\usepackage[english]{babel}  

\newcommand{\Id}{\mathds{I}}
\newcommand{\HH}{\mathbb{H}}
\title{Encoding Compact U(1) Gauge Fields in Bosonic Modes with GKP Stabilization}
\author[1]{Victor Ale,}
\author[2]{Tommaso Rainaldi,}
\author[3]{Enrique Rico,}
\author[2]{Felix Ringer,}
\author[1]{George Siopsis,}
\affiliation[1]{Department of Physics and Astronomy, The University of Tennessee, Knoxville, TN 37996, USA}
\affiliation[2]{Department of Physics and Astronomy, Stony Brook University, New York 11794, USA}
\affiliation[3]{CERN, Theoretical Physics Department, CH-1211 Geneva, Switzerland}
\emailAdd{vale@vols.utk.edu}
\emailAdd{tommaso.rainaldi@stonybrook.edu}
\emailAdd{enrique.rico.ortega@cern.ch}
\emailAdd{felix.ringer@stonybrook.edu}
\emailAdd{siopsis@tennessee.edu}
\abstract{Compact lattice gauge theories are formulated in terms of angular variables and integer electric fluxes, while bosonic quantum hardware provides oscillator modes with continuous, unbounded quadratures.  We bridge this gap with a one-to-one encoding.  After Gauss's law is solved, each remaining gauge degree of freedom is carried by a single oscillator mode, with its interactions built from trigonometric gates, and a Gottesman--Kitaev--Preskill (GKP)-type stabilizer provides the compactness that the hardware does not. The encoding becomes exact in the limit of infinite squeezing, and at finite squeezing, the leading imperfections act as small, computable shifts of physical observables rather than uncontrolled leakage. We apply the construction to compact QED\(_3\) and derive the error budget at finite squeezing, characterizing the leading errors in closed form, and showing that they can be corrected, subtracted, or extrapolated away. We construct syndrome-extraction protocols that detect and remove the displacement component of photon loss, delimit the noise it does not reach, compare two choices of dynamical variables, and collect the scaling of mode count, gate count, and measurement cost.  A one-plaquette example reproduces the exact compact-rotor dynamics, and real-time spectroscopy with controlled extrapolations recovers the exponentially small energy splitting between charge sectors, the seed of the monopole physics of the theory, at the percent level against its exact value.
}
\begin{document}
\maketitle
\newpage
\section{Introduction}
\label{sec:Introduction}
The real-time dynamics of lattice gauge theories is one of the most compelling frontiers in quantum simulation, offering access to regimes that remain intractable for classical computation~\cite{Banuls:2019bmf,PRXQuantum.2.017001,PRXQuantum.4.027001,PRXQuantum.5.037001}. Since Jordan, Lee, and Preskill showed that scattering in scalar field theory is efficiently simulable by a quantum computer~\cite{Jordan:2012xnu}, the Kogut--Susskind Hamiltonian formulation of lattice gauge theory~\cite{Kogut:1974ag} has been used to devise quantum algorithms studying Abelian and non-Abelian lattice gauge theory, confinement, and particle production, aiming at simulating real-time dynamics of QED and QCD, relevant to high-energy physics, nuclear physics, and condensed-matter physics.  Yet there is an asymmetry between what the hardware can implement and what the theory contains. Matter is fermionic, and hence discrete, and it can be easily mapped to qubits using Jordan--Wigner or Bravyi--Kitaev transformations~\cite{Jordan:1928wi,Bravyi:2000vfj}. Gauge fields take values on group manifolds, e.g.\ $U(1)$ or $SU(N)$, which can be described by a qubit register only if the link Hilbert space is truncated to a finite dimensional subspace or if one considers discrete subgroups of $U(1)$ or $SU(N)$~\cite{Chandrasekharan:1996ih,Zohar:2015hwa,Martinez2016,PhysRevA.98.032331,Kokail2019,Alexandru:2019nsa,Haase:2020kaj,Bauer:2021gek,Ciavarella:2021nmj,Davoudi:2022xmb,Grieninger:2023ehb,Halimeh:2023lid,Yao:2026rya}. This truncation is systematic but comes at a cost in resources and introduces a discretization error that increases away from the continuum limit. Instead, a CV processor can start from an infinite-dimensional space per mode~\cite{Lloyd:1998jk} and represent the gauge field without any discretization~\cite{Marshall:2015mna}. 

The literature on quantum simulations of lattice gauge theories is rich and fast-developing. Quantum digital and analog real-time dynamics and string breaking have been simulated on trapped-ion and superconducting processors in 1+1D~\cite{Martinez2016,Kokail2019}, and analog and hybrid schemes have been proposed in cold-atom, trapped-ion, and circuit-QED architectures~\cite{PhysRevA.105.023322,Muschik_2017,alcainecuervo2026compactu1latticegauge}. Continuous variable encodings have been suggested and developed for the O(3) model~\cite{Jha:2023ecu}, scalar field theory~\cite{Briceno:2023xcm,Abel:2024kuv,Abel:2025pxa,Abel:2025zxb,Gupta:2025xti}, SU(2) lattice gauge theory~\cite{Ale2025}, QED and other field theories~\cite{Crane:2024tlj,Miranda-Riaza:2025fus,k9p6-c649}, $Z_2$-Higgs~\cite{Saner:2025nrq,Varona:2024sgc,Domanti:2024zyg,Bazavan:2023oce,Schuckert:2025iko,Cobos:2025krn}, $Z_2$ lattice gauge theory~\cite{Saner:2025nrq}, and Yukawa theory~\cite{Than:2025gso}. Hybrid platforms coupling qubits~\cite{RevModPhys.75.281,wineland1998experimentalissuescoherentquantumstate,Blais:2020wjs} to bosonic modes~\cite{Stavenger:2022wzz,Liu:2024mbr,Araz:2024dcy,6prx-zmdz} seem to be the natural choice for these theories. These systems offer a native gate set with Gaussian operations, non-Gaussian phase gates, and conditional displacements~\cite{Liu:2024mbr}. Yet, every CV encoding of a compact gauge group must overcome the inherent compactness of the gauge variable while using a non-compact bosonic mode. One proposed way to address this issue for $U(1)$ is to embed the circle into the phase space of a pair of qumodes per link and then squeeze onto it~\cite{ Ale2025}. A similar procedure was also developed for $SU(2)$ in Ref.~\cite{Ale:2025sxz}. 

In this paper, we address the Abelian $U(1)$ gauge group case, and we completely bypass the problem in a novel way, while retaining the minimal resource count, i.e., one mode per link. We show that the Hilbert space of a free harmonic oscillator, $L^2(\mathbb R)$, can be seen as an exact direct integral of copies of the Hilbert space of a rotor, each copy labeled by a quasimomentum fiber, and each fiber itself a compact rotor with twisted boundary conditions. A GKP-type stabilizer~\cite{Gottesman:2000di} chooses a specific fiber, enforcing compactness by wrapping the real line onto the circle instead of projecting onto a compact subset of the full plane.

From the exactness of this correspondence, we organize the error theory of the encoding and show how it connects to the framework of bosonic error correction and mitigation. Because the fiber label is a boundary twist, any imperfect stabilization is not generic leakage out of a code space, but a distribution of background ``Aharonov--Bohm fluxes", whose effect on the electric energy we obtain in closed form. Importantly, the stabilizer commutes with the system's dynamics, so its syndrome, which can be extracted by a qunaught ancilla and phase estimation~\cite{Kitaev:1995qy,Terhal:2015fcw}, is non-demolition and provides a natural interface to bosonic error correction. Small displacement errors from photon loss or imperfect Gaussian operations are fully correctable, while we prove that the finite-energy effects that remain and the envelope distortion can be removed by extrapolation based on a known error model rather than an empirical fit. 

On the physics side, we note that once the charge-dependent linear electric term is removed by a displacement, static charges leave behind twisted boundary conditions for the wavefunctions, and the associated twist energy $\Delta_{\rm tw}$ has a vanishing perturbative series, generated at weak coupling by the monopole-instantons of the Polyakov mechanism~\cite{Polyakov:1975rs}. We show that $\Delta_{\rm tw}$ is a Fourier coefficient of the ground-state band over the twist torus, that it is known exactly in terms of Mathieu characteristic values for a single plaquette, and that it can be read off the lowest line of a pair-addition correlator once the classically computable electrostatic and mass contributions have been demodulated away.

Furthermore, the reduced (Gauss' law solved) theory admits two natural choices of dynamical variables, a loop frame and a link frame, which describe identical physics but differ in gate content and in finite-squeezing bias, and we give closed-form counts for both. Independently of the frame, we show that the time evolution compiles into Gaussian gates, conditional displacements, and controlled trigonometric gates~\cite{Chalermpusitarak:2025cod,Rainaldi:2025ymn}, and the same primitives carry the readout: every observable is obtained from a single-ancilla Hadamard test with controlled trigonometric operators interleaved with the evolution, giving the complex one- and two-point functions of the encoded observables (Appendix~\ref{app:measurement}). Throughout this work, we limit ourselves to classical simulations of the developed quantum algorithms, although small-scale demonstrations appear feasible on existing hybrid platforms in the near term. The framework extends naturally to larger lattices, and similar techniques can be adapted to fermion-boson problems in quantum chemistry, condensed-matter, and polaritonic settings~\cite{Kang:2023xfb,Vu:2025aub}.

The remainder of this paper is organized as follows. In Sec.~\ref{sec:QED3} we present the Kogut--Susskind Hamiltonian for U(1) in 2+1 dimensions after Gauss's law has been solved, together with the fermion mapping, the displacement that removes the linear electric term, and the twist energy it leaves behind. In Sec.~\ref{sec:encoding} we construct the single-mode encoding and its stabilizer, and in Sec.~\ref{sec:errorcorrection} we give the finite-energy error budget together with the physical-noise channels and the two ways of dealing with them. In Sec.~\ref{sec:prep} we validate the construction on one plaquette, comparing the compact encoded theory against the exact compact one and then using pair-addition spectroscopy to expose the twist energy, and in Sec.~\ref{sec:scaling} we collect the resource estimates and the choice of computational frame. We conclude with Sec.~\ref{sec:conc} where we make our final remarks and address prospects. The longer derivations are in the appendices: the electric Hamiltonian after Gauss's law, the one-plaquette twist energy, the finite-energy error model, the frames together with the gate synthesis, and the measurement circuits.

\section{Compact QED$_3$ and charge-induced twists}\label{sec:QED3}
\subsection{The QED$_3$ Hamiltonian}
\label{sec:hamiltonian}
\begin{figure}[t]
    \centering
    \includegraphics[width=0.62\textwidth]{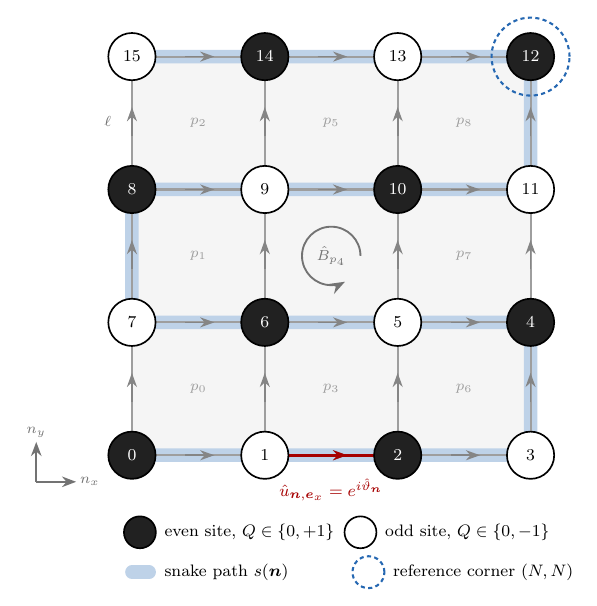}
    \caption{The lattice used throughout, drawn for $N=3$. Filled and hollow sites
    carry the staggered charges of Eq.~\eqref{eq:charge_def}, each labeled by
    its snake index $s(\bm{n})$ of Eq.~\eqref{eq:snake}, with the shaded band
    tracing the path that fixes the Jordan--Wigner string. Arrows orient the
    links $\ell$, and the highlighted link carries the link element
    $\hat{u}_{\bm{n},\bm{e}_x}=e^{i\hat{\vartheta}_{\bm{n}}}$ of the hopping
    term. Plaquettes are indexed by $\pi(p)=p_xN+p_y$, the circular arrow on
    $p_4$ giving the orientation of $\hat{B}_p$, and the dashed ring marks the
    reference corner $(N,N)$ where Gauss's law closes.}
    \label{fig:lattice_snake}
\end{figure}

We work with a system of $N\times N$ plaquettes with open boundary conditions. Sites $\bm{n}=(n_x,n_y)$ with $0\le n_x,n_y\le N$ are indexed by the snake path that also fixes the Jordan--Wigner string of Sec.~\ref{sec:fermions},
\begin{equation}
    s(\bm{n}) = n_y(N+1) +
    \begin{cases} n_x & n_y \text{ even},\\ N-n_x & n_y \text{ odd}.\end{cases}
    \label{eq:snake}
\end{equation}
Plaquettes are $p=(p_x,p_y)$ with $0\le p_x,p_y\le N-1$, indexed by $\pi(p)=p_xN+p_y$, and there are $2N(N+1)$ links, labeled with $\ell$. These conventions are collected in Fig.~\ref{fig:lattice_snake}.

The Kogut--Susskind Hamiltonian for $N\times N$ plaquettes with dynamical matter, after Gauss's law has been solved, is \cite{Kogut:1974ag, Crippa:2024cqr, Ale:2025sxz}
\begin{equation}
    \begin{split}
        \hat{H}_E &= \frac{g^2}{2}\Big[\hat{\eta}_i\HH^{(2)}_{ij}\hat{\eta}_j + \hat{\eta}_i\HH^{(1)}_{ij}\hat{Q}_j + \hat{Q}_i\HH^{(0)}_{ij}\hat{Q}_j\Big],\\
        \hat{H}_B &= \frac{1}{g^2}\sum_{p}\big(1-\cos\hat{B}_p\big),\\
        \hat{H}_M  &= m_0\sum_{\bm{n}}(-)^{n_x + n_y}\hat{\Psi}^{\dagger}_{\bm{n}}\hat{\Psi}_{\bm{n}},\\
        \hat{H}_K  &= \frac{1}{2} \sum_{\bm{n}} \Big[i\hat{\Psi}^{\dagger}_{\bm{n}}\hat{u}^{\dagger}_{\bm{n},\bm{e}_{x}}\hat{\Psi}_{\bm{n}+\bm{e}_{x}} - (-)^{n_x+n_y}\hat{\Psi}^{\dagger}_{\bm{n}}\hat{\Psi}_{\bm{n}+\bm{e}_{y}}\Big] +\text{h.c.}
    \end{split}
    \label{eq:U1_ham_dynam}
\end{equation}
After Gauss's law is solved, only $N^2$ dynamical rotors remain with conjugate pairs $(\hat{\chi}_i,\hat{\eta}_i)$. Which $N^2$ pairs these are is a choice of frame, which we defer to Sec.~\ref{sec:frames} and Appendix~\ref{app:gates}. The symbols $\HH^{(2,1,0)}$ are frame-agnostic: they stand for the blocks of the electric quadratic form before any choice of dynamical variables is made, so that every statement written in terms of them holds in both frames. Details about the derivations are in Appendix~\ref{app:GL}. Note that $\HH^{(2)}$ is an $N^2\times N^2$ matrix, $\HH^{(1)}$ a rectangular $N^2\times(N+1)^2$ block and $\HH^{(0)}$ a square $(N+1)^2\times(N+1)^2$ one. Once a frame is chosen, they take concrete entries, and we write them as $\mathcal{H}^{(2,1,0)}$ for the loop basis and $\tilde{\mathcal{H}}^{(2,1,0)}$ for the link basis. The $N^2$ modes obey the rotor algebra
\begin{equation}
    \hat{U}_i = e^{i\hat{\chi}_i},
    \qquad
    [\hat{\eta}_i,\hat{U}_j] = \hat{U}_j\delta_{ij}
    \qquad\big(\,[\hat{\chi}_i,\hat{\eta}_j] = i\delta_{ij}\,\big),
    \label{eq:rotor_algebra}
\end{equation}
where $\hat{\eta}_i=-i\partial_{\chi_i}$ has $\mathrm{Spec}(\hat\eta_i)=\mathbb{Z}$ and $\hat{U}_i$ raises the flux. Throughout the paper, we use $(\hat{\chi}_i,\hat{\eta}_i)$ and $\hat{U}_i$ to denote the generic mode variables of whichever frame we use, and every other angle in Eq.~\eqref{eq:U1_ham_dynam} is built from them. We denote the magnetic flux $\hat{B}_p$ through plaquette $p$, and the horizontal-link angle $\hat{\vartheta}_{\bm{n}}$ of the link leaving site $\bm{n}$, through which the hopping term carries its link element,
\begin{equation}
    \hat{u}_{\bm{n},\bm{e}_x} = e^{i\hat{\vartheta}_{\bm{n}}}.
    \label{eq:Theta}
\end{equation}
Both composites are fixed integer linear combinations of the $\hat{\chi}_i$, unimodularly related to them, with the explicit frame form given in Appendix~\ref{app:gates} and Fig.~\ref{fig:two_frames} there: in the loop frame the flux is itself a mode angle, $\hat{B}_p=\hat{\chi}_p$, and the link angle is the column sum $\hat{\vartheta}_{\bm{n}}=\sum_{j\ge n_y}\hat{\chi}_{(n_x,j)}$, while in the link frame the link angle is itself the mode, $\hat{\vartheta}_{\bm{n}}=\hat{\chi}_{\bm{n}}$, and the flux is the difference of the two adjacent link angles. Furthermore, the system contains $(N+1)^2$ staggered charges $\hat{Q}_{s(\bm{n})}$ with
\begin{equation}
    \hat{Q}_{\bm{n}} = \hat\Psi^{\dagger}_{\bm{n}}\hat\Psi_{\bm{n}} - \frac{1-(-)^{n_x+n_y}}{2}\,\Id,
    \label{eq:charge_def}
\end{equation}
so $Q_{\bm{n}}\in\{0,+1\}$ on even sites and $\{0,-1\}$ on odd sites. As shown in Appendix~\ref{app:GL}, Gauss's law closes at the reference corner $(N,N)$ if and only if $\sum_{\bm{n}}Q_{\bm{n}}=0$. That shows that neutrality is a consistency condition of the reduction, not an extra assumption.
\subsection{Fermion mapping}
\label{sec:fermions}
With $X^\pm\equiv(X\pm iY)/2$ we take the Jordan-Wigner map
\begin{equation}
    \hat\Psi^\dagger_{\bm{n}} \mapsto P_{L(\bm{0},\bm{n})}X^-_{\bm{n}},
    \qquad
    \hat\Psi_{\bm{n}} \mapsto P_{L(\bm{0},\bm{n})}X^+_{\bm{n}},
    \qquad
    P_{L(\bm{n},\bm{n}')} = \!\!\prod_{\bm{n}''\in L\setminus\{\bm{n}'\}}\!\! Z_{\bm{n}''},
    \label{eq:JW}
\end{equation}
with $L$ the snake path~\eqref{eq:snake}, endpoint excluded. Then $\hat\Psi^\dagger\hat\Psi=X^-X^+=(\Id-Z)/2$, i.e.\ occupied $\leftrightarrow Z=-1$, and
\begin{equation}
    \hat{Q}_{s(\bm{n})} = \frac{(-1)^{n_x+n_y}\Id - Z_{s(\bm{n})}}{2}.
    \label{eq:Q_pauli}
\end{equation}
Since $X^-Z=X^-$ and $ZX^+=X^+$, the string cancels between snake-adjacent sites in either traversal direction, $\hat\Psi^\dagger_j\hat\Psi_{j\pm1}\mapsto X^-_jX^+_{j\pm1}$. Horizontal hops are string-free in both even and odd rows.

\subsection{Static charges as fractional offset fluxes and boundary twists}
\label{sec:displacement}

Throughout this subsection and Sec.~\ref{sec:twist_energy}, we fix a static charge configuration $Q$. Dynamical matter is taken up at the end of the subsection and later in the paper.

The linear term in $\hat\eta$ can be removed by a charge-dependent displacement producing $\hat\eta_i\to\hat\eta_i+d_i$,
\begin{equation}
    \mathcal{U} = \exp\!\big(+i\hat{\chi}_i d_i\big),
    \qquad
    \mathcal{U}^\dagger \hat{\eta}_i \mathcal{U} = \hat{\eta}_i + d_i.
    \label{eq:displacement_op}
\end{equation}
Canceling the terms linear in $\hat\eta$ fixes the displacement to
\begin{equation}
        d = -\tfrac{1}{2}(\HH^{(2)})^{-1}\HH^{(1)}Q,
        \label{eq:Delta_op}
\end{equation}
and leaves the transformed electric Hamiltonian
\begin{equation}
    \hat{H}'_E = \frac{g^2}{2}\Big[\hat{\eta}_i \HH^{(2)}_{ij}\hat{\eta}_j + Q_i \HH^{(0)\rm eff}_{ij}Q_j\Big],
    \qquad
    \HH^{(0)\rm eff}\equiv\HH^{(0)} - \tfrac{1}{4}\HH^{(1)\mathsf{T}}(\HH^{(2)})^{-1}\HH^{(1)}.
    \label{eq:Hprime_E}
\end{equation}
The price is a change of boundary conditions for the gauge wavefunctions. Denoting by $\lbrace\chi\rbrace$ the set of all the dynamical local degrees of freedom, we have $\Psi'(\{\chi\})=\braket{\{\chi\}|\mathcal{U}^\dagger|\Psi}=e^{-i\chi_id_i}\Psi(\{\chi\})$, which spoils the $2\pi$ periodicity required for integer $\hat\eta_i$
\begin{equation}
    \Psi'(\chi_i + 2\pi) = e^{\,i\theta_i}\,\Psi'(\chi_i),
    \quad
    \partial_{\chi_i}\Psi'(\chi_i + 2\pi) = e^{\,i\theta_i}\,\partial_{\chi_i}\Psi'(\chi_i),
    \quad
    \theta_i \equiv -2\pi d_i = \pi\big[(\HH^{(2)})^{-1}\HH^{(1)}Q\big]_i ,
    \label{eq:twisted_bc}
\end{equation}
the second condition being what keeps $\hat\eta_i^2$ self-adjoint on the twisted domain. Equivalently, the shifted $\hat\eta_i$ has the shifted spectrum $\mathbb{Z}-d_i$. These ``twist" angles are thus fixed by the static charges. Equation~\eqref{eq:twisted_bc} can be interpreted as a flat connection on the rotor's configuration space, so $\theta_i$ is the Aharonov--Bohm holonomy picked up on transporting the wavefunction once around the $i$th configuration-space circle $\chi_i\in[0,2\pi)$. In the loop frame, where $\hat\chi_p=\hat{B}_p$ is itself the magnetic angle, $\theta_p$ is a holonomy in the variable conjugate to the magnetic flux through plaquette $p$, and is a distinct quantity from that flux. The extra phase is invisible only when $d_i$ is integer and the shift reduces to a genuine symmetry. However, in the charge-neutral sector only the vacuum configuration gives an integer, and indeed zero, twist displacement, so every other physical configuration carries a nontrivial twist.

When matter is allowed to be dynamical, the charges become operators and, by Eq.~\eqref{eq:Delta_op}, $d$ becomes an operator acting on the matter register. The displacement
Eq.~\eqref{eq:displacement_op} is then $\mathcal{U}[\hat{Q}]$, which commutes with $\hat{H}_E$ and $\hat{H}_M$ but not with the hopping term $\hat{H}_K$ of Eq.~\eqref{eq:U1_ham_dynam}, since $\hat{H}_K$ moves charge between sites and therefore changes $\hat{d}$. Conjugating the full Hamiltonian would transform $\hat{H}_K$ as well, and the twisted boundary condition Eq.~\eqref{eq:twisted_bc} would acquire an operator-valued twist. Although this is a viable route, we do not take it since it is an unnecessary complication. The construction below uses the displaced form for static charges, where it is exact, and the undisplaced form whenever $\hat{H}_K$ is present, a choice we cost in Sec.~\ref{sec:frames} and use in the numerics of Sec.~\ref{sec:prep}.

\paragraph{One plaquette:}
For $N=1$ the choice of frame is irrelevant, and with the snake ordering $(0,0)$,$(1,0)$,$(1,1)$,$(0,1)$ we have $\HH^{(2)}=4$ and $\HH^{(1)} = (-4,\,+2,\,0,\,-2)$, the vanishing third entry belonging to the reference corner, so that
\begin{equation}
    d = \tfrac{1}{2}Q_{(0,0)} - \tfrac14 Q_{(1,0)} + \tfrac{1}{4}Q_{(0,1)} .
    \label{eq:d_1plaq}
\end{equation}
Here, the subscript denotes the coordinate of the charge on the lattice. A $q\bar{q}$ pair on the vertical link ($Q_{(0,0)}=+1$, $Q_{(0,1)}=-1$) gives $d=1/4$ and $\theta=-\pi/2$, on the horizontal bottom link ($Q_{(0,0)}=+1$, $Q_{(1,0)}=-1$) $d=3/4$ and $\theta\equiv+\pi/2$, both with electrostatic energy $E_{\rm cl}=3g^2/8$ (see Eq.~\eqref{eq:Ecl} later). The equal magnitudes and opposite signs are the diagonal reflection $x\leftrightarrow y$. The neutral doubly-occupied state $Q_{(0,0)}=Q_{(1,1)}=+1$, $Q_{(0,1)}=Q_{(1,0)}=-1$ gives $d=1/2$ and $\theta=-\pi$, an antiperiodic boundary condition. These are the configurations used throughout Sec.~\ref{sec:prep}. In Appendix~\ref{app:1plaq} we work the horizontal pair out explicitly and identify the fractional part of $d$ as the gap between where the quadratic form would place the flux and where compactness permits it.

\subsection{Twist energy from winding-sector tunneling}\label{sec:twist_energy}
The twisted boundary conditions of the previous section can be shown to carry the nonperturbative content of the compact
gauge sector. To keep the classical and the nonperturbative pieces apart, we fix the gauge Hamiltonian with the electrostatic constant removed,
\begin{equation}
    \hat{H}_{\rm gauge}[\bm{\theta}] \equiv \hat{H}'_E - E_{\rm cl} + \hat{H}_B
    = \frac{g^2}{2}\hat{\eta}_i\HH^{(2)}_{ij}\hat{\eta}_j
    + \frac{1}{g^2}\sum_{p}\big(1-\cos\hat{B}_p\big),
    \label{eq:Hgauge}
\end{equation}
with $\bm{\theta} = (\theta_1,\dots,\theta_{N^2})$, $\hat{H}'_E$ given by Eq.~\eqref{eq:Hprime_E}, and
\begin{equation}
    E_{\rm cl}[Q]=\tfrac{g^2}{2}Q_i\,\HH^{(0)\rm eff}_{ij}\,Q_j
    \label{eq:Ecl}
\end{equation}
the electrostatic energy of the charge configuration $Q$ with the fluxes treated
as continuous, equivalently
$E_{\rm cl}[Q]=\tfrac{g^{2}}{2}\min_{\eta\in\mathbb{R}^{N^{2}}}\|\mathcal{K}^{\mathsf T}\eta-\mathcal{C}Q\|^{2}$
in the notation of Appendix~\ref{app:GL}.
Let $\epsilon_0[\bm{\theta}]$ be the ground-state
energy of Eq.~\eqref{eq:Hgauge} acting on wavefunctions obeying
Eq.~\eqref{eq:twisted_bc}. We define the sector energy and the twist energy as
\begin{equation}
    E_0[Q] = E_{\rm cl}[Q] + \epsilon_0[\bm{\theta}(Q)],
    \qquad
    \Delta_{\rm tw}[Q;g] \equiv \epsilon_0[\bm{\theta}(Q)] - \epsilon_0[\bm{0}].
    \label{eq:Delta_tw_def}
\end{equation}
Equation~\eqref{eq:Delta_tw_def} is exact at all couplings.
Importantly, the twist enters $\hat{H}_{\rm gauge}[\bm{\theta}]$ not through the operator itself,
which contains no $\bm{\theta}$, but through the boundary condition
Eq.~\eqref{eq:twisted_bc} that fixes its domain, so $\epsilon_0[\bm{\theta}]$ is a
continuous, even, periodic function on
$\bm{\theta}\in(\mathbb{R}/2\pi\mathbb{Z})^{N^2}$, smooth wherever the ground state
remains gapped. Its structure follows from the Bloch character of that condition. The
magnetic potential is minimized when every $\hat{B}_p\in2\pi\mathbb{Z}$, so the
problem is a tight-binding model on the winding lattice $\mathbb{Z}^{N^2}$ with
$\bm{\theta}/2\pi$ as quasimomentum, and expanding in tunneling events of winding
vector $\bm{m}$,
\begin{equation}
    \epsilon_0[\bm{\theta}] = \bar\epsilon
    - \sum_{\bm{m}\neq\bm{0}} t_{\bm{m}} \cos(\bm{m}\cdot\bm{\theta}),
    \label{eq:band}
\end{equation}
with $\bar\epsilon$ the average of $\epsilon_0$ over the twist torus and $t_{\bm{m}}$
its Fourier coefficients, real and even. This is not an approximation, but the Fourier
series of a periodic function, so it holds at all couplings. The coupling enters only through the decay of the coefficients. Evaluating it at the charge-induced twist and at zero
twist, the band center cancels in the difference and
\begin{equation}
    \Delta_{\rm tw}[Q]
    =\sum_{\bm{m}\neq\bm{0}} t_{\bm{m}}\big[1-\cos(\bm{m}\cdot\bm{\theta})\big]
    = 2\sum_{\bm{m}\neq\bm{0}} t_{\bm{m}}
      \sin^2\!\Big(\frac{\bm{m}\cdot\bm{\theta}}{2}\Big),
    \label{eq:Delta_tw_band}
\end{equation}
even in $\bm{\theta}$, with $\epsilon_0[\bm{0}]$ the band bottom, which is the
charge-free vacuum selected by the untwisted stabilizer of Sec.~\ref{sec:penalty}. The twist energy
$\Delta_{\rm tw}\ge0$ for every $\bm{\theta}$, vanishing at
$\bm{\theta}\in2\pi\mathbb{Z}^{N^2}$ and, given a unique band minimum, nowhere else.

A $\bm{m}=\pm\bm{e}_p$ event shifts the compact magnetic angle $\hat\chi_p$ by
$2\pi$, moving the configuration from one winding well to its neighbor, i.e. the
lattice counterpart of a monopole-instanton of the Polyakov
mechanism~\cite{Polyakov:1975rs}. At weak coupling, the coefficients fall
exponentially, $t_{\bm{m}}\sim e^{-S(\bm{m})}$ with $S=4/g^{2}$ per unit of winding
on one plaquette, so $\Delta_{\rm tw}$ has a vanishing perturbative series and is
instanton-dominated. And since Eq.~\eqref{eq:Delta_tw_band} vanishes only at trivial
twist, every physical $q\bar{q}$ state carries a monopole-generated
$\Delta_{\rm tw}$. We develop the monopole reading and the instanton estimate in Appendix~\ref{app:monopole}.
\paragraph{One plaquette in closed form:}
Here $\HH^{(2)}=4$ and the Hamiltonian is a compact rotor
$2g^2\hat{\eta}^2+g^{-2}(1-\cos\hat{\chi})$, whose ground band is known exactly. With
$z=\chi/2$ the Schr\"odinger equation is the Mathieu equation at $q=-g^{-4}$, the two
sectors having characteristic exponents $\nu_{\rm M}=0$ and $\theta/\pi$. Writing
$a_{n,\nu_{\rm M}}(q)$ for the Mathieu characteristic value at exponent
$\nu_{\rm M}$, 
\begin{equation}
    \Delta_{\rm tw}(g) = \tfrac{g^2}{2}\Big[a_{0,\,\theta/\pi}(-g^{-4}) - a_{0,0}(-g^{-4})\Big].
    \label{eq:Dtw_mathieu}
\end{equation}
For one plaquette this is the Cooper-pair-box Hamiltonian with $E_J/E_C=2/g^{4}$,
and $\Delta_{\rm tw}$ is its charge dispersion~\cite{Koch:2007cmy}.
Equation~\eqref{eq:Dtw_mathieu} is the reference value against which we measure in Sec.~\ref{sec:prep}.

The action $S_0=4/g^2$ is that of a phase slip on a single plaquette, which is the minimal
finite-volume counterpart of compact monopole tunneling, and it is not the Polyakov
monopole action of the many-plaquette theory. One plaquette carries no monopole gas, no Wilson-loop area law,
no string tension and no dependence on charge separation. Therefore, it can only test the mechanism by
which a twist acquires an exponentially small energy, not confinement itself.
We use it in Sec.~\ref{sec:prep} as a quantitative benchmark for compact tunneling,
and treat the confinement observables as the proposed application for future work.

\section{Encoding compact rotors in oscillator Hilbert space}
\label{sec:encoding}

We left Sec.~\ref{sec:QED3} with $N^2$ compact rotors and a twist angle per
plaquette. We now ask how a bosonic mode can carry one of those rotors. The answer is that it cannot carry just one exactly, but it carries all of them at once, as a direct
integral over twisted fibers, and that a GKP stabilizer selects the one we want. In the
rest of this section we make that statement more precise and we address the price to pay for the physically realizable finite
energy encoding. Furthermore, what follows is frame agnostic: by
Appendix~\ref{app:gates} either choice of dynamical variables satisfies the same
algebra and has the same integer spectrum so everything below applies to both.
\subsection{Representing the compact algebra}

\begin{figure}[t]
    \centering
    \includegraphics[width=0.95\linewidth]{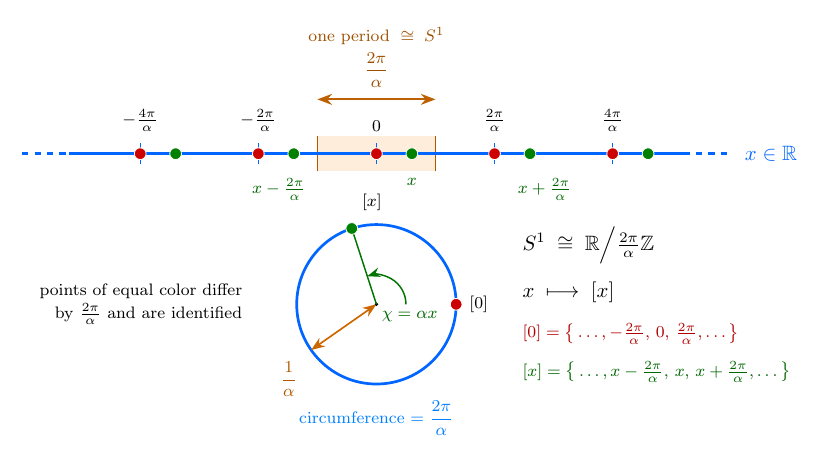}
    \caption{Wrapping the non-compact mode onto the rotor. The stabilizer
    Eq.~\eqref{eq:stab} enforces $2\pi/\alpha$-periodicity in $x$, identifying the
    line with the circle through $\chi=\alpha[x]$ and quantizing the conjugate
    quadrature to the grid $p\in\alpha\mathbb{Z}$, i.e.\ $\mathrm{Spec}(\hat\eta)=\mathbb{Z}$.}
    \label{fig:R_to_S1}
\end{figure}

A bosonic mode comes equipped with the Hilbert space $L^{2}(\mathbb{R})$ with
$[\hat{x},\hat{p}]=i$, where $\hat{x}=(\hat{a}+\hat{a}^\dagger)/\sqrt{2}$ and
$\hat{p}=(\hat{a}-\hat{a}^\dagger)/(i\sqrt{2})$. We write
$D(\beta)=e^{\beta\hat{a}^\dagger-\beta^*\hat{a}}$ for the displacement, so that
$e^{-is\hat{p}}=D(s/\sqrt{2})$ shifts $x$ by $s$ and $e^{is\hat{x}}=D(is/\sqrt{2})$
shifts $p$ by $s$, and
$S(r,\varphi)=\exp[\tfrac12(re^{-i\varphi}\hat{a}^2-re^{i\varphi}\hat{a}^{\dagger2})]$
for the squeezing gate, so that $\varphi=\pi$ squeezes $p$. For each
dynamical rotor $i$ we adopt the natural representation
\begin{equation}
  \hat{\chi}_{i}\mapsto\alpha_i\hat{x}_{i},
  \qquad
  \hat{\eta}_{i}\mapsto\frac{\hat{p}_{i}}{\alpha_i},
  \qquad
  \hat{U}_{i}\mapsto e^{i\alpha_i\hat{x}_{i}},
  \label{eq:CCR_rep_alpha}
\end{equation}
with $\alpha_i\neq0$ free. This satisfies the commutation relation in Eq.~\eqref{eq:rotor_algebra} for any
$\alpha_i$, but it does not satisfy compactness: on the line
$\mathrm{Spec}(\hat{p}/\alpha)\cong\mathrm{Spec}(\alpha\hat{x})\cong\mathbb{R}$,
whereas the compact theory needs $\mathrm{Spec}(\hat{\eta})\cong\mathbb{Z}$,
$\mathrm{Spec}(\hat{\chi})\cong[-\pi,\pi)$. The proposed map is not an isospectral representation. 
We find that the remedy for non-isospectrality is the single GKP-type stabilizer
\begin{equation}
  \hat{S}_{\alpha}=\exp\!\left(2\pi i\,\frac{\hat{p}}{\alpha}\right),
  \qquad
  \hat{S}_{\alpha}\ket{\psi}=\ket{\psi}
  \iff
  \braket{x+\tfrac{2\pi}{\alpha}|\psi}=\braket{x|\psi},
  \label{eq:stab}
\end{equation}
demanding $2\pi/\alpha$-periodicity in $x$, so $\chi\equiv\alpha[x]\in[-\pi,\pi)$ is
well defined and $p\in\alpha\mathbb{Z}$ (Fig.~\ref{fig:R_to_S1}).
Namely, the physical states must be in the $+1$-eigenspace of $\hat{S}_{\alpha}$. The latter is not a simple subspace of
$L^{2}(\mathbb{R})$, since a nonzero periodic function is never square-integrable on the
line, but below we provide its relation with the natural bosonic Hilbert space.
\paragraph{Zak--Bloch decomposition:}
\label{sec:zak}
We can always write any real momentum as $p=(n+\nu)\alpha$ with $n\in\mathbb{Z}$, $\nu\in[-\tfrac12,\tfrac12)$. With this decomposition, the
stabilizer becomes $\hat{S}_{\alpha}=e^{2\pi i\nu}$ and the Zak transform~\cite{PhysRevLett.19.1385} gives
\begin{equation}
  L^{2}(\mathbb{R})\cong\int_{-1/2}^{1/2}{}^{\!\!\oplus}\!\mathrm{d}\nu\;\mathcal{H}_{\nu},
  \qquad
  \mathcal{H}_{\nu}\cong\ell^{2}(\mathbb{Z})\cong L^{2}(S^{1}).
  \label{eq:fiber}
\end{equation}
That is, the Hilbert space of the hardware-given harmonic oscillator is a direct
integral over the fibers $\nu$, each a copy of the rotor's Hilbert space: in its
flux representation $\ell^{2}(\mathbb{Z})$ or, through the Fourier-series isomorphism,
its angle representation $L^{2}(S^{1})$. Both generators are block diagonal in $\nu$:
$\hat{\eta}=\hat{p}/\alpha$ is diagonal with spectrum $\{n+\nu\}$, while
$\hat{U}=e^{i\alpha\hat{x}}$ translates $p$ by $\alpha$ and so acts as $n\to n+1$ at
fixed $\nu$. Hence $[\hat{\eta},\hat{U}]=\hat{U}$ exactly. The same modular
decomposition splits a mode into a logical qubit and a gauge mode in the bosonic
subsystem codes of Ref.~\cite{Pantaleoni:2020nkh}, and compact configuration spaces
play the analogous role in the molecular codes of Ref.~\cite{Albert:2019ztu}. In our work,
the fiber itself is the simulated degree of freedom.
Three points follow. First, the
non-compact mode is not merely an approximation to a rotor, it is an exact direct integral of
rotors labeled by $\nu$, each carrying twisted boundary conditions with
$\theta=2\pi\nu$, matching Eq.~\eqref{eq:twisted_bc}, and Eq.~\eqref{eq:stab} selects
$\nu=0$. Second, any imperfect stabilization is therefore not generic leakage but a
distribution of background fluxes: since $\hat{H}'_{E}$ is quadratic in $\hat{\eta}$, a
fixed $\nu$ acts exactly like the displacement $\hat\eta_i\to\hat\eta_i+d_i$ of
Sec.~\ref{sec:displacement}. Third, the physical fiber is a property of the
variables chosen, not of the charges: in the undisplaced variables it is always
$\nu=0$, whatever the charge content, while after the displacement of
Sec.~\ref{sec:displacement}, the option taken for static charges, $\hat\eta_i$ has
spectrum $\mathbb{Z}-d_i$, so the same physical states sit on the fiber
$\nu_i=-d_i$ with $\theta_i=-2\pi d_i$. Which stabilizer to impose thus depends on whether we work with the displaced or undisplaced Hamiltonian, see Sec.~\ref{sec:penalty}.
Separately, $\alpha$ is only a coordinate choice. Any physical observable is independent of it and later we always choose the symmetric choice corresponding to $\alpha = \sqrt{2\pi}$.
\subsection{Regularization I: states inside the rotor fiber}
\label{sec:regI}
Let us work, for simplicity, in the untwisted $\mathcal{H}_{0}\cong\ell^{2}(\mathbb{Z})$ with flux basis
$\{\ket{n}\}$, $\hat{\eta}\ket{n}=n\ket{n}$, $\hat{U}\ket{n}=\ket{n+1}$. Here,
regularization is not needed since any $\ket{\psi}=\sum_nc_n\ket{n}$ with
$\sum_n|c_n|^2<\infty$ is normalizable, and the flux basis is the
orthonormal computational basis of the encoded theory.
The only non-normalizable states are the angular eigenstates
\begin{equation}
  \ket{\chi}\equiv\sum_{n\in\mathbb{Z}}e^{-in\chi}\ket{n},
  \qquad
  \hat{U}\ket{\chi}=e^{i\chi}\ket{\chi},
  \qquad
  \braket{\chi'|\chi}=2\pi\,\delta_{2\pi}(\chi-\chi'),
  \label{eq:chi_basis_ideal}
\end{equation}
distributions on the circle $S^{1}$ for the same reason a coordinate state $\ket{x}$ is one on $\mathbb{R}$. Formally, they live in the rigged
extension and give $\Id=\tfrac{1}{2\pi}\int_{-\pi}^{\pi}\!\mathrm{d}\chi\,\ket{\chi}\!\bra{\chi}$.
Normalizable wavepackets are built from smearing the flux coefficients with a Gaussian
\begin{equation}
  \ket{\chi;\sigma}\equiv\mathcal{N}_{\sigma}\sum_{n}e^{-\sigma^{2}n^{2}/2}e^{-in\chi}\ket{n},
  \qquad
  \braket{\chi';\sigma|\chi;\sigma}=\mathcal{N}_{\sigma}^{2}\,
   \vartheta_{3}\!\left(\tfrac{\chi'-\chi}{2},e^{-\sigma^{2}}\right).
  \label{eq:chi_sigma}
\end{equation}
The envelope must depend on $n$ alone, so
that the overlap depends only on $\chi-\chi'$ and when $\sigma\to0$ we recover
Eq.~\eqref{eq:chi_basis_ideal}. Note that $\sigma$ is a choice of state, not an error, and it survives the ideal-hardware limit.
\subsection{Regularization II: encoding the fiber in $L^{2}(\mathbb{R})$}
\label{sec:regII}
The hardware-induced regularization concerns the encoding map
$\mathcal{E}_{0}\ket{n}=\ket{p=n\alpha}_{p}$, whose image consists of Dirac combs
outside $L^{2}(\mathbb{R})$. Before regularizing, we name those combs.
Applying $\mathcal{E}_0$ to the angle eigenstates of
Eq.~\eqref{eq:chi_basis_ideal} and using
$\hat{p}\ket{p=n\alpha}_p=n\alpha\ket{p=n\alpha}_p$,
\begin{align}
    \mathcal{E}_{0}\ket{\chi}
    &= \sum_{n\in\mathbb{Z}}e^{-in\chi}\ket{p=n\alpha}_p
    = e^{-i\chi\hat{p}/\alpha}\sum_{n\in\mathbb{Z}}\ket{p=n\alpha}_p
    = D\!\left(\frac{\chi}{\alpha\sqrt{2}}\right)\ket{\varnothing},\nonumber\\
    \ket{\varnothing}&\equiv\!\sum_{n\in\mathbb{Z}}\!\ket{p=n\alpha}_p ,
    \label{eq:chi_qunaught}
\end{align}
where $\ket{\varnothing}$ is the GKP qunaught state, the grid state that carries no
logical information. Every vector of the encoded angular basis is therefore a displaced
qunaught, the displacement being a rigid translation of the comb by $\chi/\alpha$ along
$x$, which by Eq.~\eqref{eq:CCR_rep_alpha} is the angle itself. The comb has period
$\alpha$ in $p$ and $2\pi/\alpha$ in $x$, so at the symmetric point
$\alpha=\sqrt{2\pi}$ it is the square qunaught. More explicitly, the comb is the frame that makes the quadrature compact, and it is the same for
every $\chi$, while the logical state of the rotor sits entirely in the coefficients
$c_n$ and never in the grid.

Since $\ket{\varnothing}$ is not normalizable, each tooth must acquire a finite width,
for which we use the standard GKP envelope~\cite{Gottesman:2000di,Royer:2020jva}. Writing
\begin{equation}
  \hat{n}_{\alpha}\equiv S^{\dagger}(r_{\alpha})\,\hat{n}\,S(r_{\alpha}),
  \qquad
  r_{\alpha}\equiv-\ln(\alpha/\sqrt{2\pi}),
  \label{eq:n_alpha}
\end{equation}
the symmetric GKP choice is
\begin{equation}
  \mathcal{E}_{\Delta}\equiv e^{-\Delta^{2}\hat{n}_{\alpha}}\mathcal{E}_{0},
  \qquad
  \ket{n;\Delta}\equiv\mathcal{E}_{\Delta}\ket{n},
  \qquad
  \ket{\chi;\sigma,\Delta}\equiv\mathcal{E}_{\Delta}\ket{\chi;\sigma},
  \label{eq:encoder}
\end{equation}
so that $\mathcal{E}_{\Delta}\ket{\chi}$ is the finite-energy displaced qunaught. 
Here $\Delta$ is set
by the achievable squeezing and is the sole encoding error, independent of $\sigma$,
with the limits taken as $\Delta\to0$ first and then $\sigma\to0$.

All overlaps follow from the Mehler kernel~\eqref{eq:Mehler_kernel}, evaluated in
Appendix~\ref{app:mehler}. In particular, neighboring
teeth are not fully orthogonal with their overlap being
Eq.~\eqref{eq:orthogonality}. Therefore, the encoder is an isometry only up to
$\mathcal{O}(e^{-\pi/2\Delta^{2}})$. Furthermore, the same envelope imposes a spurious flux
envelope $\|\ket{n;\Delta}\|^{2}\propto e^{-2\pi\Delta^{2}n^{2}}$, silently suppressing
$|n|\gtrsim n_{\max}\simeq1/(\Delta\sqrt{2\pi})$, a distortion, since in the symmetric GKP state the tooth width and the flux envelope
are locked together. An independent envelope is available in the two-parameter state
$\ket{n;\Delta,\kappa}$, in which $\kappa\to0$ at fixed $\Delta$ decouples the two.

\begin{figure}
    \centering
    \includegraphics[width=0.70\linewidth]{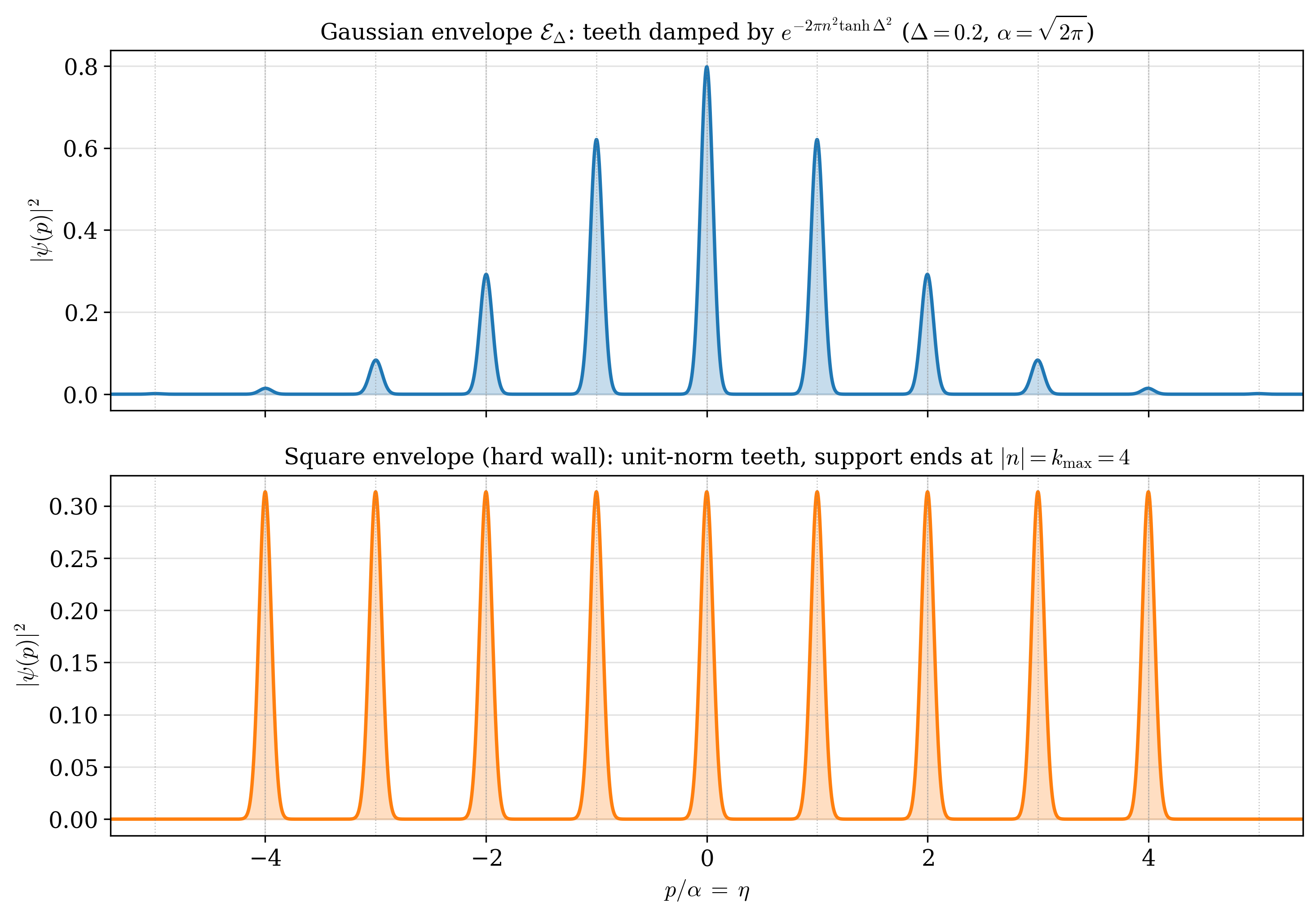}
    \caption{Momentum-space probability density of the non-ideal GKP-like qunaught
    state $\ket{\varnothing}$, which encodes the zero-angle state $\ket{\chi=0}$ of
    the compact rotor, at $\alpha=\sqrt{2\pi}$ and $\Delta=0.2$. The abscissa is
    $p/\alpha=\eta$, so the teeth sit at integer flux. Top panel: Gaussian envelope,
    Eq.~\eqref{eq:encoder}, whose teeth are damped by $e^{-2\pi n^2\tanh\Delta^2}$,
    the spurious flux envelope of Sec.~\ref{sec:err_classification}(C). Bottom panel:
    square envelope, Eq.~\eqref{eq:wall_state}, whose teeth are unit-norm and whose
    support ends abruptly at $|n|=k_{\rm max}=4$.}
    \label{fig:qunaught_density}
\end{figure}

In practice, no stabilized state is strictly normalizable, but momentum eigenstates are approximated
by squeezed vacua displaced in momentum,
\begin{equation}
    \lim_{r\to+\infty} D\!\Big(\frac{i\,k\alpha}{\sqrt{2}}\Big)S(r,\pi)\ket{n=0}
    = \ket{p=k\alpha}_p ,
    \label{eq:momentum_eigenstate}
\end{equation}
since $S(r,\pi)$ squeezes $\hat{p}$ to variance $e^{-2r}/2$ and
$D(is/\sqrt{2})=e^{is\hat{x}}$ shifts $p$ by $s$. Summing these under a hard flux cutoff
gives the case where the envelope becomes a square, a series of displaced squeezed vacua,
\begin{equation}
    \ket{\Psi_{\rm wall}} = \mathcal{N}\!\!\sum_{|n|\le k_{\rm max}}\!\! c_n\ket{\phi_n},
    \qquad
    \ket{\phi_n}\equiv D\!\Big(\tfrac{i n\alpha}{\sqrt2}\Big)S(r,\pi)\ket0,
    \qquad
    e^{-2r}=\frac{\alpha^2\Delta^2}{2\pi},
    \label{eq:wall_state}
\end{equation}
a square envelope (a top-hat window on $n$) replacing the Gaussian
$e^{-2\pi\Delta^2 \hat{n}^2}$ that $\mathcal{E}_\Delta$ imposes through the $\tanh$ term of
Eq.~\eqref{eq:Mehler_kernel}.
In Fig.~\ref{fig:qunaught_density} we show the probability density in momentum space of the GKP qunaught state $\ket{\varnothing}$ under these two encodings.

\subsection{Enforcing the code space}
\label{sec:penalty}
Under the map~\eqref{eq:CCR_rep_alpha}
the Hamiltonian~\eqref{eq:U1_ham_dynam} is built from $\hat{p}_i$, from
$\hat{p}_i\hat{p}_j$ and from cosines of integer combinations of $\alpha_i\hat{x}_i$,
all invariant under $x_i\to x_i+2\pi/\alpha_i$, so $[H,\hat{S}_{\alpha_i,i}]=0$ mode by
mode and a state supported on one fiber stays there under time evolution. That is, nothing needs to be enforced on the dynamics.  Had we been handed the
compact variable itself there would be nothing further to do, since the rotor Hilbert
space is a single fiber and the stabilizer is satisfied by construction. However, we are forced to work in $L^{2}(\mathbb{R})$, which carries every fiber at once. The only outstanding task is thus shifted to the state preparation.

We could prepare a state already concentrated on the intended fiber, which is
what the encoded momentum eigenstates of Eq.~\eqref{eq:momentum_eigenstate} do and what
the numerics of Sec.~\ref{sec:prep} use. Similarly, we can add an energetic term that
vanishes on the code space,
\begin{equation}
\hat{H}_{J} = -\frac{J}{2}\sum_{i} \Big[\big(\hat{S}_{\alpha_i,i}+\hat{S}_{\alpha_i,i}^\dagger\big)-2\Big]
 = -J\sum_{i}\cos\!\Big(\frac{2\pi \hat{p}_i}{\alpha_i}\Big) + N^2 J,
\label{eq:penalty}
\end{equation}
which, on a state with $p=(n+\nu)\alpha$, becomes $J(1-\cos2\pi\nu)\ge0$, exactly
zero at $\nu=0$ and at most $2J$ elsewhere. Practically, because $\nu$ is continuous that minimum is not isolated in $L^{2}(\mathbb{R})$, but $\hat{H}_J$ heavily penalizes all the fibers away from the intended one.

Leaving the model unstabilized costs more than accuracy: the twist is lost altogether. On the full line the shift
$\hat\eta_i\to\hat\eta_i+d_i$ is the genuine unitary
$\mathcal{U}=e^{-i d_i\alpha_i\hat{x}_i}$, so $\hat{H}[d]$ and $\hat{H}[0]$ are
isospectral and an unconstrained state relaxes to the band bottom of
Eq.~\eqref{eq:Delta_tw_band}, making $\Delta_{\rm tw}$ vanish identically.

The penalty term in Eq~\eqref{eq:penalty} is written for the untwisted sector. Once the displacement
of Sec.~\ref{sec:displacement} has been applied the physical fiber becomes $\nu_i=-d_i$, so
the penalty must be twisted to match, and the two options that will be used throughout the paper
are
\begin{align}
    \text{undisplaced:}\ &
    \hat{H}_E\ \text{of Eq.~\eqref{eq:U1_ham_dynam}},
    & \hat{H}_J &= -J\sum_i\cos\!\Big(\frac{2\pi\hat{p}_i}{\alpha_i}\Big) + N^2 J,
    \label{eq:undisplaced}\\
    \text{displaced:}\ &
    \hat{H}'_E\ \text{of Eq.~\eqref{eq:Hprime_E}},
    & \hat{H}'_J &= -J\sum_i\cos\!\Big(\frac{2\pi\hat{p}_i}{\alpha_i}-\hat{\theta}_i\Big) + N^2 J.
    \label{eq:displaced}
\end{align}
Pairing $\hat{H}'_E$ with the untwisted penalty would project onto the wrong sector and
set $\Delta_{\rm tw}$ to zero by construction. Which of the two is cheaper is a resource
question, that we settle in Sec.~\ref{sec:frames}.

\section{Finite-energy GKP encoding and error model}
\label{sec:errorcorrection}

The encoding of Sec.~\ref{sec:encoding} is exact in the limit $\Delta\to0$. At finite
squeezing, it deviates from the compact theory in ways fully characterized by the
single-tooth moments of the encoded momentum eigenstate, and those moments depend on
which envelope regularizes the comb. In this work we describe the effects of both the
Gaussian envelope (the Mehler comb $\ket{n;\Delta}=\mathcal{E}_\Delta\ket{n}$ of
Eq.~\eqref{eq:encoder}), which is the textbook finite-energy GKP
state~\cite{Gottesman:2000di,Royer:2020jva}, and the square
envelope (the displaced squeezed vacuum $\ket{\phi_n}$ of Eq.~\eqref{eq:wall_state}),
which the preparation of Eq.~\eqref{eq:momentum_eigenstate} builds and which we use for the numerics of Sec.~\ref{sec:testing}. Their moments are collected in
Table~\ref{tab:envelopes}. In a real simulation we would also encounter errors of a second kind like photon loss, dephasing, and imperfect gates. These are not
finite-$\Delta$ effects and do not vanish with better squeezing. We treat both kinds
here, since they share a single correction primitive.

\begin{table}[t]
\centering
\small
\caption{Single-tooth moments of the two finite-energy encodings, at flux label $n$
and grid spacing $\alpha$ (exact at $\alpha=\sqrt{2\pi}$, and to the quoted order
otherwise). The Gaussian envelope is the Mehler comb $\mathcal{E}_\Delta\ket{n}$ of
Eq.~\eqref{eq:encoder}. The square envelope is the displaced squeezed vacuum
$\ket{\phi_n}$ of Eq.~\eqref{eq:wall_state}. Both share the same momentum width to
leading order, but only the square envelope leaves the flux label and its parity
undistorted. At $n\neq0$ the Gaussian envelope adds a phase
$e^{-2\pi in(1-\mathrm{sech}\,\Delta^{2})}$ to $\braket{\hat S_\alpha}$, of order
$n\Delta^{4}$.}
\label{tab:envelopes}
\begin{tabular}{lcc}
\toprule
quantity & Gaussian envelope & square envelope \\
\midrule
$\braket{\hat{\eta}}$
  & $n\,\mathrm{sech}\,\Delta^{2}$
  & $n$ \\[2pt]
$\mathrm{Var}(\hat{\eta})=\delta\nu^{2}$
  & $\dfrac{\tanh\Delta^{2}}{4\pi}$
  & $\dfrac{\Delta^{2}}{4\pi}$ \\[6pt]
$\braket{\nu}$ (offset from grid)
  & $n(\mathrm{sech}\,\Delta^{2}-1)=-\tfrac12 n\Delta^{4}+\dots$
  & $0$ \\[4pt]
$|\braket{\hat{S}_{\alpha}}|$
  & $e^{-\pi\Delta^{2}/2}\,[1+\mathcal{O}(\Delta^{6})]$
  & $e^{-\pi\Delta^{2}/2}$ \\[2pt]
tooth overlap $\braket{n\!+\!\delta|n}$
  & $e^{-\pi\delta^{2}/\sinh 2\Delta^{2}}$
  & $e^{-\pi\delta^{2}/2\Delta^{2}}$ \\
\bottomrule
\end{tabular}
\end{table}

\subsection{Single-tooth moments and the energy bias}
\label{sec:err_budget}
A finite tooth is neither an exact $\hat{S}_\alpha$ eigenstate nor sharply localized
at $p=n\alpha$. Writing $\hat{p}_i=(n_i+\nu_i)\alpha_i$, the fractional part
$\nu_i\in[-\tfrac12,\tfrac12)$ is the twist offset of mode $i$, and the tooth is a
distribution $P(\bm{\nu})$ over $\nu_i$. Throughout the section, $\braket{\,\cdot\,}$ denotes the average over
this single-tooth distribution. Its first two moments are collected in Table~\ref{tab:envelopes}, with widths
\begin{equation}
  \delta\nu^{2} = \mathrm{Var}(\hat\eta) =
  \begin{cases}
    \tanh(\Delta^{2})/4\pi, & \text{Gaussian},\\[2pt]
    \Delta^{2}/4\pi, & \text{square},
  \end{cases}
  \qquad
  \delta\nu \xrightarrow[\Delta\ll1]{} \frac{\Delta}{2\sqrt{\pi}}\ \text{(both)}.
  \label{eq:dnu}
\end{equation}
However, the square envelope sits exactly
on the grid, $\braket{\hat\eta}=n$ and $\braket{\nu}=0$, whereas the Gaussian envelope
is pulled toward the origin,
\begin{equation}
  \braket{\hat\eta}_{n;\Delta} = \frac{n}{\cosh\Delta^{2}},
  \qquad
  \braket{\nu} = n\big(\mathrm{sech}\,\Delta^{2}-1\big) = -\tfrac12 n\Delta^{4}+\mathcal{O}(\Delta^{8}).
  \label{eq:mehler_moments}
\end{equation}
The contraction by $\mathrm{sech}\,\Delta^{2}$ is the same $\tanh$ term of the Mehler kernel~\eqref{eq:Mehler_kernel} that produces the flux envelope,
now acting on the tooth's center of mass. It vanishes at $n=0$ and grows with the flux
label. Correspondingly the stabilizer expectation is $e^{-\pi\Delta^2/2}$ for the
square envelope exactly, and its modulus for the Gaussian one up to
$\mathcal{O}(\Delta^6)$, with an $\mathcal{O}(n\Delta^{4})$ phase from the label
contraction.

Since $\hat{H}'_{E}$ is quadratic in $\hat{\eta}$, the offset acts as a coherent
background flux $\hat{\eta}_{i}\to\hat{\eta}_{i}+\nu_{i}$, a twist. Expanding about the integer-rotor value,
\begin{equation}
  \delta E = \frac{g^{2}}{2}\Big[\,
    \underbrace{2\,n_i\,\HH^{(2)}_{ij}\,\braket{\nu_j}}_{\text{label shift}}
    + \underbrace{\HH^{(2)}_{ij}\,\big(\braket{\nu_i}\braket{\nu_j}+\mathrm{Cov}(\nu_i,\nu_j)\big)}_{\text{blur}}
    \Big].
  \label{eq:dE_shot}
\end{equation}
For the square envelope $\braket{\nu}=0$, the label shift vanishes, and only the blur
survives with $\mathrm{Cov}=\delta\nu^2\Id$ for independently prepared teeth, giving
\begin{equation}
  \braket{\delta E}_{\rm square}
  = \frac{g^{2}}{2}\,\delta\nu^{2}\,\mathrm{tr}\,\HH^{(2)}
  = \frac{g^{2}\Delta^{2}}{8\pi}\,\mathrm{tr}\,\HH^{(2)} .
  \label{eq:dE_trace}
\end{equation}
This is a pure $\mathcal{O}(\Delta^{2})$ shift, independent of the flux content of the
state, and we refer back to it in Secs.~\ref{sec:err_scope} and~\ref{sec:scaling}. For the
Gaussian envelope the label shift does not vanish, and using
Eq.~\eqref{eq:mehler_moments},
\begin{equation}
  \braket{\delta E}_{\rm Gauss}
  = \frac{g^{2}\Delta^{2}}{8\pi}\,\mathrm{tr}\,\HH^{(2)}
  - \frac{g^{2}\Delta^{4}}{2}\,\big\langle n_i\,\HH^{(2)}_{ij}\,n_j\big\rangle
  + \mathcal{O}(\Delta^{6}),
  \label{eq:dE_trace_gauss}
\end{equation}
carrying an extra $\mathcal{O}(\Delta^{4})$ piece that grows with the populated flux
and is negative, since $\HH^{(2)}$ is positive definite and the contracted labels lower
the electric energy. The contraction runs over the full quadratic form, so the
off-diagonal entries of $\HH^{(2)}$ contribute as soon as more than one plaquette is
present and the flux labels are correlated across modes, while on one plaquette
$\HH^{(2)}$ reduces to the single number $4$. The square envelope is the cleaner of the
two, its bias being a single known $\mathcal{O}(\Delta^2)$ number, and the
$\mathcal{O}(\Delta^{4})$ term is specific to the symmetric comb, in which the tooth
width and the flux envelope share one parameter, being absent for any grid-centered
alternative (Sec.~\ref{sec:err_classification}).

The bias is the one figure of merit here that is not frame-agnostic, since the GKP
error is pinned to each physical mode through Eq.~\eqref{eq:dnu}, so the same hardware
$\Delta$ feeds an offset vector into whichever quadratic form the frame supplies. The
two frame values and their ratio are given in Sec.~\ref{sec:frames}. 

\subsection{Classification of errors}
\label{sec:err_classification}
The moments above sort into three finite-$\Delta$ errors that differ in whether
feedback removes them.

\paragraph{(A) Stabilizer violation (correctable):} A finite tooth overlaps its
neighbors (last row of Table~\ref{tab:envelopes}) and equivalently carries a small
random momentum displacement $\delta$ off the grid $\alpha\mathbb{Z}$. This is a
coherent, detectable error: $\delta$ commutes with the logical phase $\hat\chi$ and can
be measured and removed without disturbing the encoded
state~\cite{Gottesman:2000di} (Sec.~\ref{sec:err_active}).
It is common to both envelopes, differing only in the overlap width of
Table~\ref{tab:envelopes}. We can detect whether the circle is ``unrolling" to the real line and correct it back to the intended topology. 

\paragraph{(B) Twist spread (mitigable, not correctable):} Even on the grid, the tooth
occupies a band of fibers of width $\delta\nu$, Eq.~\eqref{eq:dnu}, a coherent
background flux $\hat\eta_i\to\hat\eta_i+\nu_i$. It is not removable by a grid
measurement, since it lives within a cell, but being a known deformation it can be
subtracted or extrapolated (Sec.~\ref{sec:err_active}). It is the origin of the
energy bias~\eqref{eq:dE_trace}. Because $\hat\nu$ commutes with the Hamiltonian the
distribution over fibers is itself conserved, so (B) is static and on a frequency observable it appears as a lineshape and not as
a shift. We work this out in Sec.~\ref{sec:fiber_dist}.

\paragraph{(C) Envelope distortion (avoidable by design):} In the symmetric comb one
parameter does two jobs, the $\tanh$ term of Eq.~\eqref{eq:Mehler_kernel} both
suppressing flux sectors $|n|\gtrsim n_{\max}=1/(\Delta\sqrt{2\pi})$ and, through the
same term, contracting the flux label itself, Eq.~\eqref{eq:mehler_moments}, which
gives the nonzero $\braket{\nu}$ and the $\mathcal{O}(\Delta^{4})$ tail of
Eq.~\eqref{eq:dE_trace_gauss}. This is a faithfulness limit of the prepared state, not an operator error, and it is cured by centering the teeth, whatever the shape of the envelope. Any state whose teeth are unit-norm, grid-centered and parity-symmetric
has $\braket{\nu}=0$, so the label shift of Eq.~\eqref{eq:dE_shot} vanishes and only
(A) and (B) remain. The square envelope of Eq.~\eqref{eq:wall_state} is one such state
and is the one assumed by Eq.~\eqref{eq:dE_trace}, at the price of a hard cutoff at
$|n|=k_{\max}$ that must be chosen wide enough to hold the flux content of the state.
The two-parameter state $\ket{n;\Delta,\kappa}$ of Sec.~\ref{sec:regII} is another,
since the factor $e^{-\kappa^{2}(n\alpha)^{2}/2}$ reweights the teeth without
displacing them and $\kappa\to0$ at fixed $\Delta$ decouples the envelope from the
tooth width. Only the symmetric comb, in which the two are locked together, is obliged
instead to keep $\Delta\ll1/(\sqrt{2\pi}\,\eta_{\max})$. Freedom from (C) is moreover a
property of the prepared state, because the deterministic
contraction under photon loss (Sec.~\ref{sec:err_loss}) grows with the flux label and afflicts
both envelopes equally, so what sustains $\braket{\nu}=0$ across an evolution is the
active correction of Sec.~\ref{sec:err_active}, which returns the teeth to the grid.

In one picture, (A) is a shift along the grid, detectable off-lattice, (B) a spread
within a cell, invisible to a lattice measurement but shifting energies, and (C) a
displacement of the teeth off the grid, present in the symmetric comb and absent
whenever the envelope is decoupled from the tooth width. The full budget, including the
physical and operational channels that do not vanish as $\Delta\to0$ and which the
rotor stabilizer alone does not address, is collected in Table~\ref{tab:qec_errors} of
Appendix~\ref{app:errors}.

\subsection{Action on \texorpdfstring{$\hat\eta$}{eta}- and \texorpdfstring{$\hat\chi$}{chi}-observables}
\label{sec:err_operators}
The finite-energy effects of the encoding reflect differently for the two conjugate classes of observable. Let
$\ket{\Psi}=\sum_n c_n\ket{n;\Delta}$ be a generic encoded state with support inside the
unclipped band. For a unit-norm tooth centered on the grid a local $f(\hat\eta)$ is
measured tooth by tooth,
\begin{equation}
    \braket{\Psi|f(\hat\eta)|\Psi} = \sum_n |c_n|^2\,f(n)
    + \tfrac12\,\delta\nu^2\sum_n |c_n|^2\,f''(n) + \mathcal{O}(\Delta^4),
    \label{eq:eta_obs}
\end{equation}
with $\delta\nu=\Delta/2\sqrt\pi$ the tooth width of Eq.~\eqref{eq:dnu}. The first
term is the exact rotor value, the second a spectral blur set by the curvature of
$f$, vanishing for linear $f$. For the Gaussian envelope the label contraction of
Eq.~\eqref{eq:mehler_moments} adds $-n^2\tanh^2\Delta^2$ to $f=\hat\eta^2$, the
operator-level face of the energy bias~\eqref{eq:dE_trace_gauss}, absent for the
square envelope.

Any $2\pi$-periodic $g(\{\hat\chi\})$ is a Fourier series in
$\hat U^k=e^{ik\hat\chi}$, and the Mehler overlaps computed in
Appendix~\ref{app:errors} collapse the sum to its diagonal term,
\begin{equation}
    \braket{\Psi|\hat U^k|\Psi} = \gamma_k\sum_n c_{n+k}^* c_n
    + \mathcal{O}\!\big(e^{-\pi/2\Delta^2}\big),
    \qquad
    \gamma_k = 1 - \frac{\pi}{8}\,k^2\Delta^6 + \mathcal{O}(\Delta^{10}),
    \label{eq:chi_obs}
\end{equation}
the collapse holding for $\Delta\lesssim0.5$, about $6$~dB of squeezing, and nearly
independently of the harmonic. The magnetic term is $\cos\hat\chi$ at $k=1$, damped
only at $\mathcal{O}(\Delta^{6})$, the mildest error of any observable in the
encoding. For the unit-norm teeth of Sec.~\ref{sec:testing_wall} the multiplicative factor is
exactly one and even this residual disappears (Appendix~\ref{app:wall}).

\subsection{The fiber distribution and the twist lineshape}
\label{sec:fiber_dist}
Error (B) is different from the others. Crucially, the fiber label is a constant of motion: since
$\hat{S}_\alpha=e^{2\pi i\hat\nu}$ with $\hat\nu=(\hat{p}/\alpha)\bmod1$, the
commutation $[\hat{H},\hat{S}_{\alpha_i,i}]=0$ of Sec.~\ref{sec:penalty} makes
$\hat\nu_i$ conserved, and it is conserved term by term.
The electric term is built from $\hat{p}_i$, the magnetic and kinetic terms from
cosines and link elements that translate $\hat{p}_i$ by integer multiples of
$\alpha_i$, and the mass term acts on the matter register alone. Every factor of a
product formula therefore commutes with $\hat{S}_\alpha$ separately, so the
conservation survives Trotterization exactly, at any order and any step size, and it
is unaffected by the presence of $\hat{H}_K$.

The two consequences below are the main structural argument for encoding a rotor in an oscillator at all.

\paragraph{(i) Finite squeezing is a reparametrization, not a decoherence channel:}
Coherences between different $\nu$ are inaccessible to the encoded rotor algebra of
Eq.~\eqref{eq:rotor_algebra}, generated by $\hat\eta_i$ and $\hat{U}_i$, which is
exactly the observable algebra of the compact theory. Within it $P(\bm{\nu})$ acts as a
classical weight over sectors that are individually exact, so a finite-$\Delta$
simulation is not an approximate simulation of one compact rotor but an exact
simulation of an ensemble of them, with a twist distribution the experimenter fixes at
preparation and can compute. Operators outside that algebra do connect fibers, and we discuss them in the rest of the section. A momentum displacement by a non-integer
multiple of $\alpha$ changes $\nu$ and is therefore visible to the syndrome of
Sec.~\ref{sec:err_active}, which is error (A), while a displacement by an integer
multiple is $\hat{U}^{k}$ and leaves the fiber untouched.

\paragraph{(ii) Squeezing and depth are independent budgets, and the syndrome is
non-demolition:} The distribution at the end of an evolution is the one prepared at
the start, so the encoding error does not accumulate with depth and time-evolving for longer times
never requires squeezing harder. Gate infidelity and photon loss are then the only
depth-limiting quantities. The stabilizer read out in Sec.~\ref{sec:err_active} is
the conserved quantity itself, so measuring it repeatedly cannot perturb the dynamics
under study. This is a stronger statement than the usual requirement that a syndrome commute
with the logical operators. Both statements are exact for the untruncated Hamiltonian
on $L^{2}(\mathbb{R})$. Photon loss does more than broaden $P(\bm\nu)$: its
deterministic part shifts each fiber in proportion to the flux label
(Sec.~\ref{sec:err_loss}), correlating $\nu$ with $n$ and dephasing flux coherences
until the correction of Sec.~\ref{sec:err_active} re-centers the comb. A finite Fock
truncation breaks $[\hat H,\hat S_\alpha]=0$ at the truncation edge, which the
penalty absorbs.

For an expectation value only the first two moments of $P$ enter, so in Sec.~\ref{sec:err_budget} we could compress (B) into the single number
$\braket{\delta E}$ of Eq.~\eqref{eq:dE_trace}. A frequency carries the whole
distribution, because each fiber propagates at its own rate. In a charge sector of
twist $\theta$ the fiber $\nu$ contributes at
\begin{equation}
    f(\nu) = \epsilon_0[\theta+2\pi\nu]-\epsilon_0[2\pi\nu],
    \label{eq:fiber_line}
\end{equation}
equal to $\Delta_{\rm tw}$ only at $\nu=0$. Expanding with the grid-centered moments
$\braket{\nu}=0$ and $\braket{\nu^{2}}=\delta\nu^{2}$ of Eq.~\eqref{eq:dnu}, and using
$\epsilon_0'[0]=0$ because $\bm{\theta}=\bm{0}$ is the band minimum,
\begin{equation}
    \braket{f} = \Delta_{\rm tw}
      + \tfrac12(2\pi\delta\nu)^{2}\big(\epsilon_0''[\theta]-\epsilon_0''[0]\big),
    \qquad
    \sigma_f = 2\pi\delta\nu\,\big|\epsilon_0'[\theta]\big| ,
    \label{eq:lineshape_moments}
\end{equation}
with $(2\pi\delta\nu)^{2}=\pi\Delta^{2}$ for either envelope. Keeping the lowest
harmonic of Eq.~\eqref{eq:band}, so that
$\epsilon_0[\theta]=\bar\epsilon-2t_1\cos\theta$ and
$\Delta_{\rm tw}=2t_1(1-\cos\theta)$, both moments become universal ratios,
\begin{equation}
    \frac{\braket{f}-\Delta_{\rm tw}}{\Delta_{\rm tw}} = -\frac{\pi\Delta^{2}}{2},
    \qquad
    \frac{\sigma_f}{\Delta_{\rm tw}} = \sqrt{\pi}\,\Delta\,\cot\frac{\theta}{2} .
    \label{eq:lineshape_ratios}
\end{equation}
The centroid shift is the same in every charge sector and is second order in
$\Delta$. The width depends on the sector through $\cot(\theta/2)$ and is first
order, so for this observable it is the larger of the two.
Both agree with the exact one-plaquette band at $g=0.8$: at $\theta=\pi/2$ and
$\Delta=0.2$ the shift is $-5.7\%$ against a predicted $-6.3\%$ and the width $35.1\%$
against $35.4\%$. At the band edge $\theta=\pi$ the leading width vanishes with
$\cot(\theta/2)$ and the residual is second order,
$\sigma_f/\Delta_{\rm tw}\simeq(\pi/\sqrt{2})\Delta^{2}$, which the exact band confirms
at $2.2\%$ for $\Delta=0.1$.

\paragraph{Dynamical matter:} Once the charges are dynamical the twist is no longer
ours to set, since $\hat{Q}$ is an operator and the state spreads over charge
configurations. The fiber distribution is untouched by this. $\hat\nu$ commutes
with the full Hamiltonian including $\hat{H}_K$, so $P(\bm{\nu})$ remains conserved
and fixed by preparation, and Eqs.~\eqref{eq:lineshape_moments}
and~\eqref{eq:lineshape_ratios} describe each charge sector separately. The measured line, however, becomes a weighted sum over the sectors a probe populates, so
the operating point is set by probe design instead of state preparation.

\paragraph{The fiber distribution is measurable:} $P(\bm{\nu})$ is fixed by
preparation, conserved, and measurable on the same hardware that runs the
simulation. The stabilizer powers are its Fourier coefficients, since
$\hat{S}_{\alpha}^{k}=e^{2\pi ik\hat\nu}$, so
\begin{equation}
    P(\nu) = \sum_{k\in\mathbb{Z}}\braket{\hat{S}_{\alpha}^{k}}\,e^{-2\pi ik\nu},
    \qquad
    \braket{\hat{S}_{\alpha}^{-k}}=\braket{\hat{S}_{\alpha}^{k}}^{*},
    \qquad
    \braket{\hat{S}_{\alpha}^{0}}=1,
    \label{eq:P_reconstruct}
\end{equation}
and truncating at $|k|\le K$ resolves $P$ on a scale $1/K$. Equivalently, the
analog syndrome of Sec.~\ref{sec:err_active} returns $\alpha\nu$ itself, so
repeated rounds histogram $P(\nu)$ directly, and since $\hat\nu$ is conserved they
can be taken during the run itself. For the grid-centered teeth of
Eq.~\eqref{eq:wall_state} the coefficients are Gaussian,
$\braket{\hat{S}_{\alpha}^{k}}=e^{-\pi k^{2}\Delta^{2}/2}$, summing to a wrapped
Gaussian of width $\delta\nu=\Delta/2\sqrt{\pi}$ and recovering
Eq.~\eqref{eq:dnu}. Any correlator measured on the encoded state is then the average
$\tilde{C}(t)=\int\!\mathrm{d}\nu\,P(\nu)\,C(t;\nu)$ of its fiber-resolved
counterpart over a verified distribution, each fiber being an exact compact
theory at twist $2\pi\nu$. When a single line of frequency $f(\nu)$ is isolated
from the rest of the signal, as the pair-addition spectroscopy of
Sec.~\ref{sec:pair_add} does for the twist line, its contribution to the kernel
is the phase $e^{-if(\nu)t}$ up to its spectral weight. The frequency is then
recovered by fitting that phase not just by locating a centroid, without bias
and at the cost of Fisher information alone, the residual shift of
Eq.~\eqref{eq:lineshape_ratios} falling to the extrapolation of
Sec.~\ref{sec:err_active}. 

\subsection{Physical noise: photon loss and dephasing}
\label{sec:err_loss}
Errors (A)--(C) are properties of the prepared state and are fixed once $\Delta$ is
fixed. A real simulation would add dynamical noise, of which the leading channel on
bosonic hardware is normally photon loss. Over a correction interval $\tau$ at loss rate
$\kappa$ a single oscillator undergoes the attenuation channel
\begin{equation}
    \hat a \longrightarrow \sqrt{T}\,\hat a + \sqrt{1-T}\,\hat e,
    \qquad
    T=e^{-\kappa\tau},
\end{equation}
with $\hat e$ an environmental vacuum mode. We write the transmittance as $T$ rather
than the customary $\eta$, which is reserved throughout this work for the electric
field of Eq.~\eqref{eq:rotor_algebra}. In quadratures
$\hat x\to\sqrt{T}\hat x+\xi_x$ and $\hat p\to\sqrt{T}\hat p+\xi_p$, so a tooth
prepared on the target fiber $\nu^\star$ at $p=\alpha(n+\nu^\star)$ acquires, relative
to its intended grid point, the momentum error
\begin{equation}
    \delta_n = \big(\sqrt{T}-1\big)\alpha(n+\nu^\star) + \xi_p .
    \label{eq:loss_displacement}
\end{equation}
(For the untwisted sector $\nu^\star=0$, and with static charges $\nu^\star=-d$, by the
third point of Sec.~\ref{sec:zak}.) Two features distinguish this from (A). The
stochastic piece $\xi_p$ merely adds to the tooth width already accounted for in
Eq.~\eqref{eq:dnu}. The deterministic piece $(\sqrt{T}-1)\alpha n$ is a phase-space
contraction that grows with the flux label, so the outermost occupied teeth leave the
grid first. It is the dynamical counterpart of the envelope contraction (C) of
Eq.~\eqref{eq:mehler_moments}, and unlike (C) it afflicts both envelopes equally.

Nearest-tooth correction requires $|\delta_n|<\alpha/2$ for every occupied $n$ with
high probability, which for $\kappa\tau\ll1$ and support on $|n|\le \eta_{\max}$ gives the
cadence condition
\begin{equation}
    \kappa\tau\, \eta_{\max} \ll 1 ,
    \label{eq:loss_cadence}
\end{equation}
up to the stochastic noise margin. Equation~\eqref{eq:loss_cadence} ties the allowed
correction interval to the loss rate and to the flux support the gauge-theory state
requires. If the known attenuation is first compensated, by a calibrated gain or by
tracking the contraction in a classical displacement frame, the residual channel is
approximately an additive Gaussian shift~\cite{Albert:2018reb,Noh:2018ogu} with
effective syndrome variance
\begin{equation}
    \sigma_{\rm eff}^2
    = \sigma_{\rm loss}^2 + \sigma_{\rm anc}^2 + \sigma_{\rm gate}^2 + \sigma_{\rm meas}^2 ,
    \label{eq:sigma_eff}
\end{equation}
the terms arising from photon loss, finite squeezing of the ancilla resource state,
imperfect syndrome gates, and homodyne readout.

Loss also injects noise into the conjugate quadrature. Since $\hat\chi=\alpha\hat x$ by
Eq.~\eqref{eq:CCR_rep_alpha}, that noise is a logical gauge-angle error rather than a
grid error, and the momentum-comb syndrome of Sec.~\ref{sec:err_active} is blind to it.
The same is true of intrinsic oscillator dephasing. This is taken up in
Sec.~\ref{sec:err_scope}.

\subsection{Correction and mitigation}
\label{sec:err_active}
Error (A), and the loss-induced displacement~\eqref{eq:loss_displacement}, are
correctable because the stray displacement $\delta$ is a syndrome that can be read out
without measuring the logical state. Two protocols realize this, both correcting any
$|\delta|<\alpha/2$ in a single shot and both leaving the encoded state untouched, and we give them in full in Appendix~\ref{app:errors}. The analog one entangles the mode with a
GKP qunaught ancilla through the momentum-sum gate
$\hat{U}_{p}=\exp(i\hat{p}_s\hat{x}_a)$ and reads the ancilla by
homodyne~\cite{Glancy:2005aac,Duivenvoorden:2016rdq}, returning
the syndrome modulo the target residue while learning nothing about the flux label. The
digital one applies $M$ controlled stabilizers to a $\ket{+}^{\otimes M}$ register and
recovers $\delta/\alpha$ as a binary fraction by phase
estimation~\cite{Kitaev:1995qy,Terhal:2015fcw}, to precision $\alpha/2^{M}$. In either case the
feedback displacement $e^{-i\varepsilon\hat{x}_s}$ re-centers the comb on the intended
fiber while preserving the logical $U(1)$ coherence. The analog protocol has one cost to record here: a finite-energy ancilla of squeezing $\Delta_a$ leaves
which-path information about the flux label, damping coherence between sectors
separated by $n$ as $e^{-\pi\Delta_a^{2}n^{2}}$ per round, so it is faithful only while
$\Delta_a\,\eta_{\max}\ll1$.

\paragraph{Repeated correction and failure probability:} Either protocol is interleaved
with the simulation, one cycle being a short block of Trotter layers, a loss
compensation step, syndrome extraction on each gauge mode against its own
target residue $r_i^\star=\alpha_i\nu_i^\star\bmod\alpha_i$, and a feedback displacement or an equivalent update of the classical
displacement frame. For Gaussian syndrome noise of variance $\sigma_{\rm eff}^2$ from
Eq.~\eqref{eq:sigma_eff}, the probability of assigning the state to the wrong tooth is
\begin{equation}
    P_{\rm fail} \simeq \operatorname{erfc}\!\left(\frac{\alpha}{2\sqrt{2}\,\sigma_{\rm eff}}\right),
    \label{eq:pfail}
\end{equation}
a wrong assignment being a logical electric-flux error $\ket{n}\to\ket{n\pm1}$ on the
affected rotor, and over $N^2$ gauge modes and $N_{\rm round}=t_{\rm phys}/\tau$ rounds
a union bound gives $P_{\rm logical}\lesssim N^2 N_{\rm round}P_{\rm fail}$. In
practice the continuous syndrome values should be retained and passed to an analog
decoder, since a syndrome near the edge of the Voronoi cell carries a larger
probability of a flux jump. Interleaving either protocol holds
$\braket{\hat{S}_\alpha}$ near unity over a run, converting the static squeezing floor
of (A) into a dynamically stabilized quantity, at the cost of a fresh ancilla per round.
The number of rounds a run can absorb is itself bounded, since each analog round damps
coherence across the flux support by
$\exp[-\pi N_{\rm round}\,\Delta_a^{2}\eta_{\max}^{2}]$, so the correction remains a
net gain only while
\begin{equation}
    N_{\rm round} \;\lesssim\; \frac{1}{\pi\,\Delta_a^{2}\,\eta_{\max}^{2}} ,
    \label{eq:round_budget}
\end{equation}
which together with the cadence condition~\eqref{eq:loss_cadence} brackets the
correction interval from both sides. The digital protocol replaces this particular
bound with the fidelity of the $M$-qubit register. The momentum-comb syndrome stabilizes the flux structure of the encoded rotor against displacement errors, and
statements about how long a full simulation survives require the remaining channels of
Table~\ref{tab:qec_errors} together with a noisy-channel simulation carrying all of
them, which we do not attempt here.

\paragraph{Passive mitigation:} Error (B) is not detectable by a grid measurement, but
it is a coherent, computable shift, so it can be removed in post-processing. Every finite-$\Delta$ bias is analytic in $\Delta^2$, the spectral
blur~\eqref{eq:eta_obs} and the energy bias~\eqref{eq:dE_trace} entering at leading
order $\Delta^{2}$ and the magnetic contrast loss only at $\Delta^{6}$, so running at
two or three squeezing levels and extrapolating $\Delta\to0$ cancels the leading term
and needs no ancilla. Where the bias is known in closed form it can simply be
subtracted, the blur as $\tfrac12\delta\nu^2\braket{f''}$ and a measured
$\braket{g(\hat\chi)}$ by rescaling with $\gamma_k^{-1}$ harmonic by harmonic. Both
require only classical knowledge of the tooth profile. An asymmetric tooth reinstates
the label-shift term of Eq.~\eqref{eq:dE_shot} and must be calibrated on a known
reference state before extrapolation, or it biases the $\Delta\to0$ limit.

\paragraph{Scope of the comb syndrome:}\label{sec:err_scope}
The syndrome above measures $p\bmod\alpha$ and therefore corrects exactly those
errors that move the comb off its target fiber. Three classes escape it, and because of them the present construction should be read as an encoding with a natural
bosonic-QEC interface and not as a complete threshold architecture.

\paragraph{Conjugate-quadrature errors:} Because $\hat\chi=\alpha\hat x$, noise in $x$
is a logical gauge-angle error affecting Wilson-loop and magnetic observables, and it
is invisible to a momentum-comb measurement without additional redundancy. Shorter
correction intervals, echo and calibration protocols, or concatenation with an outer
bosonic or qubit code are required. A stronger variant of the primitive itself is GKP
teleportation refresh~\cite{Walshe:2020vwp}, in which the logical rotor state is teleported into a freshly
prepared resource mode by Gaussian entangling gates, homodyne measurement and
feedforward. This extracts displacement syndromes and refreshes the finite-energy
envelope in one step, at higher resource cost, and is the more natural primitive for
long-depth simulation.

\paragraph{Finite-energy bias:} Error (B) lives inside a single Voronoi cell, where the
penalty~\eqref{eq:penalty} is quadratic and small, so neither the penalty nor the
syndrome removes it. It is a deterministic coherent bias and belongs to mitigation (Sec.~\ref{sec:err_active}). Its frame dependence
through $\mathrm{tr}\,\HH^{(2)}$, quantified in Sec.~\ref{sec:frames}, therefore has direct
error-correction relevance: at fixed squeezing the frame with the smaller electric-kernel
trace carries the smaller bias.

\paragraph{Matter sector and Gauss's law:} In the reduced variables of
Appendix~\ref{app:GL} the local constraints are solved before the oscillator encoding
is introduced, which removes gauge-violating degrees of freedom kinematically for
pure-gauge and static-charge simulations. With dynamical matter, qubit faults and
imperfect matter--gauge hopping gates can still leave the state inconsistent with the
intended charge sector, and the rotor stabilizer does not detect this. A scalable
implementation therefore needs encoded matter qubits and verified matter--gauge
operations. In an unreduced formulation one would instead enforce $G_x\ket{\psi}=0$ by
repeated gauge checks or a penalty $H_G=J_G\sum_x G_x^2$.

\section{Validation on one-plaquette compact QED$_3$}
\label{sec:prep}
In this section, we test the above construction on the smallest system: a 
one-plaquette system with four staggered sites and one bosonic mode. We fix the state preparation,
compare the compact encoded theory against the exact compact one, and finally use the
pair-addition correlator to expose the twist energy. The hardware protocols for the
observables used here are collected in Appendix~\ref{app:measurement}.

Two calculations of the same compact theory are compared throughout the section. 
We also write $\ket{\rm GS}$ for the interacting ground state of
$\hat{H}$ in the charge sector under discussion, with energy $E_0$, and reserve the symbol
$\Omega_s$ for the transition frequencies of Sec.~\ref{sec:pair_add}.

Diagonalizing Eq.~\eqref{eq:U1_ham_dynam} under the map~\eqref{eq:CCR_rep_alpha}
without the penalty gives the wrong spectrum. On the line the one-plaquette problem is
a Mathieu equation whose solutions are Bloch waves of quasimomentum $k$, so the
spectrum is a band $\mathcal{E}_n(k)$ not a set of levels, and the fiber label
is $\nu=k/\alpha$. The true compact theory admits no bands and requires $k=0$. The penalty $\hat{H}_{J}$ of Eq.~\eqref{eq:penalty} repairs exactly this, costing
$J[1-\cos(2\pi k/\alpha)]$ on the Bloch state of quasimomentum $k$ and so lifting the
whole band except its $k=0$ point. In Appendix~\ref{app:1plaq} we give the band explicitly and recover Eq.~\eqref{eq:Dtw_mathieu} from it.

Following Sec.~\ref{sec:frames}, for static charges we take
\begin{equation}
    \hat{H}'_E = \frac{g^2}{2}\frac{\hat{p}_i}{\alpha_i}\HH_{ij}^{(2)}\frac{\hat{p}_j}{\alpha_j} + E_{\rm cl}[Q],
    \quad
    \hat{H}_B = \frac{1}{g^2}\sum_p\big(1-\cos\hat{B}_p\big),
    \quad
    \hat{H}'_J\ \text{of Eq.~\eqref{eq:displaced}},
    \label{eq:final_static_ham}
\end{equation}
with c-number twists $\theta_i$ from Eq.~\eqref{eq:twisted_bc}. With dynamical
matter we use the undisplaced Eq.~\eqref{eq:undisplaced} and the untwisted
penalty~\eqref{eq:penalty}.
The algorithm must preserve the stabilizer. Every operator applied to a stabilized
input has to commute with $\hat{S}_{\alpha}$ of Eq.~\eqref{eq:stab} on every mode, so
that the state never leaves the intended fiber, and the states we use are the
squeezed-and-displaced approximations of Eq.~\eqref{eq:momentum_eigenstate}.

\subsection{Testing the encoding}
\label{sec:testing}
Every quantity below is a one- or two-point function of $\hat\eta$ or $\hat\chi$, and
on hardware each reduces to a Hadamard test with a single ancilla, one circuit for a
one-point function and four per time for the full complex correlator, with
Appendix~\ref{app:measurement} giving the protocols and the circuit count quoted in
Sec.~\ref{sec:scaling_meas}. Here we evaluate them directly, so that in what follows
we compare the two theories directly.

\begin{figure}[h!]
    \centering
    \includegraphics[width=0.49\linewidth]{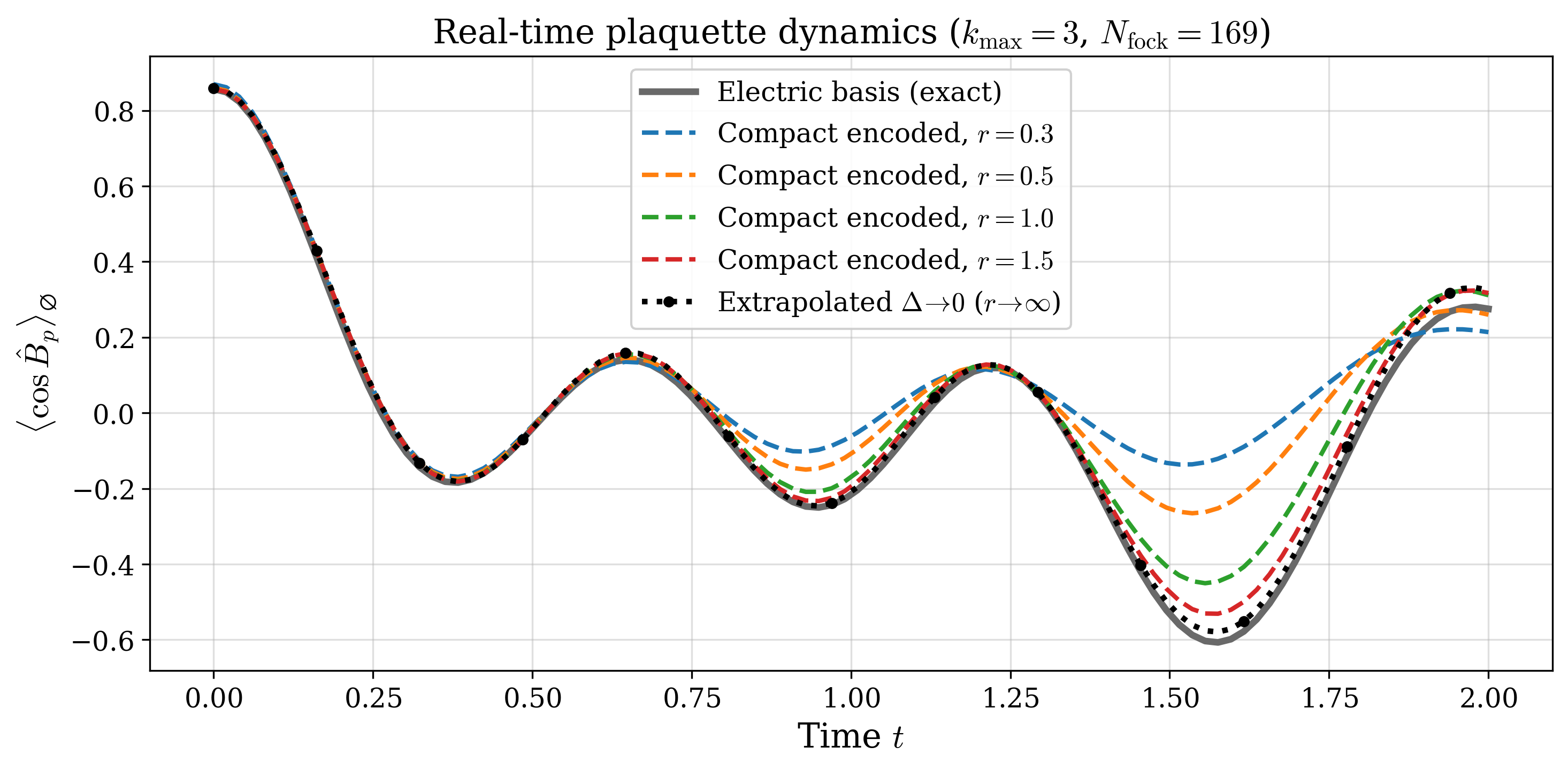}\hfill
    \includegraphics[width=0.49\linewidth]{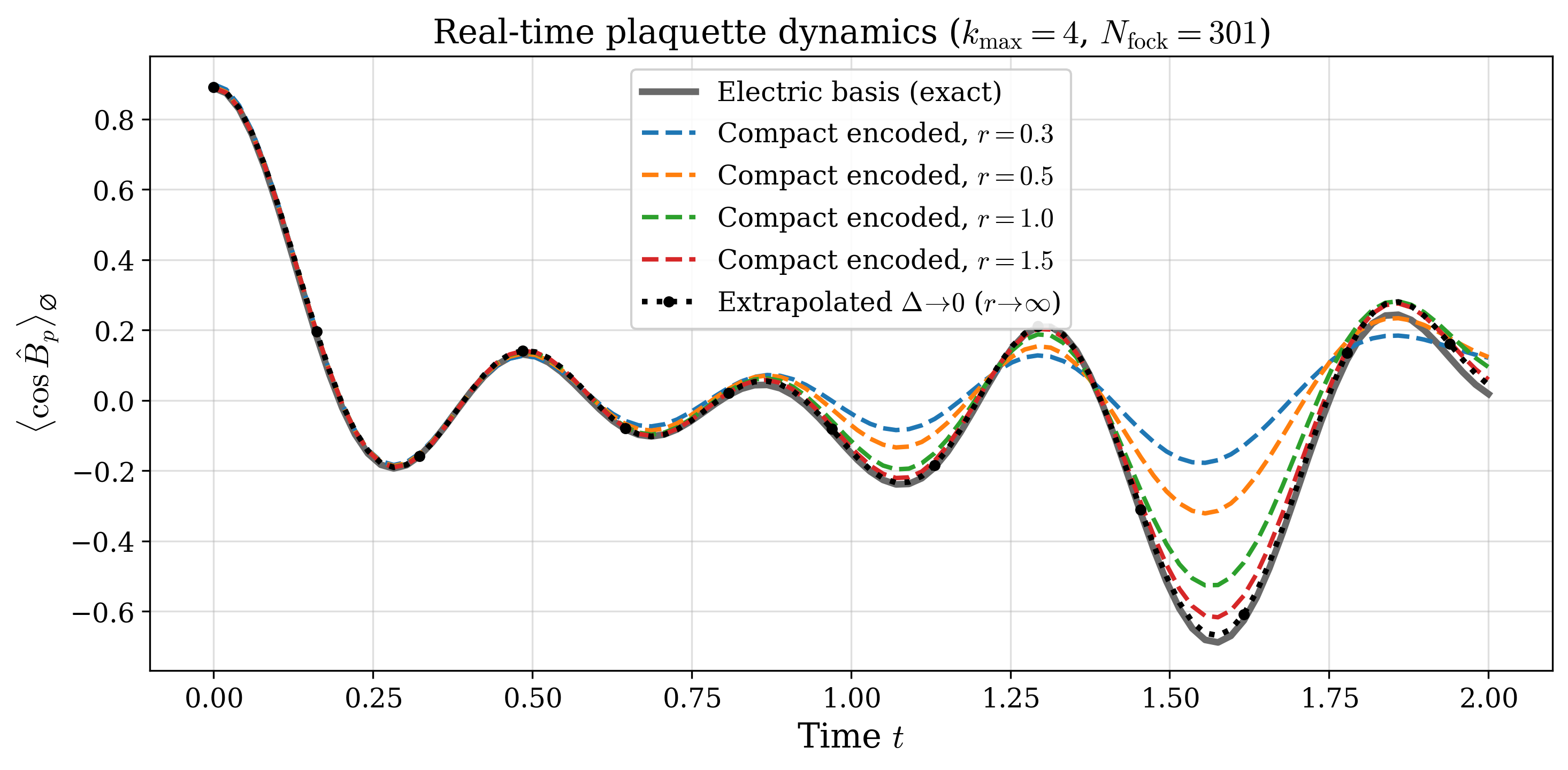}
    \caption{Plaquette expectation value for a single plaquette with no charges,
    comparing the compact theory $\bra{\chi=0}e^{itH}\tfrac{\hat U+\hat
    U^\dagger}{2}e^{-itH}\ket{\chi=0}$ with its encoded counterpart
    $\bra{\varnothing}e^{itH}\cos(\alpha\hat x)e^{-itH}\ket{\varnothing}$, using the
    hard-wall states of Eq.~\eqref{eq:test_states}. Each panel shows four squeezing
    levels, $r=0.3,0.5,1,1.5$, together with the extrapolation to $r\to\infty$
    ($\Delta\to0$). Left: $k_{\rm max}=3$, $N_{\rm fock}=169$. Right:
    $k_{\rm max}=4$, $N_{\rm fock}=301$. Here $g=1$ and $\alpha=\sqrt{2\pi}$.}
    \label{fig:plaq_time_GKP}
\end{figure}

\begin{figure}[h!]
    \centering
    \includegraphics[width=0.49\linewidth]{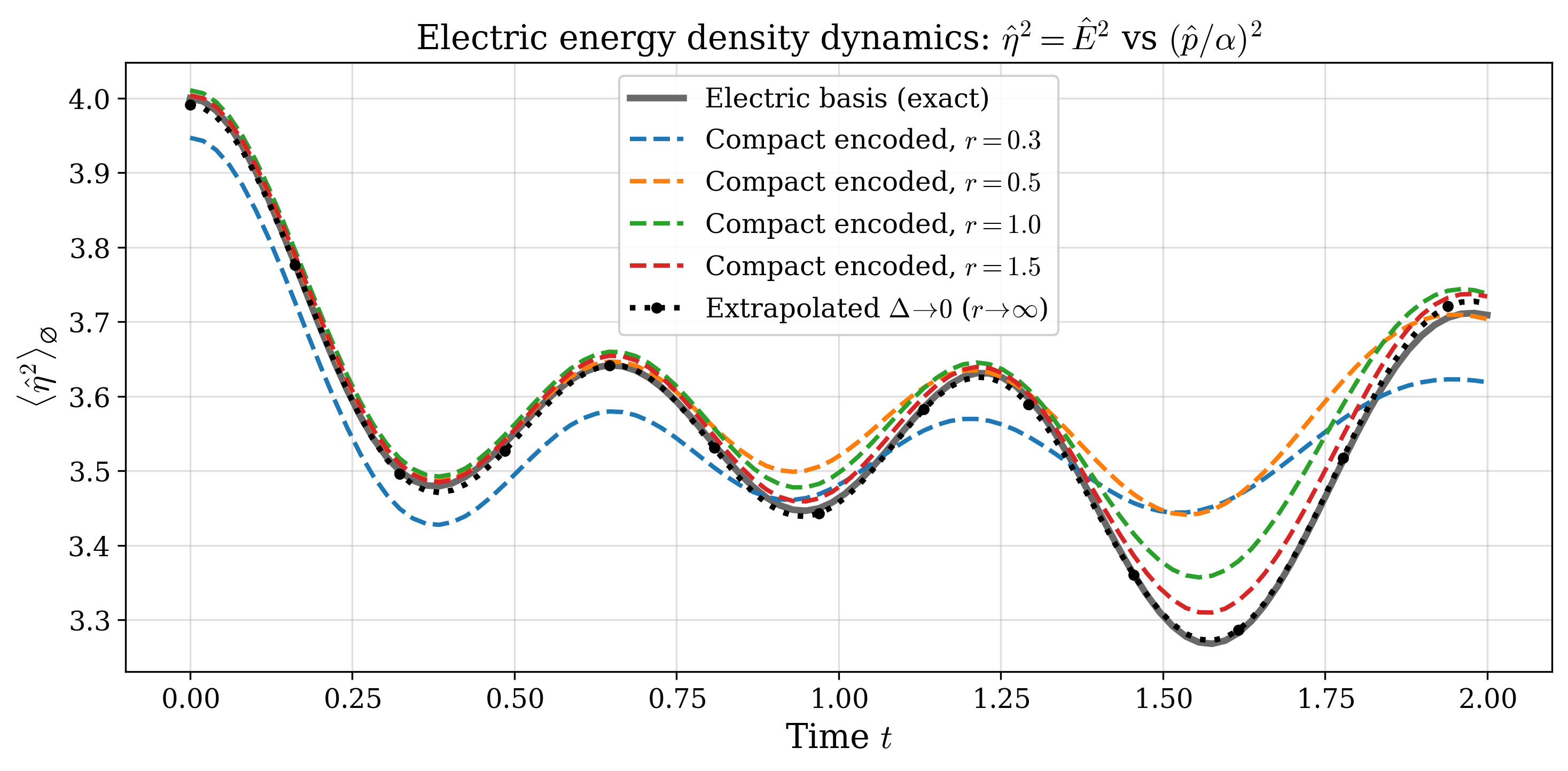}\hfill
    \includegraphics[width=0.49\linewidth]{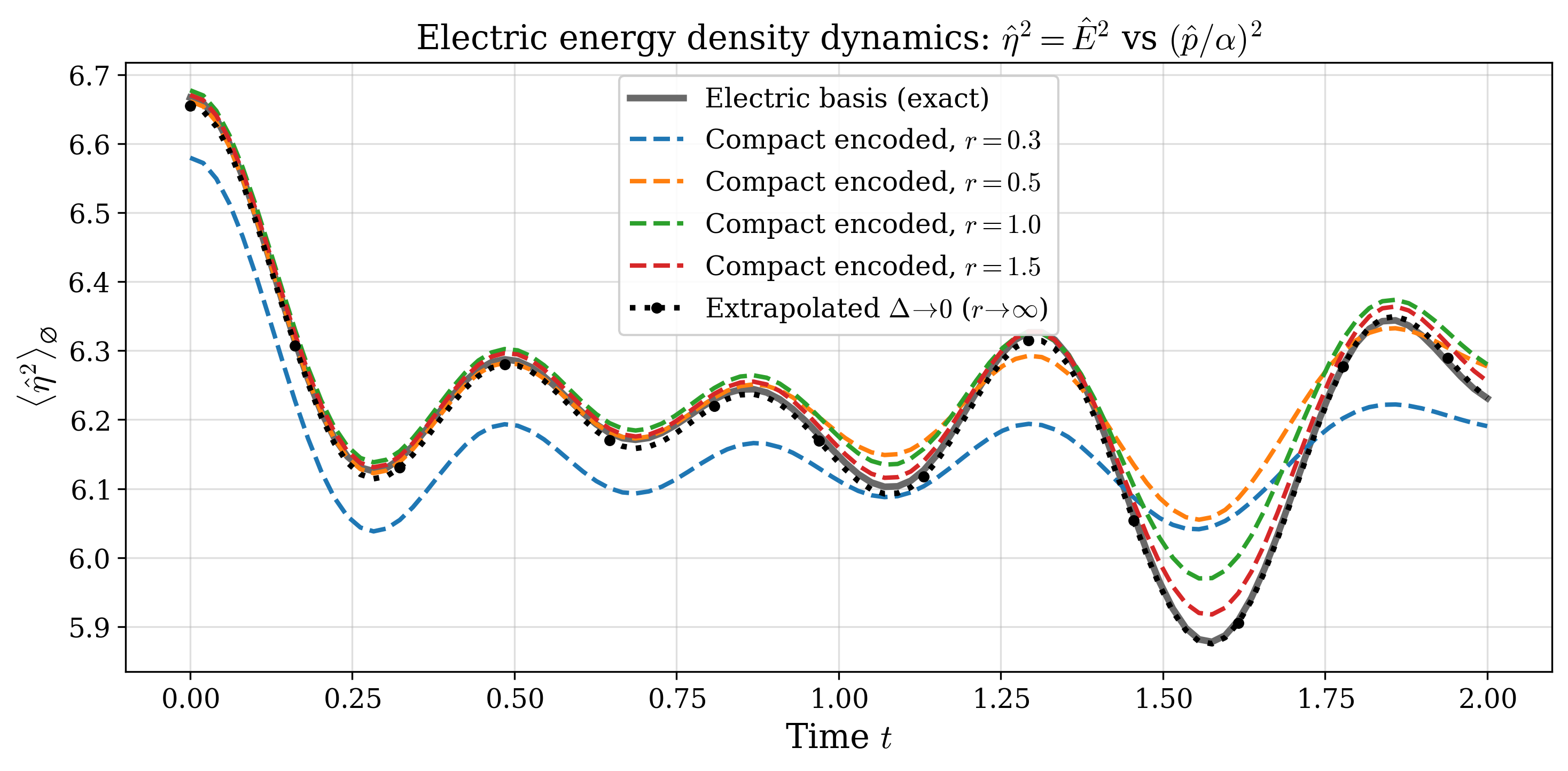}
    \caption{As Fig.~\ref{fig:plaq_time_GKP}, for the electric energy
    $\hat\eta^2\mapsto(\hat p/\alpha)^2$, at $r=0.3,0.5,1,1.5$ and extrapolated to
    $r\to\infty$ ($\Delta\to0$). Left: $k_{\rm max}=3$, $N_{\rm fock}=169$. Right:
    $k_{\rm max}=4$, $N_{\rm fock}=301$. Here $g=1$ and $\alpha=\sqrt{2\pi}$.}
    \label{fig:elec_ener_time_GKP}
\end{figure}

For a single plaquette with no charges we compare
$\bra{\varnothing}e^{itH}\cos(\alpha\hat{x})e^{-itH}\ket{\varnothing}$ with its
compact counterpart
$\bra{\chi=0}e^{itH}\tfrac{\hat{U}+\hat{U}^\dagger}{2}e^{-itH}\ket{\chi=0}$, the two
Hamiltonians being the encoded and compact theories. Using
Eq.~\eqref{eq:momentum_eigenstate} we approximate the ideal states by
\begin{equation}
        \ket{\chi = 0, k_{\rm max}} \equiv \!\!\sum_{k = -k_{\rm max}}^{k_{\rm max}}\!\! \ket{\eta = k},
        \qquad
        \ket{\varnothing, k_{\rm max}} \equiv \!\!\sum_{k = -k_{\rm max}}^{k_{\rm max}}\!\! D\!\Big(\frac{i\,k\alpha}{\sqrt{2}}\Big)S(r,\pi)\ket{0},
    \label{eq:test_states}
\end{equation}
normalized. These are the square-envelope states of Table~\ref{tab:envelopes},
so the finite-$\Delta$ budget in force is the right-hand column throughout this
section, and being prepared on the fiber directly these runs take the preparation
route with $J=0$. Figures~\ref{fig:plaq_time_GKP} and~\ref{fig:elec_ener_time_GKP}
show the plaquette and electric energy expectation value, respectively, at $g=1$
and $k_{\rm max}=3,4$, generated at $\alpha=\sqrt{2\pi}$. In each panel we overlay four
squeezing levels together with the $r\to\infty$ ($\Delta\to0$) extrapolation, and
the extrapolated curves land on the compact theory to within the truncation. This
is the operational content of Sec.~\ref{sec:errorcorrection} and a central point
of this work: the finite-energy encoding error is not noise but a controlled
systematic, a known function of the squeezing, so a handful of runs at accessible
$r$ removes it by extrapolation. The same logic returns in
Sec.~\ref{subsec:qed3_plaquette} for the damping bias in $\Gamma^{2}$ and the
fermionic dressing in $1/m_0$.
\paragraph{Hard-wall states and the exact magnetic contrast:}
\label{sec:testing_wall}
The states of Eq.~\eqref{eq:test_states} are the wall states of
Eq.~\eqref{eq:wall_state} not the Mehler-encoded teeth, and the preparation of Eq.~\eqref{eq:momentum_eigenstate} builds precisely these. Because each tooth
is a displaced squeezed vacuum undistorted by a post-displacement damping, the magnetic
contrast is exact up to the tooth overlap, Eq.~\eqref{eq:wall_contrast}, so
$\cos(\alpha\hat{x})$ contributes no polynomial error. The only finite-$\Delta$ errors
entering the comparison are the additive electric blur $\mathcal{O}(\Delta^2)$
propagated by $\hat{H}_E$ during the evolution and the neglected tail
$\sum_{|n|>k_{\rm max}}|c_n|^2$ of the hard cutoff, both removed by taking
$k_{\rm max}$ past the flux support, so the compact and encoded curves track to within the truncation. The derivation is in Appendix~\ref{app:wall}.

\begin{figure}
    \centering
    \includegraphics[width=0.49\linewidth]{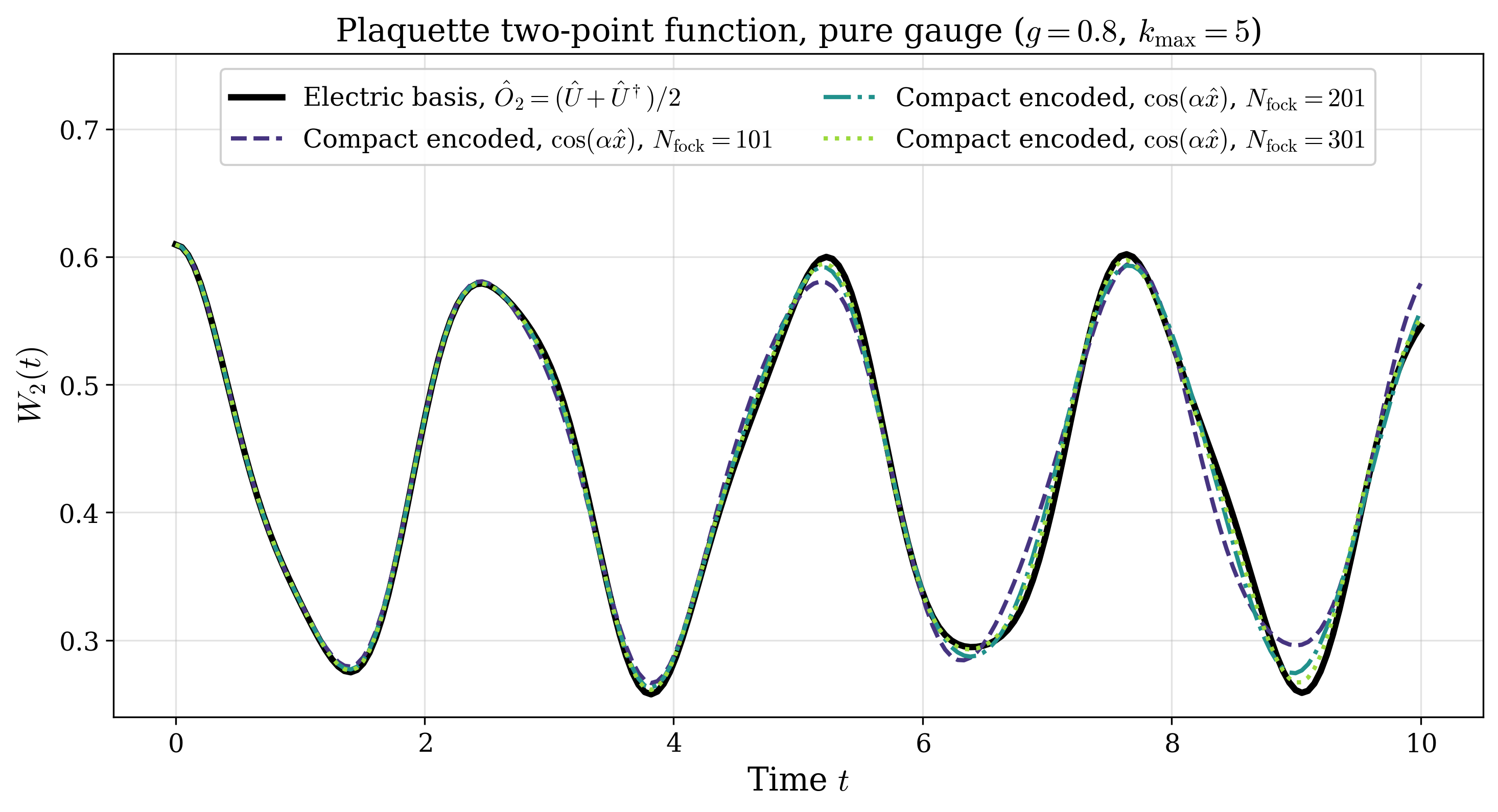}\hfill
    \includegraphics[width=0.49\linewidth]{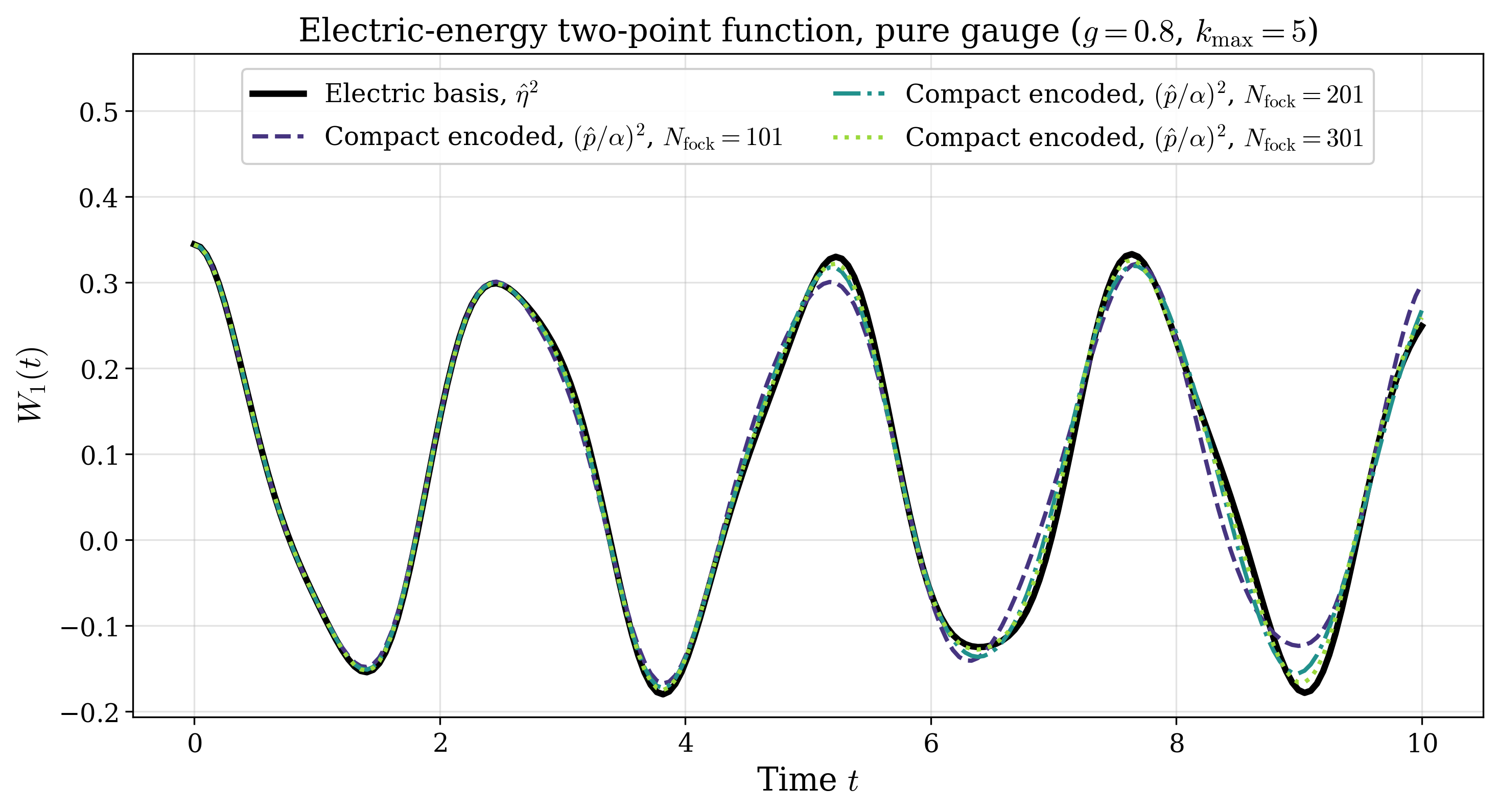}
    \caption{Real part of the ground-state two-point functions
    $W_{1,2}(t)=\bra{\rm GS}\hat{O}_{1,2}(t)\hat{O}_{1,2}(0)\ket{\rm GS}$ for one
    plaquette in the pure-gauge sector, with
    $\hat{O}_1=\hat\eta^2\mapsto(\hat p/\alpha)^2$ and
    $\hat{O}_2=(\hat U+\hat U^\dagger)/2\mapsto\cos(\alpha\hat x)$. Left: the magnetic
    correlator $\operatorname{Re}W_2$. Right: the electric correlator
    $\operatorname{Re}W_1$. Each panel overlays the electric-basis and compact-encoded
    results, which agree to within the truncation, and the two observables are
    related by Eq.~\eqref{eq:W1W2}. Here $g=0.8$ and $k_{\rm max}=5$.}
    \label{fig:two_point_time}
\end{figure}

\begin{figure}[h!]
    \centering
    \includegraphics[width=0.49\linewidth]{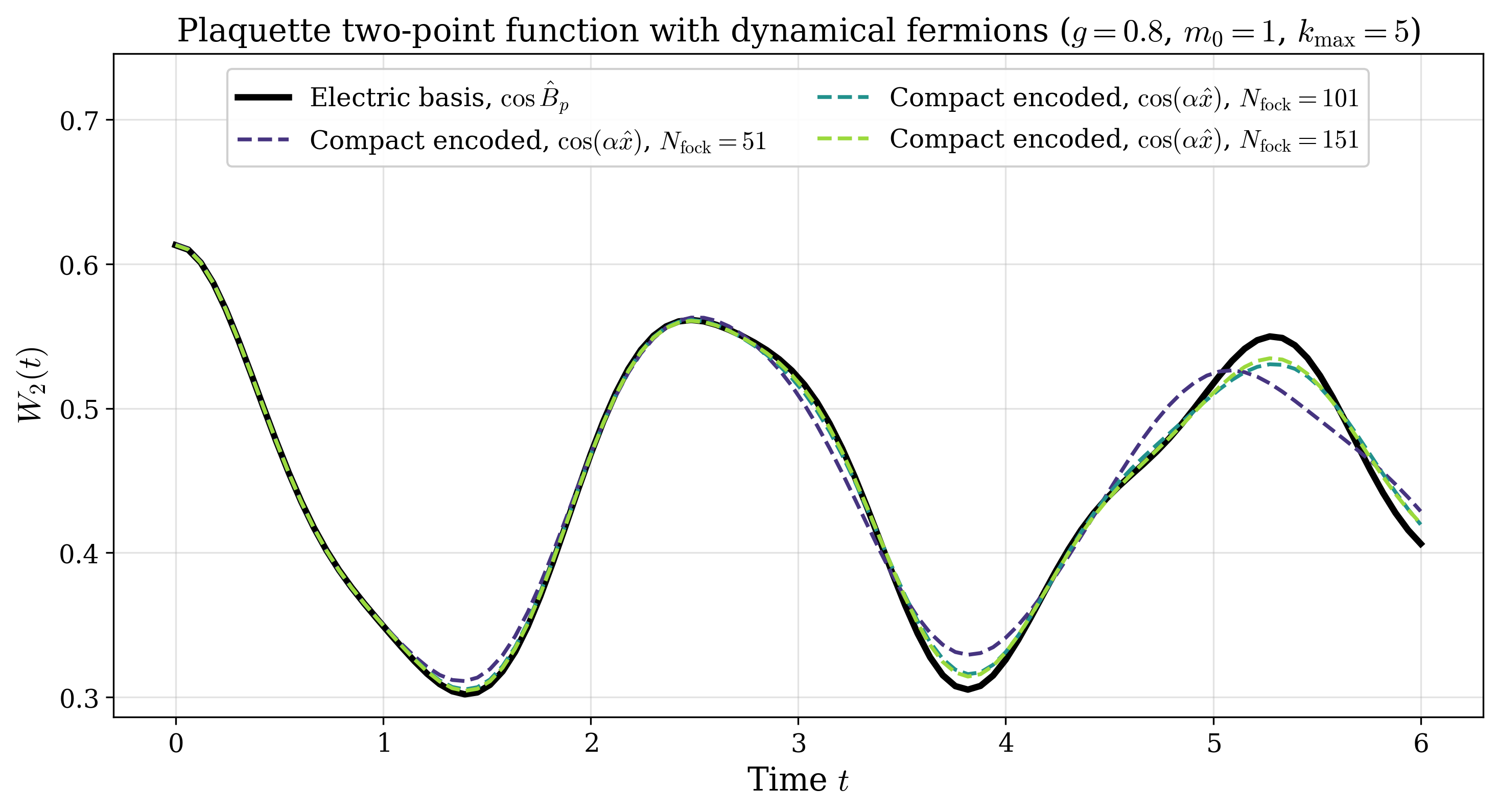}\hfill
    \includegraphics[width=0.49\linewidth]{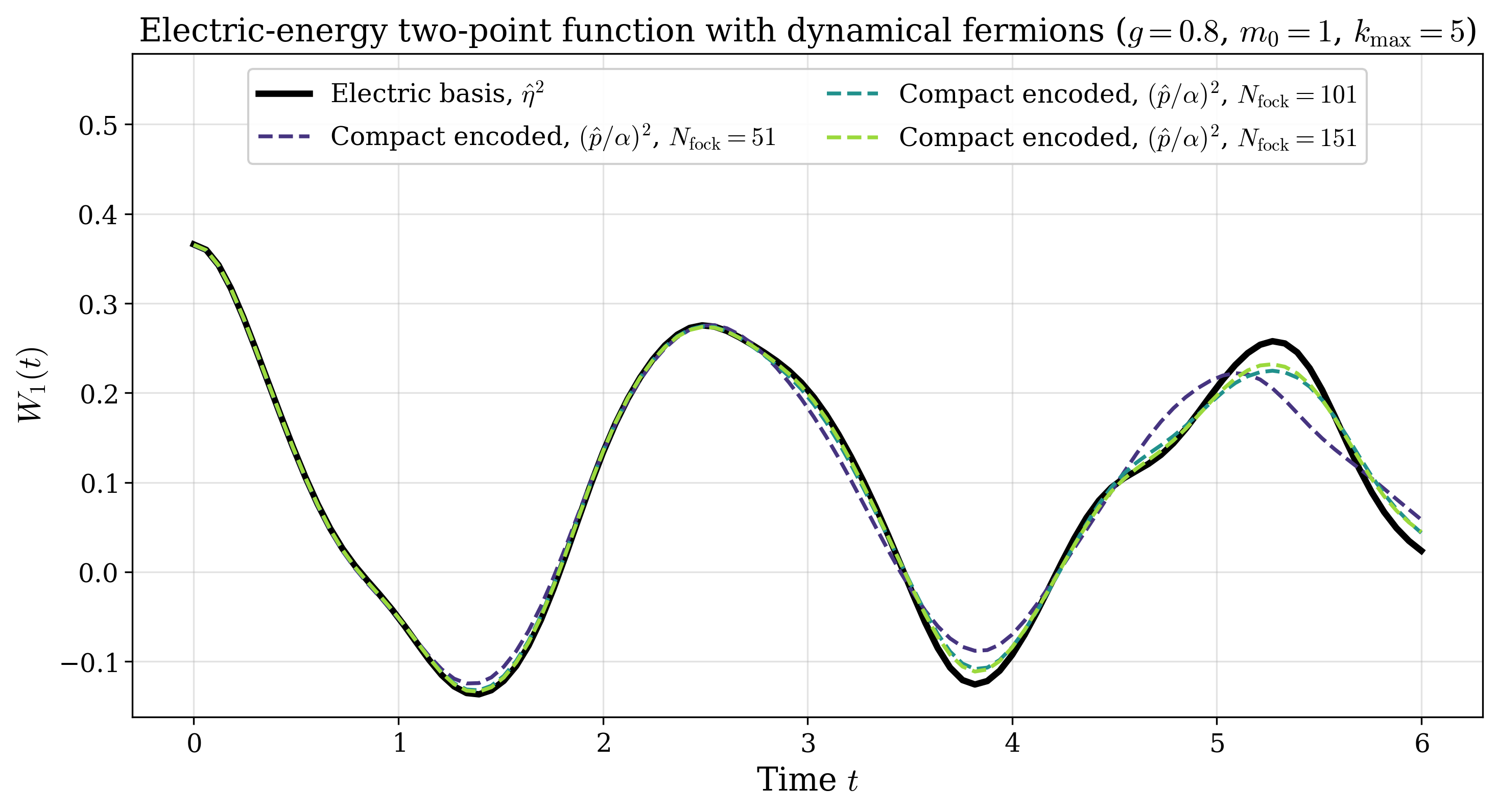}
    \caption{As Fig.~\ref{fig:two_point_time}, now with dynamical staggered fermions:
    real part of the ground-state two-point functions of the magnetic operator
    $\hat{O}_2$ (left) and the electric operator $\hat{O}_1$ (right). Electric-basis
    and compact-encoded results are overlaid, the latter at several $N_{\rm fock}$, so
    that the residual spread between compact curves measures the truncation error
    not a systematic difference between representations. The electric
    Hamiltonian is built from the physical charge $\hat{Q}_j$ of
    Eq.~\eqref{eq:charge_def}, not from the bare occupation. Here $g=0.8$, $m_0=1$
    and $k_{\rm max}=5$.}
    \label{fig:dyn_E2_corr_g0.8_m01_Nmax5_Nfock_overlay}
\end{figure}

\paragraph{Two-point functions:}
We next compute the ground state two-point functions $W_{1,2}(t)\equiv\bra{\rm GS}\hat{O}_{1,2}(t)\hat{O}_{1,2}(0)\ket{\rm GS}$
for $\hat{O}_1=\hat{\eta}^2\mapsto(\hat{p}/\alpha)^2$ and
$\hat{O}_2=(\hat{U}+\hat{U}^\dagger)/2\mapsto\cos(\alpha\hat{x})$. The two are
linearly dependent: from $\hat{O}_1=\tfrac{1}{2g^2}(\hat{H}+g^{-2}\hat{O}_2-g^{-2})$
acting on the stationary $\ket{\rm GS}$, together with
$\braket{\hat{O}_2(t)}=\braket{\hat{O}_2(0)}$,
\begin{equation}
   4g^8W_1(t) = W_2(t) + 2(g^2 E_0 - 1)\bra{\rm GS}\hat{O}_2\ket{\rm GS} + (g^2 E_0 - 1)^2,
   \label{eq:W1W2}
\end{equation}
confirmed in Fig.~\ref{fig:two_point_time} along with the equivalence of the two
representations. Figure~\ref{fig:dyn_E2_corr_g0.8_m01_Nmax5_Nfock_overlay} repeats
this with dynamical fermions.

\subsection{Pair-addition real-time spectroscopy}
\label{sec:pair_add}
We now put the displaced Hamiltonian of Sec.~\ref{sec:displacement} to work. The
observable is the response to the sudden creation of a gauge-invariant
particle--antiparticle pair, whose lowest line isolates $\Delta_{\rm tw}$ and hence,
by Sec.~\ref{sec:twist_energy}, the monopole physics.

For the smallest system, the pair-addition operator on the bottom link is
\begin{equation}
    \hat{O}_b = \hat\Psi^\dagger_{(0,0)}\,\hat{u}^{\dagger}_{(\bm{0},\bm{e}_x)}\,\hat\Psi_{(1,0)},
    \label{eq:Ob}
\end{equation}
where $\hat{u}^\dagger$ lowers the flux by one unit (Eq.~\eqref{eq:rotor_algebra}:
it is $\hat{U}$ that raises), so that $\hat{O}_b\ket{\rm GS}$ respects Gauss's law.
At $N=1$ the frames coincide and $\hat{u}_{(\bm{0},\bm{e}_x)}=\hat{U}_{(0,0)}$. The
induced charge configuration is $Q_{(0,0)}=+1$, $Q_{(1,0)}=-1$, i.e.\ in the snake
ordering is neutral
\begin{equation}
    Q = \big(Q_{(0,0)},Q_{(1,0)},Q_{(1,1)},Q_{(0,1)}\big) = (1,-1,0,0).
    \label{eq:Q_bottom}
\end{equation}
Equation~\eqref{eq:d_1plaq} gives $d=3/4$ and
$\theta\equiv+\pi/2$.
Acting with $\hat{O}_b$ places the gauge sector in a twisted sector, and applying the
displacement of Eq.~\eqref{eq:displacement_op} at $\hat{Q}\to Q$ leaves the state
quasi-periodic, $\psi_{\rm tw}(\chi+2\pi)=e^{-2\pi id}\psi_{\rm tw}(\chi)$, which in
the CV encoding is implemented by the twisted penalty $\hat{H}'_J$ of
Eq.~\eqref{eq:final_static_ham}. The lowest line therefore decomposes as
\begin{equation}
    \Omega_0 = 2m_0 + E_{\rm cl}[Q] + \Delta_{\rm tw}
    + \delta_{\rm dyn}(m_0,\kappa,N),
    \label{eq:Omega0}
\end{equation}
with a dynamical term
$\delta_{\rm dyn}$ that vanishes for static charges, $\kappa\in[0,1]$ being the
strength of the pair-addition hopping channel used below, distinct from the loss
rate of Sec.~\ref{sec:err_loss} and the envelope parameter of
Sec.~\ref{sec:regII}. Its origin is the one identified in
Sec.~\ref{sec:displacement}. With dynamical fermions $\hat{Q}$ is an operator, so
$d\to\hat{d}$ and the displacement becomes a controlled one entangling gauge and matter,
which dresses the pair created by $\hat{O}_b$ instead of leaving it in a sharp twisted
sector. The first three terms are exact in the pure-gauge case. At finite mass and
hopping the lowest line does not isolate $\Delta_{\rm tw}$ on its own, and only a
controlled $\kappa\to0$ and $1/m_0\to0$ extrapolation removes $\delta_{\rm dyn}$, which we carry out in Sec.~\ref{subsec:qed3_plaquette}, with the leading $1/m_0$ form derived in Appendix~\ref{app:dressing}.

\paragraph{Correlator and demodulation:}
The time-domain correlator of $\hat{O}_b$, written out with its spectral function in
Appendix~\ref{app:demod}, is a sum $\sum_s|Z_s|^2e^{-i\Omega_s t}$ over transitions
and concentrates its weight on $\Omega_0$ of
Eq.~\eqref{eq:Omega0}, whose first two terms $2m_0$ and $E_{\rm cl}$ are classically
known from the Gauss-law matrices. Multiplying by $e^{i(2m_0+E_{\rm cl})t}$ re-centers
$\Omega_0$ near $\omega=0$ and exposes $\Delta_{\rm tw}$ as a resolvable peak instead of
a tiny offset on top of $\sim2m_0$, exactly as one demodulates a Ramsey signal by a
known local oscillator. In Appendix~\ref{app:demod} we give the spectral function and the weights, and validate the pipeline against the exact twist energy.
\subsection{QED\texorpdfstring{$_3$}{3} on one plaquette}
\label{subsec:qed3_plaquette}
We now take the full $H=\hat{H}_E+\hat{H}_B+\hat{H}_M+\hat{H}_K$ with four dynamical
staggered sites and all kinetic terms at physical strength: the three spectator
channels, the two $y$-hops and the top $x$-hop, all gauge-fixed to unity, are
comparable to the pair-creation term.
We compute $C_b(t)$ in the electric basis, with the gauge space truncated to
$|\eta|\le k_{\rm max}$ and Gauss's law exact, and in the compact encoded basis, with
$N_{\rm fock}=101$ Fock states and flux quantization enforced softly by
Eq.~\eqref{eq:penalty} at $J=2$ (Appendix~\ref{app:insens}), the dual of the magnetic
term. The compact encoded basis serves as an independent cross-check.

We choose $g=0.8$ and $m_0=5$, so $E_{\rm cl}=3g^2/8=0.240$ and
$2m_0+E_{\rm cl}=10.24$. The exact twist energy from Eq.~\eqref{eq:Dtw_mathieu} at
$|\theta|=\pi/2$ is $\Delta_{\rm tw}=0.0126$, giving the reference
$2m_0+E_{\rm cl}+\Delta_{\rm tw}=10.253$. This is the exact Mathieu value. The
one-instanton estimate would give $0.0154$, so the agreement below is with
Eq.~\eqref{eq:Dtw_mathieu} and not with the weak-coupling asymptotics.

\begin{figure}[t]
    \centering
    \includegraphics[width=0.49\linewidth]{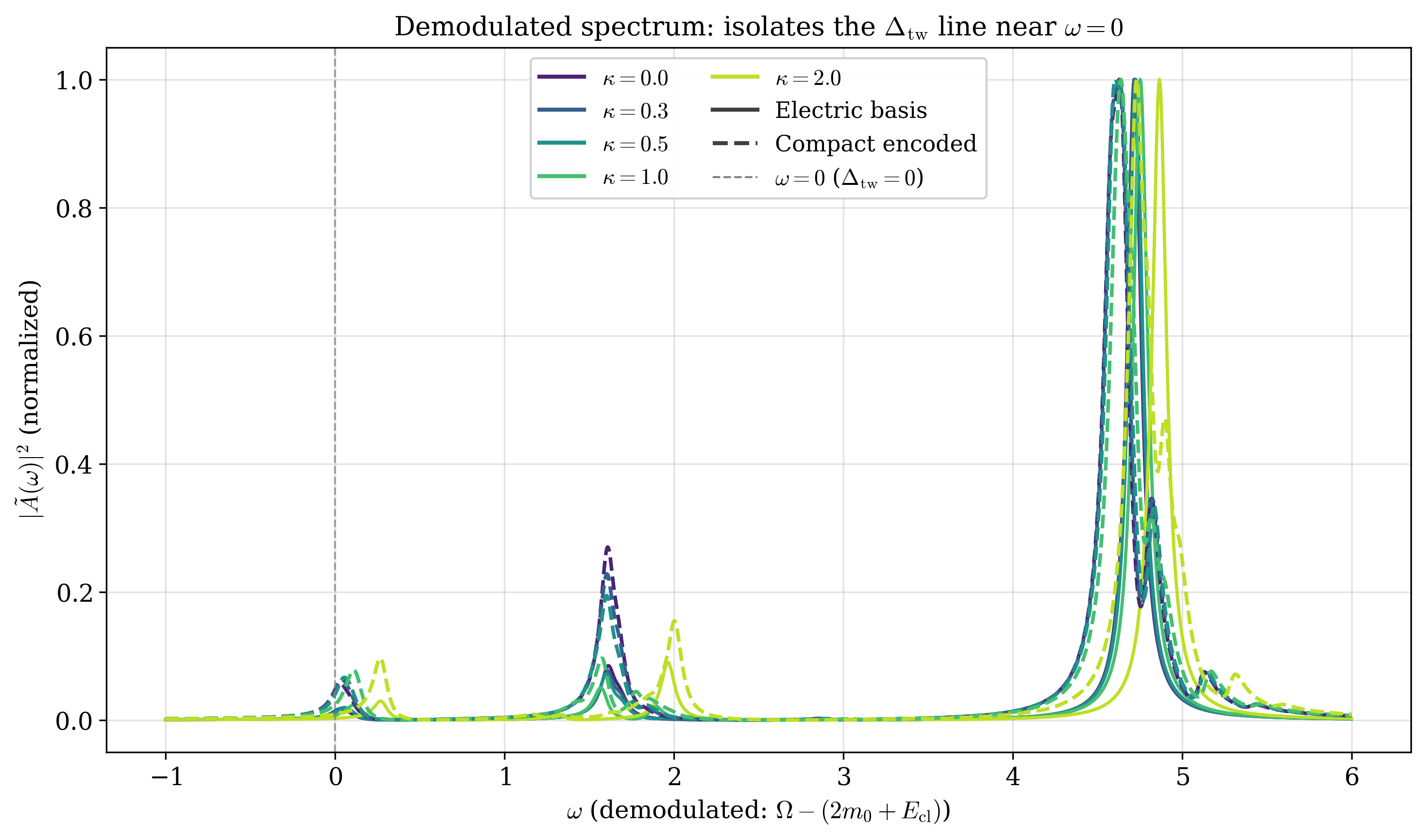}\hfill
    \includegraphics[width=0.49\linewidth]{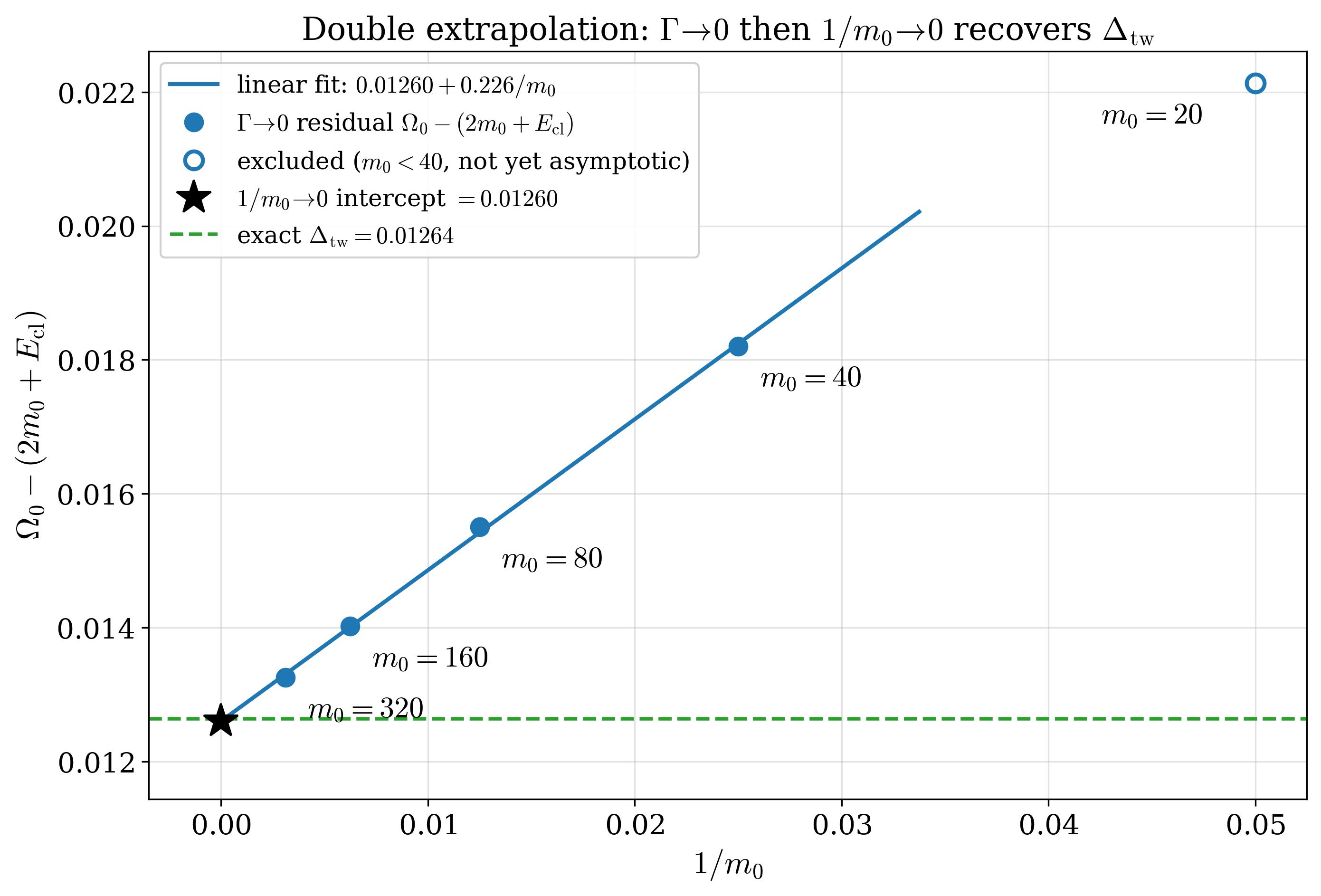}
    \caption{Left: demodulated spectral function
             $|\tilde A(\omega)|^2=|\mathcal{F}[\tilde C_b](\omega)|^2$ for
             one-plaquette QED$_3$ versus $\kappa$, in the electric (solid) and
             compact (dashed) bases. The axis is shifted by the known
             $2m_0+E_{\rm cl}=10.24$, so $\Omega_0$, which encodes
             $\Delta_{\rm tw}$ via Eq.~\eqref{eq:Omega0}, sits near $\omega=0$,
             and the multi-line structure is the dressing of the pair by the
             spectator channels ($\Gamma=0.05$, $T_{\rm max}=150$). Right: double
             extrapolation recovering the twist energy at $\kappa=0$. Each point is
             the demodulated lowest line after the $\Gamma\to0$ extrapolation. The
             fit is linear in $1/m_0$ over $m_0\ge40$, and its intercept (star)
             agrees with the exact Mathieu $\Delta_{\rm tw}$ (dashed) within the
             $\pm1.5\%$ fit-window and fit-form systematic quoted in the text. The
             open point at $m_0=20$ is not yet in the asymptotic regime and is
             excluded. Here $g=0.8$ and $m_0=5$ on the left.}
    \label{fig:twist_extrap}
\end{figure}

The spectral function (the Fourier transform of the time signal) 
$|\tilde A(\omega)|^2=|\mathcal{F}[\tilde C_b](\omega)|^2$ over the $\kappa$ sweep is shown in Figure~\ref{fig:twist_extrap} (left) in both representations. They agree to
within the compact-encoded truncation error throughout. The spectra are multi-line: a
small peak near $\omega=0$, a cluster of satellites, and a dominant peak at
$\omega\approx4.7$ carrying most of the weight. The pair created by $\hat{O}_b$
overlaps most strongly not with the lowest excited state but with one also excited in
the spectator channels. This shake-up is a leading effect at the physical couplings
of the one-plaquette system, not a perturbative correction.

\begin{figure}[t]
    \centering
    \includegraphics[width=0.45\linewidth]{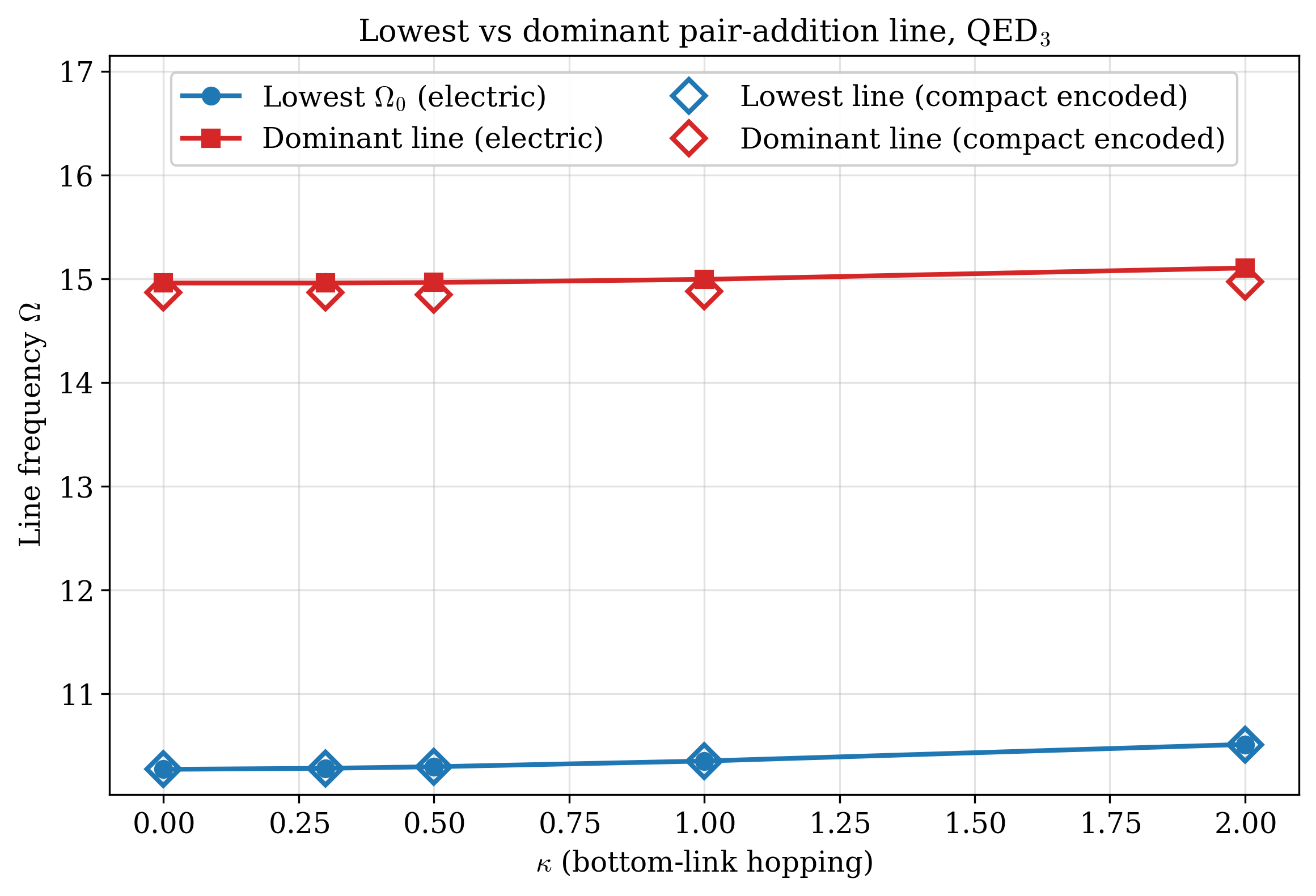}
    \includegraphics[width=0.45\linewidth]{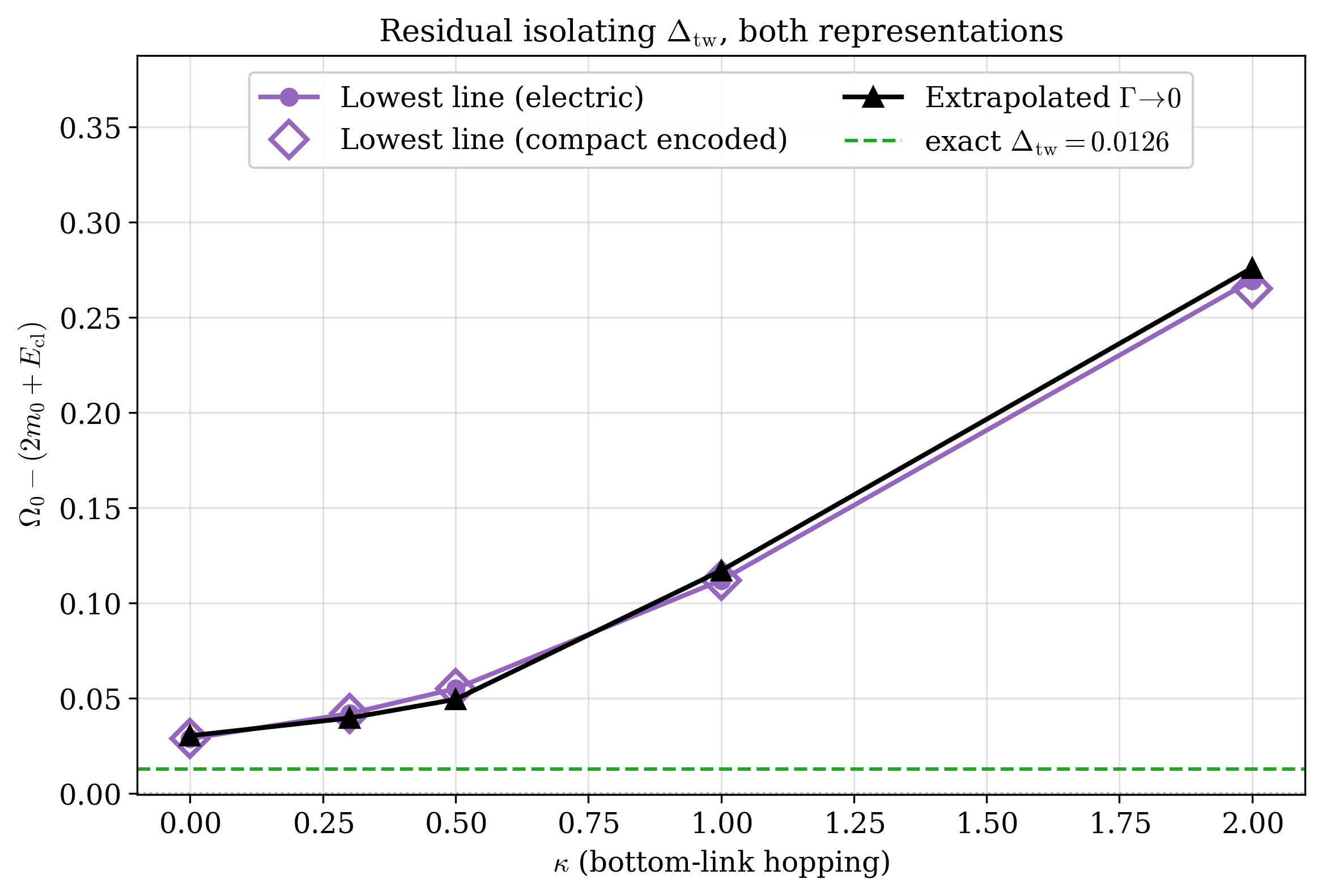}
    \caption{Left: lowest line $\Omega_0$ (circles) and dominant-weight line
             (squares) versus $\kappa$, electric basis (filled) and compact encoded basis
             (open diamonds). The dominant line is $46\%$ higher than $\Omega_0$.
             Accuracy on $\Omega_0$ itself is not informative, since the carrier
             $2m_0+E_{\rm cl}$ exceeds $\Delta_{\rm tw}$ by three orders of magnitude.
             Right: the residual $\Omega_0-(2m_0+E_{\rm cl})$, with the exact
             $\Delta_{\rm tw}(g{=}0.8)=0.0126$ marked and the $\Gamma\to0$
             extrapolation overlaid. The $\kappa=0$ intercept sits a factor $2.4$
             above $\Delta_{\rm tw}$: that excess is the $1/m_0$ fermionic dressing,
             removed in Fig.~\ref{fig:twist_extrap} (right), not a resolution effect. The
             growth with $\kappa$ is the operator-valued correction to $d$. Markers at
             $\Gamma=0.05$, $T_{\rm max}=150$.}
    \label{fig:qed3_lines}
\end{figure}

In Figure~\ref{fig:qed3_lines} (left) we show the lowest
resonance $\Omega_0$ and the dominant-weight line versus $\kappa$. It is $\Omega_0$
that carries $\Delta_{\rm tw}$, whereas the dominant line sits $46\%$ higher and
encodes kinetic hybridization with the spectators instead of the topology. Accuracies here
are quoted against $\Delta_{\rm tw}$ and not against the line position, because the
carrier $2m_0+E_{\rm cl}=10.24$ exceeds $\Delta_{\rm tw}=0.0126$ by three orders of
magnitude, so sub-percent agreement on $\Omega_0$ is compatible with an order-unity
error on the quantity of interest and only the residual $\Omega_0-(2m_0+E_{\rm cl})$
measures anything. The right panel isolates that residual, which grows with $\kappa$ as
the spectators dress the lowest state. Electric and compact results coincide at all
five $\kappa$.

That residual makes the point. At $\kappa=0$, once the damping bias is removed, it
settles at $0.0303$ against the exact $\Delta_{\rm tw}(g{=}0.8)=0.0126$, a factor
$2.4$, yet the same discrepancy is only $0.17\%$ of the line position. It is the
$\delta_{\rm dyn}$ term of Eq.~\eqref{eq:Omega0}: no accessible mass removes it, the
residual still being $11\%$ high at $m_0=160$, but it is a clean $1/m_0$, so it
extrapolates. The lowest feature is in fact a small multiplet, the twist line
accompanied by satellites spaced by $\simeq0.05$ at $m_0=5$, a spacing that
collapses as $1/m_0$ and lies below the $\Gamma$ used, so the quoted $\Omega_0$ is
the weight centroid of that multiplet, every member converging to
$\Delta_{\rm tw}$ in the double limit.
Figure~\ref{fig:twist_extrap} (right) carries out the double extrapolation in the
electric basis, the two representations coinciding to within truncation error
(Fig.~\ref{fig:qed3_lines}) and the encoded-basis systematics being bounded
separately in Appendix~\ref{app:insens}: $\Gamma\to0$ at each mass, then
$1/m_0\to0$ across $m_0=40,80,160,320$, giving
\begin{equation}
    \Omega_0-(2m_0+E_{\rm cl})\Big|_{\Gamma\to0,\;1/m_0\to0} = 0.01260,
    \qquad \Delta_{\rm tw}^{\rm exact}=0.01264,
    \label{eq:double_extrap}
\end{equation}
against the exact Mathieu value $\Delta_{\rm tw}^{\rm exact}=0.01264$. The fitted
slope $0.226/m_0$ is the dressing coefficient of that centroid, each member of the
multiplet carrying its own coefficient, and both systematics removed here are known
in form, not fitted away: the damping is $\mathcal{O}(\Gamma^2)$ by
Appendix~\ref{app:demod} and the dressing is $\mathcal{O}(1/m_0)$ by
Appendix~\ref{app:dressing}.

The number in Eq.~\eqref{eq:double_extrap} is the central value of one fit, and its
uncertainty is dominated by the choice of extrapolation and not by the data.
Varying the fit window over the asymptotic masses and replacing the linear form by one
carrying a $1/m_0^{2}$ term spreads the intercept by $\pm1.5\%$ about the exact value,
and we quote the recovery at that level rather than at the $0.3\%$ of the central fit.
In Appendix~\ref{app:extrap} we list the windows and state what the figure does and does not propagate.

\section{Scaling and resource estimates}
\label{sec:scaling}

We now collect the scaling consequences of the encoding. The purpose is not to
claim a quantum advantage, but to identify which parts of the construction are
polynomial, which error parameter must be tightened as the lattice grows, and which
single requirement dominates the cost at weak coupling.

\subsection{Computational frames and circuit cost}
\label{sec:frames}\label{sec:scaling_gates}
The gauge fixing of Appendix~\ref{app:GL} retains $N^2$ links but does not fix which
$N^2$ conjugate pairs become the modes, and two natural choices exist. In the loop
frame each mode is a plaquette, with $\hat{\chi}_p=\hat{B}_p$ the magnetic flux
through it and $\hat{\eta}_p$ the conjugate loop electric field. In the link frame
each mode is a retained link, with $\hat{\chi}_\ell=\hat{\vartheta}_\ell$ the link
angle and $\hat{\eta}_\ell=\hat{e}_\ell$ the physical electric field on it. The two
are related by an integer unimodular transformation, so they describe the same theory
and coincide at $N=1$, where all numerics of this work are performed. The frames differ in $\HH^{(2,1,0)}$ and hence in the gate content, each frame making one of the two
non-Gaussian terms elementary and the other composite. That the choice of operator
basis changes what a simulation costs is established in the qubit lattice-gauge
literature~\cite{Haase:2020kaj,Bender:2020ztu,Grabowska:2022uos,Kane:2022ejm}, and the dictionary, the transformed matrices and the circuits are collected in Appendix~\ref{app:gates}.

One quantity is not merely traded between the frames. The GKP error is pinned to each
physical mode, and the transformation is unimodular but not orthogonal, so the trace
that sets the bias~\eqref{eq:dE_trace} is not preserved,
\begin{equation}
    \mathrm{tr}\,\mathcal{H}^{(2)} = 4N^{2},
    \qquad
    \mathrm{tr}\,\tilde{\mathcal{H}}^{(2)} = N^{2}(N+3),
    \qquad
    \frac{\braket{\delta E}_{\rm link}}{\braket{\delta E}_{\rm loop}} = \frac{N+3}{4}.
    \label{eq:bias_ratio}
\end{equation}
At fixed squeezing the link basis is biased more, by a factor growing linearly in $N$:
equal at $N=1$, $3/2$ at $N=3$, $5/2$ at $N=7$. Matching requires about
$10\log_{10}[(N+3)/4]$~dB of extra squeezing, which is $1.8$~dB at $N=3$ and $4.0$~dB
at $N=7$.

Whether to apply the displacement of Eq.~\eqref{eq:displacement_op} is a separate
resource question, with different answers for static and dynamical charges. The two
options are Eqs.~\eqref{eq:undisplaced} and~\eqref{eq:displaced}. For static charges
$\hat{\theta}_i$ is a c-number and
$\hat{H}'_J$ is a phase-shifted cosine costing exactly what $\hat{H}_J$ costs, so
displacing removes all $\mathrm{nnz}(\HH^{(1)})$ displacements and collapses the
charge block to the free c-number $E_{\rm cl}[Q]$. With dynamical charges
$\hat{\theta}$ becomes an operator supported on all $(N+1)^2-1$ non-reference qubits
and the charge--charge block densifies, so one prefers the undisplaced
Eq.~\eqref{eq:undisplaced} whenever $\hat{H}_K$ is present. The counts are in Appendix~\ref{app:gates}.

Table~\ref{tab:bases} collects the gate content of one Trotter step frame by
frame. Setting aside the charge--charge block, which is frame invariant and itself
$\mathcal{O}(N^4)$, the frame-dependent gate content totals $\mathcal{O}(N^3)$ in
the loop basis against $\mathcal{O}(N^4)$ in the link basis under uniform
weighting. Write $C_{\rm step}(N)$ for that per-step cost, so that a first-order
product formula gives
\begin{equation}
    U(t)\simeq\Big[\prod_a e^{-iH_a\delta t}\Big]^{r},
    \qquad
    r=t/\delta t,
    \qquad
    C_{\rm sim}=r\,C_{\rm step}(N),
\end{equation}
with $r=\operatorname{poly}\!\big(t,N,\eta_{\max},g,m_0,\epsilon_{\rm Trot}^{-1}\big)$.
The commutator norms of the terms in Eq.~\eqref{eq:U1_ham_dynam} fix $r$, together with
the energy window the state occupies. Higher-order formulas or qubitization improve the
dependence on $t$ and $\epsilon_{\rm Trot}$, without changing the conclusion that the
cost is polynomial at fixed precision and cutoff. The correction rounds of
Sec.~\ref{sec:err_active} add to this, at a cadence fixed by
Eq.~\eqref{eq:loss_cadence} and $N^2$ ancillas per fully parallel round.

\subsection{Hardware size versus classical dimension}
\label{sec:scaling_size}
On an $N\times N$ lattice with open boundaries the reduction of Appendix~\ref{app:GL}
leaves $N^2$ compact rotors, one bosonic mode each, and staggered matter contributes
$(N+1)^2$ qubits, so the register is
\begin{equation}
    N_{\rm mode}=N^{2},
    \qquad
    N_{\rm q}=(N+1)^{2},
    \qquad
    N_{\rm mode}+N_{\rm q}=2N^{2}+2N+1 ,
\end{equation}
and only $N^2$ modes in the pure-gauge or static-charge sector. Classically the same
problem carries, at flux cutoff $|\eta_i|\le\eta_{\max}$,
\begin{equation}
    d_{\rm gauge}=(2\eta_{\max}+1)^{N^{2}},
    \qquad
    d_{\rm tot}=(2\eta_{\max}+1)^{N^{2}}\,2^{(N+1)^{2}} ,
\end{equation}
before exploiting further symmetries: at $\eta_{\max}=3$ already
$d_{\rm tot}\simeq2.6\times10^{12}$ for $N=3$ and $1.1\times10^{21}$ for $N=4$, against
$9+16$ and $16+25$ physical degrees of freedom. The encoded register is therefore
polynomial in the volume where the classical state vector is exponential. This is a
statement about representation only. State preparation, simulation, correction and
measurement are costed below and do not inherit that separation.

The comparison against truncated encodings should be stated with equal care, because
the truncation does not disappear here, it moves. Classically, the encoding is the
more expensive representation, and by how much depends on the state: a tooth at flux
$n$ sits at momentum $n\alpha$, so the photon cost is set by the flux support the
state actually occupies, $\bar n \simeq \pi\braket{\hat\eta^{2}}+\sinh^{2}r$, not by
the reference cutoff. The wall-state tests of Sec.~\ref{sec:testing}, preparing
unit-weight teeth across the full window $|n|\le k_{\rm max}$, are the worst case at
$N_{\rm fock}\simeq3\alpha^{2}k_{\rm max}^{2}$, whereas the physical ground sector at
$g=0.8$ carries $\braket{\hat\eta^{2}}=0.28$ and converges already at
$N_{\rm fock}\simeq101$--$151$ (Appendix~\ref{app:insens}), the support growing
toward weak coupling as $\braket{\hat\eta^{2}}\simeq1/(4g^{2})$. In exchange, the mode carries the group manifold itself, so every group operation,
the link element, the Wilson-loop cosine, the electric quadratic, is a fixed
hardware primitive whose count is independent of the flux support, with no register that grows with the cutoff. The costs that do grow with
$\eta_{\max}$ are energetic, photon number and squeezing through the requirement
$\Delta\ll1/(\sqrt{2\pi}\,\eta_{\max})$, which is the precise content of the phrase
``without truncation''.

\subsection{Finite squeezing at scale}
\label{sec:scaling_squeezing}
The finite-energy bias~\eqref{eq:dE_trace} is the one error parameter that does not
merely persist but accumulates with volume, and its behavior separates the two
frames more sharply than the fixed-$N$ ratio of Eq.~\eqref{eq:bias_ratio} suggests.
Dividing the traces of Eq.~\eqref{eq:bias_ratio} by the plaquette count, the bias
per plaquette is
\begin{equation}
    \frac{\braket{\delta E}_{\rm loop}}{N^{2}} = \frac{g^{2}\Delta^{2}}{2\pi},
    \qquad
    \frac{\braket{\delta E}_{\rm link}}{N^{2}} = \frac{g^{2}\Delta^{2}}{8\pi}(N+3),
    \label{eq:bias_density}
\end{equation}
both up to $\mathcal{O}(\Delta^{4})$. The loop-basis bias is therefore intensive,
while the link-basis bias is superextensive: the same hardware squeezing gives a
constant error density in the loop frame and one growing linearly in $N$ in the link
frame. Consequently, at fixed error density the loop basis admits $\Delta=\mathcal{O}(1)$
whereas the link basis requires $\Delta^{2}=\mathcal{O}(N^{-1})$. At fixed absolute
accuracy $\braket{\delta E}\le\epsilon$ the requirements are
\begin{equation}
    \Delta^{2}\le\frac{2\pi\epsilon}{g^{2}N^{2}}\ \ \text{(loop)},
    \qquad
    \Delta^{2}\le\frac{8\pi\epsilon}{g^{2}N^{2}(N+3)}\ \ \text{(link)},
\end{equation}
that is $\Delta^{2}=\mathcal{O}(N^{-2})$ against $\mathcal{O}(N^{-3})$. In the decibel convention of Sec.~\ref{sec:frames} the loop frame costs
$20\log_{10}N$~dB of additional squeezing to hold absolute accuracy from $N=1$ to $N$,
some $17$~dB at $N=7$, and the link frame half again as much. Both figures assume the
bias is left in place. Subtracting it with Eq.~\eqref{eq:dE_trace} or extrapolating in
$\Delta^{2}$ (Sec.~\ref{sec:err_active}) removes the leading term and relaxes the
requirement to the residual $\mathcal{O}(\Delta^{4})$, so mitigation is the practical route to large $N$. The requirement is softened in two further ways. The bound carries $g$ explicitly, so what fixes $\Delta$ is the
combination $\epsilon/(g^{2}N^{2})$, and demanding a fixed relative rather than
absolute accuracy on the electric energy holds that combination constant, at which
point the squeezing requirement stops growing with either $N$ or $g$. Furthermore the
square-envelope bias is independent of the flux content of the state by
Eq.~\eqref{eq:dE_trace}, so it is the same additive shift in every charge sector and
cancels identically in any energy difference, $\Delta_{\rm tw}$ included. This
cancellation concerns the leading electric bias of Eq.~\eqref{eq:dE_trace} alone:
the lineshape effects of Sec.~\ref{sec:fiber_dist}, second order in $\Delta$ on the
centroid and first order on the width, act on $\Delta_{\rm tw}$ directly and do not
cancel. The cancellation and the frame dependence are therefore statements about
different quantities: absolute energies retain the frame-dependent bias, sector
energy differences shed it.

\subsection{Measurement and spectral resolution}
\label{sec:scaling_meas}
Estimating an observable $O$ to statistical uncertainty $\epsilon_{\rm meas}$ costs
$N_{\rm shot}=\mathcal{O}\big(\mathrm{Var}(O)/\epsilon_{\rm meas}^{2}\big)$
repetitions. The twist energy needs more care, because it is exponentially small at
weak coupling and it is natural to assume that resolving it requires an exponentially
long evolution. It does not, and the reason is the demodulation of
Sec.~\ref{sec:pair_add}. Both other terms in $\Omega_0$ of Eq.~\eqref{eq:Omega0} are
classically known, so multiplying by $e^{i(2m_0+E_{\rm cl})t}$ places the twist line at
$\omega=\Delta_{\rm tw}$ against an exactly known origin. Its neighbors are not twist-scale features. The nearest lines are the satellites of the multiplet of
Sec.~\ref{subsec:qed3_plaquette}, a distance set by the fermionic dressing,
$\simeq0.05$ at $m_0=5$ and collapsing as $1/m_0$, while the first large-weight
excitation sits an $\mathcal{O}(1)$ distance away at fixed volume (the harmonic gap
closes as $1/N$ with lattice size, Appendix~\ref{app:meas_scaling}). The window
required is set by the smaller of these two scales, here the satellite spacing,
still free of the naive Rayleigh estimate $2\pi/\Delta_{\rm tw}$: at $g=0.8$,
$m_0=5$ it demands $T_{\max}\gtrsim2\pi/0.05\simeq126$, met by the
$T_{\max}=150$ used, and the demodulated spectrum has its
twist multiplet at $\omega\approx0.03$, the $m_0=5$ dressed position whose
double limit is $\Delta_{\rm tw}=0.0126$, with the dominant shake-up line at
$\omega\approx4.7$.

Once isolated, locating that line is parameter estimation, and
the estimator matters. The demodulated correlator is not a phase but a sum of lines,
of which the twist line contributes $w\,e^{-i\Delta_{\rm tw}t}$ with $w$ its
spectral weight, and the Hadamard tests of Appendix~\ref{app:measurement} measure
the $\mathrm{Re}$ and $\mathrm{Im}$ of that sum directly. Separating the twist
term from the others is cheap, a handful of time points or a spectral filter,
Appendix~\ref{app:meas_scaling}, and once it is separated, estimating its phase
at a single long time costs
\begin{equation}
    N_{\rm shot}^{\rm tot} \;\sim\; \frac{1}{\big(w\,\epsilon\,\Delta_{\rm tw}\,T\big)^{2}} ,
    \label{eq:shot_phase}
\end{equation}
with no spectral reconstruction, and it is unambiguous while the accumulated phase
stays inside one branch,
\begin{equation}
    T \;<\; \frac{\pi}{\Delta_{\rm tw}} ,
    \label{eq:wrap_ceiling}
\end{equation}
a ceiling set by the observable itself and equal to $249$ at $g=0.8$. At
$\epsilon=3\%$ and $T=150$ this is $3\times10^{2}$ shots at unit weight. The
alternative of reconstructing the whole spectral function on a Nyquist grid and
locating its centroid needs nearly five orders of magnitude more on the same data,
the arithmetic carried out in Appendix~\ref{app:meas_scaling}, because the Fourier
route spends its repetitions rebuilding a whole spectral function to extract one
number from it.

The one-plaquette calculation shows that the weight, not the statistics, is the binding constraint. The probe $\hat{O}_b$ of Sec.~\ref{sec:pair_add} places only
$w=0.054$ of the correlator on the twist line, a weight read off the spectra of
Figure~\ref{fig:twist_extrap}, the bulk sitting on the shake-up
resonance of Sec.~\ref{subsec:qed3_plaquette}, and the $1/w^{2}$ in
Eq.~\eqref{eq:shot_phase} turns $3\times10^{2}$ into $1\times10^{5}$ shots. The weight is a property of the source operator alone, and any preparation with larger overlap onto the lowest state in the
charge sector reduces the budget quadratically. We have not optimized the probe here.
In Appendix~\ref{app:meas_scaling} we spell out the assumptions behind these
estimates, the inhomogeneous width the line inherits from the fiber
distribution of Sec.~\ref{sec:fiber_dist}, and why the gap to the first
excitation is set by the lattice volume and not by the coupling.


\section{Discussion and outlook}
\label{sec:conc}

We have introduced a new encoding of compact $U(1)$ gauge fields using non-compact
bosonic modes, developed its finite-energy error theory, and validated it on a
single plaquette with and without dynamical fermions. After Gauss's law is solved, the remaining gauge degree of
freedom is carried by one oscillator mode, while compactness is enforced by a
GKP-type stabilizer. The reason the
encoding works is its fiber structure. The Hilbert space of the oscillator splits into
a continuum of sectors, where each is an exact copy of the compact gauge theory with a
twisted boundary condition, labeled by a stabilizer that the encoded dynamics
conserves. Under this stabilization procedure, compactness is therefore exact. In fact, a state
prepared in the physical sector never leaves it under unitary time evolution because the Hamiltonian commutes with the stabilizer. We also show that a state prepared at finite
squeezing is not a slightly wrong compact state, but an exact ensemble of compact
theories whose twist (fiber) distribution is fixed once and for all at preparation and
can be measured without demolishing the state, thus allowing for error correction and mitigation.

Finite-energy is then akin to bookkeeping, not mere uncontrolled errors. In particular, finite
squeezing enters an observable in exactly two ways: through the width of the
twist distribution and through the envelope distortion of the trigonometric
operators, and both carry a functional form that we derive rather than model.
The encoding error is thereby a systematic error: it is monitored through the stabilizer, fitted through the known
lineshape, and removed by extrapolating in the squeezing. 

The construction comes with the complete toolkit to use it, a synthesis of the
full Hamiltonian, non-Gaussian couplings included, from trigonometric gates in
two computational frames, and qubit-based protocols measuring its one- and
two-point functions. However, here, we make no claim of full fault tolerance. The compactness stabilizer is a GKP
stabilizer, so the standard machinery of bosonic error correction applies as
is: repeated syndrome extraction corrects the small displacement errors from
photon loss and imperfect Gaussian operations, while large displacements act
as logical flux errors, the natural target for concatenation with an outer
qubit code that could also protect the matter qubits. Whether this stack
admits a threshold is an open question, for which the closed-form error model
developed here is a starting point. The immediate next steps are to test the protocols on hardware
and to move to larger systems, ladders and small two-dimensional arrays, where
tensor-network benchmarks are available.


\section*{Acknowledgments}

GS acknowledges support by NSF award DGE-2152168 and DOE awards DE-SC0024325 and DE-SC0024328. TR and FR are supported by the DOE, Office of Science, Office of Nuclear Physics, Early Career Program under contract No. DE-SC0025881. This work has been partially funded by the Eric \& Wendy Schmidt Fund for Strategic Innovation through the CERN Next Generation Triggers project under grant agreement number SIF-2023-004.

\bibliographystyle{utphys.bst}
\bibliography{bibliography}

\appendix
\addtocontents{toc}{\protect\setcounter{tocdepth}{1}}
\section{Electric Hamiltonian after Gauss's law}\label{app:GL}
Solving Gauss's law writes each link electric field as a loop part minus a
charge string, Eq.~\eqref{eq:E_link_GaussLaw} below, governed by two integer
matrices: the plaquette--link incidence matrix $\mathcal{K}$, whose row $p$
lists the links bounding plaquette $p$ with counterclockwise signs, and the
charge-string matrix $\mathcal{C}$, whose entry $\mathcal{C}_{\ell s}$ records
whether the string of the charge at site $s$ passes through link $\ell$. The
$2N(N+1)$ links fall into three families, whose orientations are not free but
are the unique choice under which the counterclockwise convention for
$\mathcal{K}$ and the positivity $\mathcal{C}\ge0$ hold simultaneously:
\begin{center}
\begin{tabular}{llll}
\hline
Family & Links & Count & Orientation \\
\hline
A & horizontal $h(n_x,n_y)$, $0\le n_x<N$, $0\le n_y<N$ & $N^2$ & $+\bm{e}_x$\\
B & horizontal $h(n_x,N)$, $0\le n_x<N$ (top row)       & $N$   & $-\bm{e}_x$\\
C & vertical $v(n_x,n_y)$, $0\le n_x\le N$, $0\le n_y<N$ & $N(N+1)$ & $-\bm{e}_y$\\
\hline
\end{tabular}
\end{center}
Family A is the ``bottom link of each plaquette'': $h(n_x,n_y)$ is the bottom edge of
plaquette $(n_x,n_y)$ and the top edge of $(n_x,n_y-1)$. Note that family C separates horizontally adjacent plaquettes and so runs in
$y$, while family A separates vertically adjacent ones and runs in $x$. Each
link borders at most two plaquettes, and the row of $\mathcal{K}$ of a
plaquette carries $+1$ on the links its counterclockwise circulation traverses
along their orientation and $-1$ on those it traverses against, giving the
nonzero entries
\begin{equation}
\begin{aligned}
    \mathcal{K}_{(n_x,n_y),\,h(n_x,n_y)} &= +1, &
    \mathcal{K}_{(n_x,n_y-1),\,h(n_x,n_y)} &= -1, &&\text{(A)}\\
    \mathcal{K}_{(n_x,N-1),\,h(n_x,N)} &= +1, &&&&\text{(B)}\\
    \mathcal{K}_{(n_x,n_y),\,v(n_x,n_y)} &= +1, &
    \mathcal{K}_{(n_x-1,n_y),\,v(n_x,n_y)} &= -1, &&\text{(C)}
\end{aligned}
\end{equation}
entries with an out-of-range plaquette index being absent, since a boundary
link borders a single plaquette. The charge-string matrix has
$\mathcal{C}_{\ell s}=1$ iff $s\in\mathrm{aq}(\ell)$, the accumulated-charge
set of link $\ell$, the sites whose charge string passes through it,
\begin{equation}
\begin{aligned}
    \mathrm{aq}\big(h(n_x,n_y)\big) &= \varnothing, && 0\le n_y<N, &&\text{(A)}\\
    \mathrm{aq}\big(h(n_x,N)\big) &= \{(i,j) : 0\le i \le n_x,\ 0\le j\le N\}, &&&&\text{(B)}\\
    \mathrm{aq}\big(v(n_x,n_y)\big) &= \{(n_x,j) : 0\le j \le n_y\}. &&&&\text{(C)}
\end{aligned}
\end{equation}
In words, the string of the charge at site $\bm{n}$ climbs its own column and
then runs along the top row to the reference corner $(N,N)$, crossing only
vertical and top-row links: on the $N=3$ lattice of
Fig.~\ref{fig:lattice_snake}, the charge at $(1,1)$ threads the vertical links
$v(1,1)$ and $v(1,2)$ up its column, then the top-row links $h(1,3)$ and
$h(2,3)$ into the corner. No interior horizontal link lies on any such path.
This is the empty set in line (A), and it is what makes family A the natural
gauge-fixing choice: there the electric field is a pure curl, and it is the
family retained in Appendix~\ref{app:gates}.
With
\begin{equation}
    \hat{E}_\ell = \big(\mathcal{K}^{\mathsf T}\hat{\eta} - \mathcal{C}\hat{Q}\big)_\ell
    \label{eq:E_link_GaussLaw}
\end{equation}
one verifies that $\nabla\cdot(\mathcal{K}^{\mathsf T}\hat{\eta})=0$ identically,
the curl part is sourceless, and that $(\nabla\cdot\hat{E})_{\bm{n}}=\hat{Q}_{\bm{n}}$
at every site except $(N,N)$, where instead
\begin{equation}
    (\nabla\cdot\hat{E})_{(N,N)} = -\!\!\sum_{\bm{n}\neq(N,N)}\!\!\hat{Q}_{\bm{n}} .
\end{equation}
Gauss's law therefore closes at the corner if and only if the total charge vanishes.
This single fact underlies three statements used in the main text: (i) $(N,N)$ lies
in no $\mathrm{aq}(\ell)$, so its columns of $\mathcal{C}$, $\mathcal{H}^{(1)}$ and
$\mathcal{H}^{(0)}$ vanish and $Q_{(N,N)}$ never enters $d$; (ii)
$\mathrm{nnz}(\mathrm{diag}\,\mathcal{H}^{(0)})=(N+1)^2-1$; (iii) neutrality is a consistency condition of the reduction. 

Substituting Eq.~\eqref{eq:E_link_GaussLaw} into
$\hat{H}_E=\tfrac{g^2}{2}\sum_\ell\hat{E}_\ell^2$ gives Eq.~\eqref{eq:U1_ham_dynam}
with
\begin{equation}
    \mathcal{H}^{(2)} = \mathcal{K}\mathcal{K}^{\mathsf T} = 4\,\Id - A, \qquad
    \mathcal{H}^{(1)} = -2\,\mathcal{K}\mathcal{C}, \qquad
    \mathcal{H}^{(0)} = \mathcal{C}^{\mathsf T}\mathcal{C},
\end{equation}
$A$ the plaquette-graph adjacency matrix. These are the loop-basis values of
$\HH^{(2,1,0)}$. Equivalently the three blocks form one Gram matrix,
\begin{equation}
    \begin{pmatrix} \mathcal{H}^{(2)} & \tfrac{1}{2}\mathcal{H}^{(1)} \\[2pt]
    \tfrac{1}{2}\mathcal{H}^{(1)\mathsf T} & \mathcal{H}^{(0)} \end{pmatrix}
    = \begin{pmatrix} \mathcal{K} \\ -\mathcal{C}^{\mathsf T} \end{pmatrix}
    \begin{pmatrix} \mathcal{K}^{\mathsf T} & -\mathcal{C} \end{pmatrix},
\end{equation}
a structure preserved by the frame change of Appendix~\ref{app:gates}. Contracting
Eq.~\eqref{eq:E_link_GaussLaw} with $\mathcal{K}$ inverts the relation,
\begin{equation}
    \hat{\eta} = \big(\mathcal{H}^{(2)}\big)^{-1}\mathcal{K}\hat{E}
    + \tfrac{1}{2}\big(\mathcal{H}^{(2)}\big)^{-1}\mathcal{H}^{(1)}\hat{Q} ,
\end{equation}
so $\hat{\eta}$ is recovered from the discrete flux divergence $\mathcal{K}\hat{E}$
by inverting the plaquette-graph Laplacian. For $N=1$, in the snake ordering,
\begin{equation}
    \mathcal{H}^{(2)} = 4,\qquad
    \mathcal{H}^{(1)} = (-4,\,+2,\,0,\,-2),\qquad
    \mathcal{H}^{(0)} = \begin{pmatrix} 2&0&0&1\\ 0&1&0&0\\ 0&0&0&0\\ 1&0&0&1\end{pmatrix},
\end{equation}
the vanishing entries belonging to the reference corner $(1,1)$.

\section{The one-plaquette twist energy}
\label{app:1plaq}
In this appendix we collect the explicit one-plaquette computations behind
Secs.~\ref{sec:displacement} and~\ref{sec:twist_energy} and behind the numerics of
Sec.~\ref{sec:prep}: the displacement worked out, the band on the line, the
large-mass dressing, the method-parameter systematics, the mass extrapolation, and
the monopole reading with its instanton estimate.

\subsection{The displacement worked out}
With the conventions of Eq.~\eqref{eq:d_1plaq}, take the $q\bar{q}$ pair on the
horizontal bottom link. Taking the horizontal pair
explicitly, $\HH^{(1)}Q=-6$ and $Q^{\mathsf T}\HH^{(0)}Q=3$, so the bracket of
Eq.~\eqref{eq:U1_ham_dynam} is the polynomial
\begin{equation}
    4\eta^{2}-6\eta+3 = 4\big(\eta-\tfrac34\big)^{2}+\tfrac34 ,
    \label{eq:1plaq_poly}
\end{equation}
whose relaxed minimum sits at $\eta=d=3/4$ and supplies $E_{\rm cl}=3g^{2}/8$. The
flux itself remains an integer. The nearest allowed value $\eta=1$ costs a further
$\tfrac{g^{2}}{2}\,4(1-\tfrac34)^{2}=g^{2}/8$, for a total of $g^{2}/2$, the value obtained by evaluating Eq.~\eqref{eq:U1_ham_dynam} directly at $\eta=1$. The fractional part of $d$
is exactly this gap between where the quadratic form would place the flux and where
compactness permits it, and it is what survives as the twist. 
\subsection{The one-plaquette band on the line}
\label{app:band}
The band quoted in Sec.~\ref{sec:prep} is obtained as follows. For one plaquette with
no charges the encoded Hamiltonian is
\begin{equation}
    H_{U(1)} = 2g^2\Big(\frac{\hat{p}}{\alpha}\Big)^2 + \frac{1}{g^2}\big[1 - \cos(\alpha\hat{x})\big],
    \label{eq:noncomp_encoded_comp_ham}
\end{equation}
whose Schr\"odinger equation is the Mathieu equation
\begin{equation}
    \psi''(x) - \frac{\alpha^2}{2g^4}\big(1-\cos\alpha x\big)\psi(x) = -\frac{\alpha^2}{2g^2}\mathcal{E}\psi(x).
    \label{eq:Schr_EQ}
\end{equation}
Since $x\in\mathbb{R}$, Bloch's theorem gives $\psi_{n,k}(x)=e^{ikx}u_{n,k}(x)$ with
$u_{n,k}(x+T_\alpha)=u_{n,k}(x)$, $T_\alpha=2\pi/\alpha$ and
$k\in[-\tfrac{\alpha}{2},\tfrac{\alpha}{2})$, so
$\psi_{n,k}(x+T_\alpha)=e^{2\pi ik/\alpha}\psi_{n,k}(x)$. In $z=\alpha x/2$ these are
Mathieu functions of exponent $\nu_{\rm M}=2k/\alpha$, and
\begin{equation}
    \mathcal{E}_n(k) = \tfrac{g^2}{2}a_{n,\nu_{\rm M}}(-g^{-4}) + \tfrac{1}{g^2},
    \label{eq:band_spectrum}
\end{equation}
independent of $\alpha$ as it must be. The connection to the rest of the paper is
$\theta=2\pi k/\alpha=\pi\nu_{\rm M}=2\pi\nu$, so that Eq.~\eqref{eq:band_spectrum} at
$n=0$ returns Eq.~\eqref{eq:Dtw_mathieu} as the difference between two sectors of the
same band. The penalty of Sec.~\ref{sec:prep} leaves this expression at
$\nu_{\rm M}=0$ untouched.

\subsection{The fermionic dressing at large mass}
\label{app:dressing}
The dynamical term $\delta_{\rm dyn}$ of Eq.~\eqref{eq:Omega0} has a definite leading
form. At large $m_0$ the mass term dominates and $\hat{H}_K$ is a perturbation that
changes the charge configuration, so its first-order expectation vanishes in every
unperturbed eigenstate. Both energies entering $\Omega_0$ then shift at second order,
\begin{equation}
    \delta E_s = -\sum_{x}\frac{|\braket{x|\hat{H}_K|s}|^{2}}{E_x-E_s},
    \label{eq:second_order}
\end{equation}
and every intermediate state differs from $s$ by a single hop, that is by one created
or annihilated pair, so $E_x-E_s=\pm2m_0+\mathcal{O}(m_0^{0})$. Hence
\begin{equation}
    \delta_{\rm dyn} = \frac{c_1}{m_0} + \mathcal{O}(m_0^{-2}),
    \label{eq:dressing_form}
\end{equation}
with $c_1$ a sum of squared hop matrix elements weighted by the sign of the crossed
gap, differing between the pair and vacuum sectors only through Pauli blocking and
the annihilation channel of the created pair. The annihilation channel exists only in
the pair sector and carries a negative denominator, so $c_1>0$. Each line of the
low-lying multiplet of Sec.~\ref{subsec:qed3_plaquette} carries its own such
coefficient, and the fit there, tracking the weight centroid of the multiplet,
returns $0.226$ at $g=0.8$. Equation~\eqref{eq:dressing_form} is the form assumed by
the $1/m_0\to0$ extrapolation of Eq.~\eqref{eq:double_extrap}.

\subsection{Penalty and truncation systematics}
\label{app:insens}
The encoded numerics of Sec.~\ref{sec:prep} carry two method parameters, the penalty
strength $J$ and the Fock truncation $N_{\rm fock}$, which by Sec.~\ref{sec:penalty} must be quoted together. The working points are as follows. The wall-state
comparisons of Sec.~\ref{sec:testing} prepare the fiber directly and run at $J=0$
with $N_{\rm fock}=169$ and $301$, the two-point overlays with dynamical fermions at
$J=0$ with $N_{\rm fock}\in\{51,101,151\}$, and the compact cross-check of the
spectroscopy at $J=2$ with $N_{\rm fock}=101$.

To bound the sensitivity we recompute the static-sector observables by exact
diagonalization of the same construction over $J\in[0.5,20]$ and
$N_{\rm fock}\in\{101,200,600\}$ at $g=0.8$. The twist energy moves little once
$N_{\rm fock}$ tracks $J$: against the exact Mathieu value it is $-1.4\%$ at the
working point $(J,N_{\rm fock})=(2,101)$, $-0.9\%$ at $(2,600)$, and $+0.1\%$ at
$(20,600)$, while at fixed $N_{\rm fock}=600$ varying $J$ from $0.5$ to $20$ spans
$-1.9\%$ to $+0.1\%$. The magnetic expectation $\braket{\cos\hat\chi}$ is insensitive
at the $10^{-3}$ level at $N_{\rm fock}=600$ for every $J$ in the range, and below
$10^{-2}$ at the smaller truncations. The absolute ground-state energy converges more slowly. In the exact theory it is
$J$-independent, since the ground state sits on the stabilized fiber where the
penalty vanishes identically and only the $\nu\neq0$ fibers are lifted
(Sec.~\ref{sec:penalty}). At finite $N_{\rm fock}$ the truncated basis cannot
represent an arbitrarily sharp fiber, so the computed energy carries a penalty
residue proportional to $J$ and to the smallest fiber variance the truncation can
hold: at $N_{\rm fock}=600$ it rises linearly in $J$, from $+1.4\%$ at $J=2$ to
$+14\%$ at $J=20$, and at every $J$ it falls monotonically with $N_{\rm fock}$. The shift is thus a truncation artifact. The
residue is common to every charge sector and cancels in the difference, which the
twist-energy column demonstrates directly, the same run being $14\%$ high in
$\epsilon_0$ and $0.1\%$ off in $\Delta_{\rm tw}$. The coupling between the two cutoffs is equally visible:
$(J,N_{\rm fock})=(20,200)$ misses the twist energy by $+34\%$, the configuration we warned against in Sec.~\ref{sec:penalty}, in which the comb is sharpened beyond what the
truncation can represent.

\subsection{The mass extrapolation}
\label{app:extrap}
The fit windows behind the $\pm1.5\%$ systematic quoted in
Sec.~\ref{subsec:qed3_plaquette} are as follows. Varying the window over the
asymptotic masses, from $m_0\in[40,80]$ to $[80,320]$, gives intercepts between
$0.01280$ and $0.01251$, and replacing the
linear form by $c_0+c_1/m_0+c_2/m_0^2$ over $m_0\ge40$ gives $0.01246$, the spread
bounding the mass-extrapolation systematic at $\pm1.5\%$. Including $m_0=20$, which is not yet asymptotic,
moves the intercept up by as much as $7\%$ depending on the window, which
is why it is excluded. Fourier-window, flux-truncation and damping-model systematics
are not propagated into this figure, so $\pm1.5\%$ bounds the mass-extrapolation
systematic alone and not the total error.

\subsection{Monopole interpretation and instanton estimate}
\label{app:monopole}
The tight-binding picture of Sec.~\ref{sec:twist_energy} admits a monopole reading.
At weak coupling the winding wells are deep, the coefficients $t_{\bm{m}}$ fall
exponentially and the lowest harmonics dominate, while at strong coupling this
hierarchy is lost. The off-diagonal entries of $(\HH^{(2)})^{-1}$ set the effective
inertia and correlate tunneling events on different plaquettes, the Hamiltonian
counterpart of the monopole Coulomb interaction. Perturbation theory around a single
well gives no $\bm{\theta}$ dependence at any order, which is why the perturbative
series of $\Delta_{\rm tw}$ vanishes, and since Eq.~\eqref{eq:Delta_tw_band}
vanishes only at trivial twist, the configurations of Sec.~\ref{sec:displacement}
with integer $d$ are exactly those of zero twist energy, in the neutral sector the
vacuum alone.

For the compact rotor of Sec.~\ref{sec:twist_energy} the Euclidean event
$\chi:0\to2\pi$ is a $0+1$-dimensional phase slip with action $S_0=4/g^2$ and
hopping $t = Ke^{-S_0}\big[1+\mathcal{O}(g^2)\big]$, $K = \tfrac{4}{g}\sqrt{2/\pi}$.
Keeping $m=\pm1$, $\Delta_{\rm tw}\simeq4t\sin^2(\theta/2)$, so a unit pair on a
link ($|\theta|=\pi/2$) has
\begin{equation}
    \Delta_{\rm tw}(g) \simeq 2t = \frac{8}{g}\sqrt{\frac{2}{\pi}}\,e^{-4/g^2}
    \big[1+\mathcal{O}(g^2)\big] + \mathcal{O}\big(e^{-8/g^2}\big).
    \label{eq:Dtw_instanton}
\end{equation}
This is only the weak-coupling asymptotics of the exact Eq.~\eqref{eq:Dtw_mathieu}
and should not be used at moderate coupling: at $g=0.8$, $|\theta|=\pi/2$ the exact
value is $\Delta_{\rm tw}=0.0126$ against $0.0154$ from one instanton, a $22\%$
overestimate.

\section{Finite-energy error model: derivations}
\label{app:errors}
In this appendix we carry out the longer computations behind Sec.~\ref{sec:errorcorrection}:
the full error budget, the Mehler overlaps that set the contrast of angle
observables, the wall states with their exact magnetic contrast, and the two
syndrome-extraction protocols.

\subsection{The full error budget}
Table~\ref{tab:qec_errors} collects every channel, the finite-energy errors
(A)--(C) of Sec.~\ref{sec:err_classification} in the first block and the physical and
operational ones in the second.

\begin{table}[p]
\centering
\footnotesize
\begin{tabular}{p{0.19\linewidth}p{0.22\linewidth}p{0.28\linewidth}p{0.21\linewidth}}
\hline
Error source & Physical origin & Effect on encoded gauge theory & Treatment \\
\hline
\multicolumn{4}{l}{Finite-energy (state preparation), Sec.~\ref{sec:err_classification}}\\[1mm]
(A) Stabilizer violation
& Finite squeezing; tooth overlap
& Spread in $p\bmod\alpha$; non-orthogonal flux states
& Syndrome correction, Sec.~\ref{sec:err_active}
\\[2mm]

(B) Twist spread
& Finite squeezing, within one cell
& Coherent twist bias; energy shift Eq.~\eqref{eq:dE_trace}
& Subtraction or extrapolation, Sec.~\ref{sec:err_active}
\\[2mm]

(C) Envelope distortion
& Gaussian envelope only
& Suppression and contraction of large $|n|$
& Square envelope; or $\kappa\to0$; or $\Delta\ll1/\sqrt{2\pi}\eta_{\max}$
\\[2mm]
\hline
\multicolumn{4}{l}{Physical and operational}\\[1mm]
Photon loss
& Oscillator attenuation
& Random displacement plus flux-dependent contraction
& Loss compensation; repeated syndrome extraction, Sec.~\ref{sec:err_loss}
\\[2mm]

Oscillator dephasing
& Frequency noise or phase diffusion
& Logical gauge-angle error
& Not caught by comb syndrome; echo, refresh, or outer code
\\[2mm]

Gaussian gate error
& Imperfect displacements, phase gates, SUM gates
& Residual and correlated displacement errors
& Frame tracking and repeated syndrome extraction
\\[2mm]

Non-Gaussian gate error
& Imperfect cosine or trigonometric-gate synthesis
& Coherent Hamiltonian error
& Calibration, randomized compiling, extrapolation
\\[2mm]

Ancilla-state error
& Finite-energy GKP/qunaught ancilla
& Noisy syndrome estimate
& Analog decoding; improved resource states
\\[2mm]

Homodyne/readout error
& Finite-resolution measurement
& Wrong feedback displacement
& Repeated syndrome rounds; analog decoding
\\[2mm]

Matter-qubit error
& Qubit decoherence and gate faults
& Incorrect charge configuration or hopping term
& Independent qubit QEC layer
\\[2mm]

Gauss-law violation
& Noisy matter--gauge operations
& Leakage from physical gauge sector
& Gauge checks or gauge-penalty terms
\\[2mm]

Fock truncation
& Finite $N_{\rm fock}$ per mode
& $[\hat H,\hat S_\alpha]\neq0$ at the truncation edge; $J$-dependent bias
& Joint $(J,N_{\rm fock})$ convergence, Appendix~\ref{app:insens}
\\[2mm]

Trotter/synthesis error
& Approximate Hamiltonian simulation
& Coherent simulation error
& Higher-order formulas, qubitization, extrapolation
\\
\hline
\end{tabular}
\caption{
Error sources in the compact-rotor encoding. The rotor stabilizer protects the
compact flux structure, but a scalable implementation must also control
finite-energy bias, photon loss, matter-qubit faults, gauge-sector leakage, and
coherent gate-synthesis errors.
}
\label{tab:qec_errors}
\end{table}

\subsection{Mehler overlaps of the encoded teeth}
\label{app:mehler}
All overlaps follow from the Mehler kernel,
\begin{equation}
K_{\beta}^{(\alpha)}(p',p)=\bra{p'}e^{-\beta\hat{n}_{\alpha}}\ket{p}
=\frac{e^{\beta/2}}{\alpha\sqrt{\sinh\beta}}
  \exp\!\left[-\frac{\pi(p'-p)^{2}}{2\alpha^{2}}\coth\frac{\beta}{2}
              -\frac{\pi(p'+p)^{2}}{2\alpha^{2}}\tanh\frac{\beta}{2}\right],
\label{eq:Mehler_kernel}
\end{equation}
so that $\braket{m;\Delta|n;\Delta}=K^{(\alpha)}_{2\Delta^{2}}(m\alpha,n\alpha)$
exhibits two distinct effects:
\paragraph{Loss of orthogonality:} The $\coth$ term controls neighboring-tooth
overlap. Normalizing and using $\coth x-\tanh x=2/\sinh(2x)$,
\begin{equation}
  \frac{\big|\braket{m;\Delta|n;\Delta}\big|}
       {\|\ket{m;\Delta}\|\,\|\ket{n;\Delta}\|}
  =\exp\!\left[-\frac{\pi (m-n)^{2}}{\sinh(2\Delta^{2})}\right]
  \;\xrightarrow[\Delta\ll1]{}\;
  e^{-\pi(m-n)^{2}/2\Delta^{2}} .
  \label{eq:orthogonality}
\end{equation}
The encoder is an isometry only up to $\mathcal{O}(e^{-\pi/2\Delta^{2}})$, and the result is
$\alpha$-independent: the scaling freedom in Eq.~\eqref{eq:CCR_rep_alpha} is a
squeezing gauge, absorbed into $r_{\alpha}$.
\paragraph{Spurious flux envelope:} The $\tanh$ term gives on the diagonal
$\|\ket{n;\Delta}\|^{2}\propto e^{-2\pi n^{2}\tanh(\Delta^{2})}\to e^{-2\pi\Delta^{2}n^{2}}$,
so $\mathcal{E}_{\Delta}$ suppresses
$|n|\gtrsim n_{\max}\simeq1/(\Delta\sqrt{2\pi})$. This is a distortion, not a
normalization: in the symmetric GKP state the tooth width and the flux envelope are
locked together, and the encoder is faithful only on $|n|\lesssim n_{\max}$.
Two tooth widths must not be confused. For $\Delta\ll1$ the amplitude and the
probability density have
\begin{equation}
  \Delta_{\rm amp} = \frac{\alpha\Delta}{\sqrt{2\pi}},
  \qquad
  \Delta_p = \frac{\alpha\Delta}{2\sqrt{\pi}} = \frac{\Delta_{\rm amp}}{\sqrt{2}},
  \label{eq:tooth_widths}
\end{equation}
$\Delta_p$ being the standard deviation of the momentum distribution. At
$\alpha=\sqrt{2\pi}$ this is the familiar $\Delta_p^2=\Delta^2/2$.

\subsection{Angle observables and the contrast factor}
 Any $2\pi$-periodic $g$ is a
Fourier series in $\hat U^k=e^{ik\hat\chi}$, and $\hat U^k=e^{ik\alpha\hat x}$ shifts
$p\to p+k\alpha$. Inserting a momentum resolution between the displacement and the
encoder, the expectation is a double sum of Mehler-kernel overlaps,
\begin{equation}
    \braket{\Psi|\hat U^k|\Psi}
    = \sum_{n',n} c_{n'}^* c_n \int\! d(\alpha m)\,
      K_{\Delta^2}(\alpha n',(m+k)\alpha)\,K_{\Delta^2}(m\alpha,n\alpha) ,
    \label{eq:chi_double}
\end{equation}
the intermediate $m\alpha$ being the momentum after the first encoder. The Gaussian
integral is elementary, and once normalized by the teeth norms it depends only on the
harmonic $k$ and on the mismatch $\delta\equiv n'-n-k$, not on $n$ separately, because
the flux envelope and the scaling gauge $\alpha$ both cancel in the ratio,
\begin{equation}
    \frac{\int d(\alpha m)\,K_{\Delta^2}(\alpha n',(m+k)\alpha)\,K_{\Delta^2}(m\alpha,n\alpha)}
         {\|\ket{n';\Delta}\|\,\|\ket{n;\Delta}\|}
    = \exp\!\left[-\frac{\pi\,(\delta-\delta_k)^2}{\sinh 2\Delta^2}\right],
    \qquad
    \delta_k \equiv k\big(\cosh\Delta^2-1\big).
    \label{eq:chi_matrix}
\end{equation}
This is a single Gaussian in the mismatch, of width $\mathcal{O}(\Delta)$ and
centered not at $\delta=0$ but at the displaced resonance
$\delta_k\simeq k\Delta^4/2$, the same $\cosh\Delta^{2}$ that contracted the flux
label in Eq.~\eqref{eq:mehler_moments} seen now in the conjugate quadrature.
Expanding the square, the cross term between $\delta$ and $\delta_k$ is
$\mathcal{O}(\Delta^2)$ but odd in $\delta$ and vanishes identically on the
diagonal, so it never enters the contrast. At $\delta=0$, the pair that $\hat U^k$
is built to connect, only $-\pi\delta_k^2/\sinh2\Delta^2$ survives and gives
\begin{equation}
    \gamma_k = \exp\!\left[-\frac{\pi k^2\big(\cosh\Delta^2-1\big)^2}{\sinh 2\Delta^2}\right]
    = 1 - \frac{\pi}{8}\,k^2\Delta^6 + \mathcal{O}(\Delta^{10}),
    \label{eq:gamma_k}
\end{equation}
whose $\mathcal{O}(\Delta^6)$ smallness is now transparent: the resonance is
displaced only at $\mathcal{O}(\Delta^4)$ while the Gaussian width is
$\mathcal{O}(\Delta)$, so the exponent scales as $\delta_k^2/\Delta^2\sim\Delta^6$.
At $\delta\neq0$ the orthogonality term dominates and recovers the tooth-overlap
factor $e^{-\pi\delta^2/\sinh2\Delta^2}\to e^{-\pi\delta^2/2\Delta^2}$ of
Eq.~\eqref{eq:orthogonality}, so the off-diagonal contributions are exponentially
suppressed and the sum collapses to the diagonal term quoted as
Eq.~\eqref{eq:chi_obs}. Below the boundary quoted in Sec.~\ref{sec:err_operators} the
$\mathcal{O}(\Delta^6)$ contrast is the leading finite-$\Delta$ error in the magnetic
sector, while above it the off-diagonal contamination takes over.


\subsection{Wall states and the magnetic contrast}
\label{app:wall}
The wall states of Eq.~\eqref{eq:wall_state} sharpen the operator analysis of
Sec.~\ref{sec:err_operators} in two ways.

First, each $\ket{\phi_n}$ has unit norm and the same momentum width $\Delta_p$ as the
Mehler tooth, but carries no flux envelope: here the squeezing precedes the
displacement, whereas $\mathcal{E}_\Delta$ damps after it, which is exactly what
generated both the envelope (C) and the $\mathcal{O}(\Delta^6)$ distortion of
Eq.~\eqref{eq:gamma_k}. The teeth overlap as
\begin{equation}
    \braket{\phi_m|\phi_n} = e^{-\pi(m-n)^2/2\Delta^2},
    \label{eq:wall_overlap}
\end{equation}
the minimum-uncertainty form of Eq.~\eqref{eq:orthogonality}, and a field observable
retains the same blur, $\bra{\phi_n}f(\hat\eta)\ket{\phi_n}=f(n)+\tfrac12\delta\nu^2
f''(n)+\mathcal{O}(\Delta^4)$ with $\delta\nu^2=\Delta^2/4\pi$, so Eq.~\eqref{eq:eta_obs} holds
as before.

Second, and most important for the magnetic observable, $\hat U^k=e^{ik\alpha\hat x}$ maps
one squeezed vacuum exactly onto the next,
\begin{equation}
    \hat U^k\ket{\phi_n}
    = D\!\Big(\tfrac{ik\alpha}{\sqrt2}\Big)D\!\Big(\tfrac{in\alpha}{\sqrt2}\Big)S(r,\pi)\ket0
    = \ket{\phi_{n+k}},
    \label{eq:wall_shift}
\end{equation}
the displacement-composition phase vanishing for collinear (purely imaginary in this case)
arguments. The aligned matrix element is therefore unity, not $\gamma_k<1$, and
\begin{equation}
    \braket{\Psi_{\rm wall}|\hat U^k|\Psi_{\rm wall}}
    = \sum_n c_{n+k}^* c_n + \mathcal{O}\!\big(e^{-\pi/2\Delta^2}\big):
    \label{eq:wall_contrast}
\end{equation}
the contrast is exactly one up to the exponentially small tooth overlap, with
no polynomial term at all. It is sharper even than the Mehler $\mathcal{O}(\Delta^6)$, because the
squeezed vacuum is not distorted by a post-displacement damping. Consequently the magnetic operator contributes no polynomial error, leaving only the
propagated electric blur and the cutoff tail already recorded in
Sec.~\ref{sec:testing_wall}.

\subsection{Syndrome-extraction protocols}
Two features of the entangling gates make the protocols of
Sec.~\ref{sec:err_active} possible: the analog gate writes the syndrome into the
ancilla while leaving the encoded state untouched, and the digital gate's
back-action is itself a stabilizer. For the analog gate
$\hat{U}_p=\exp(i\hat{p}_s\hat{x}_a)$ the Heisenberg action is
\begin{equation}
    \hat{U}_p^\dagger\,\hat{p}_s\,\hat{U}_p = \hat{p}_s,
    \qquad
    \hat{U}_p^\dagger\,\hat{p}_a\,\hat{U}_p = \hat{p}_a+\hat{p}_s,
    \qquad
    \hat{U}_p^\dagger\,\hat{x}_s\,\hat{U}_p = \hat{x}_s-\hat{x}_a ,
    \label{eq:sum_heisenberg}
\end{equation}
the first relation exact, so the flux content of $\hat{p}_s$ carrying the logical
amplitudes is untouched, and the back-action on $\hat{x}_s$ is absorbed by the
ideal comb, as is explicit in Eq.~\eqref{eq:syndrome_out}.

\paragraph{Analog (ancilla-mode) correction:} Prepare an ancilla oscillator in the
GKP qunaught state $\ket{\varnothing}_a$ and apply the momentum-sum gate
$\hat{U}_{p}=\exp(i\hat{p}_s\hat{x}_a)$~\cite{Glancy:2005aac}. Using $e^{i\phi\hat{x}}\ket{p}=\ket{p+\phi}$,
\begin{equation}
    \hat{U}_p\ket{n\alpha+\delta}_s\ket{m\alpha}_a
    =\ket{n\alpha+\delta}_s\ket{(m+n)\alpha+\delta}_a ,
\end{equation}
so on the full state, relabeling $k=m+n$ at fixed $n$, which runs over
$\mathbb{Z}$ whatever $n$ is,
\begin{equation}
    \ket{\Psi}_{\text{out}} = \ket{\psi_\delta}_s\otimes\sum_{k}\ket{k\alpha+\delta}_a .
    \label{eq:syndrome_out}
\end{equation}
A single homodyne measurement of $\hat{p}_a$ returns $P_{\text{meas}}=k\alpha+\delta$.
Because the ancilla is a coherent sum over all $k$ it carries no information about the
system index $n$ or the coefficients $c_n$. In a twisted sector the comb is centered not
on $\alpha\mathbb{Z}$ but on the target residue $r^\star=\alpha\nu^\star\bmod\alpha$, so
the centered syndrome is
\begin{equation}
    \varepsilon = \big(P_{\text{meas}}-r^\star\big)\bmod\alpha
    \in\Big[-\tfrac{\alpha}{2},\tfrac{\alpha}{2}\Big],
\end{equation}
which reduces to $\varepsilon=P_{\text{meas}}\bmod\alpha=\delta$ when $\nu^\star=0$;
the feedback displacement then acts as in Sec.~\ref{sec:err_active}.

The factorization in Eq.~\eqref{eq:syndrome_out} is exact only for an ideal resource
state, and its failure is the leading cost of the protocol. A finite-energy qunaught
carries an envelope, $\sum_m e^{-2\pi\Delta_a^2m^2}\ket{m\alpha}$ at ancilla squeezing
$\Delta_a$, and the grid shift moves that envelope to be centered on $k=n$. The ancilla state then depends on the system's flux label, the
two no longer factorize, and the syndrome carries which-path information about $n$.
Writing $\ket{A_n}$ for the ancilla branch accompanying $\ket{n\alpha+\delta}_s$,
\begin{equation}
    \braket{A_0|A_n} \simeq e^{-\pi\Delta_a^{2}n^{2}},
    \label{eq:ancilla_dephasing}
\end{equation}
so coherence between flux sectors separated by $n$ is damped at each round. The
protocol is faithful only while $\Delta_a\,\eta_{\max}\ll1$: the ancilla envelope must
be wider than the flux support of the state it is correcting, a requirement on the
resource state distinct from the tooth width that sets the correctable range
$|\delta|<\alpha/2$. In the conjugate description this damping is the random logical
angle shift left by the back-action $\hat{x}_s\to\hat{x}_s-\hat{x}_a$ of
Eq.~\eqref{eq:sum_heisenberg}, an error of the class the comb syndrome cannot itself
detect (Sec.~\ref{sec:err_scope}). It is first order in $\Delta_a^2$, with no exponential suppression, and a complete noise budget must carry it. We do not attempt
this here and leave it for future work.

\paragraph{Digital (qubit-register) correction:} Prepare $\ket{+}^{\otimes M}$ and
apply $M$ controlled stabilizers
$\hat{V}_j=\exp(i2^j\tfrac{2\pi\hat{p}}{\alpha}\otimes\ket{1}\bra{1}_j)$, reserving
$\hat{U}$ for the rotor raising operator of Eq.~\eqref{eq:rotor_algebra}. The
$n\alpha$ contributions give a trivial phase $e^{i2\pi(2^jn)}=1$, so the gauge sector
is again a spectator. The back-action in $x$ is exactly a stabilizer here:
$\hat{V}_j$ translates $\hat{x}_s$ by $2^j(2\pi/\alpha)$
conditioned on qubit $j$, an exact integer multiple of the grid period and hence the
stabilizer power $\hat{S}_\alpha^{2^j}$, which is the identity on the code space
whatever the state of the register. The register becomes
$2^{-M/2}\sum_{k=0}^{2^M-1}e^{i\phi k}\ket{k}$ with $\phi=2\pi\delta/\alpha$. Applying
$\text{QFT}^\dagger$ concentrates the phase on the $\ket{y}$ with
$y/2^M\approx\delta/\alpha$, so a single $Z$-basis measurement yields the binary
fraction $\delta/\alpha\approx0.b_{M-1}\dots b_0$~\cite{Kitaev:1995qy}. Reconstructing
$\varepsilon$ and applying $e^{-i\varepsilon\hat{x}_s}$ corrects to precision
$\alpha/2^M$.

\section{Computational frames and gate synthesis}
\label{app:gates}
In this appendix we give the two frames of Sec.~\ref{sec:frames} in full, together with the
circuits behind the gate counts of Table~\ref{tab:bases}. Formulas are written
frame-agnostically in terms of $\HH^{(2,1,0)}$, $\hat{B}_p$ and
$\hat{\vartheta}_{\bm{n}}$, which is where the two frames differ and hence where the gate
content does. The generic pairs $(\hat{\chi}_i,\hat{\eta}_i)$ of
Eq.~\eqref{eq:rotor_algebra} are realized as $(\hat{B}_p,\hat{\eta}_p)$ in the loop
frame and as $(\hat{\vartheta}_\ell,\hat{e}_\ell)$ in the link frame, and the
raising operator $\hat{U}_i=e^{i\hat{\chi}_i}$ is correspondingly the gauge-fixed
plaquette loop $\hat{U}_p$ or the link element $\hat{u}_\ell$.
\subsection{The two frames}
\label{sec:two_bases}
\begin{figure}[t]
    \centering
    \includegraphics[width=\linewidth]{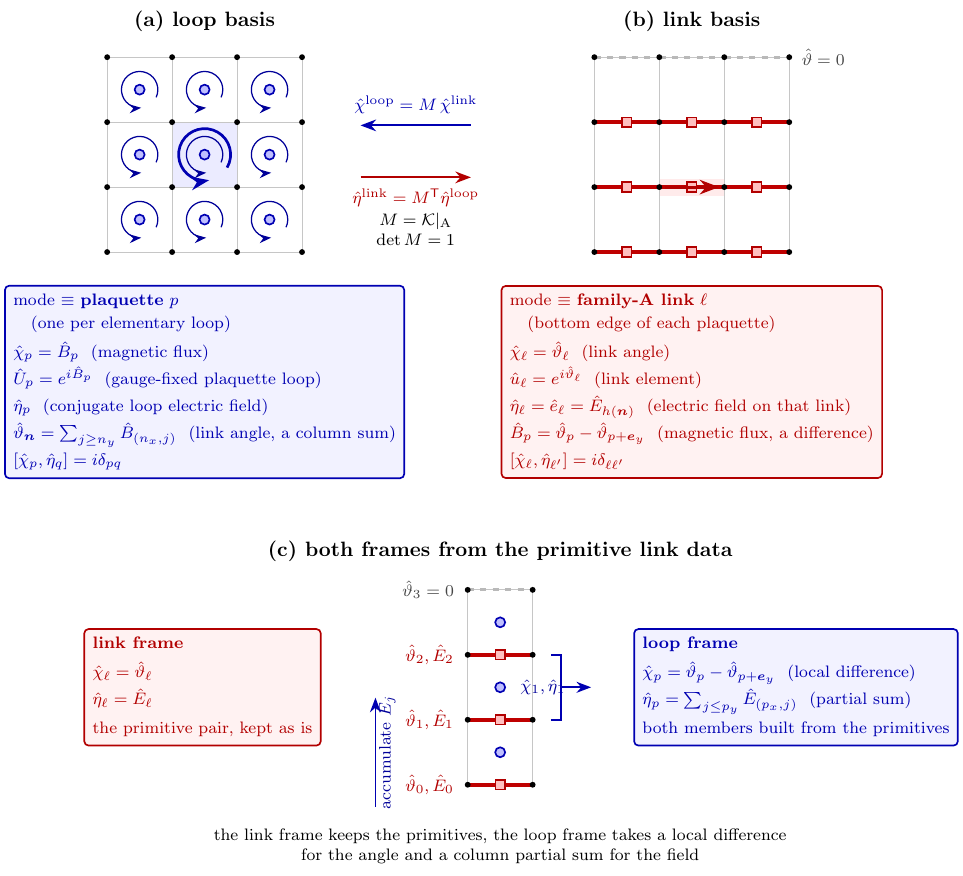}
    \caption{The two choices of dynamical conjugate pairs, drawn on the same $N=3$
    lattice. (a) Loop frame: each of the $N^2$ modes is a plaquette, with
    $\hat{\chi}_p=\hat{B}_p$ the magnetic flux through it and $\hat{\eta}_p$ the
    conjugate loop electric field. (b) Link frame: each mode is a retained family-A
    link, with $\hat{\chi}_\ell=\hat{\vartheta}_\ell$ the link angle and
    $\hat{\eta}_\ell=\hat{e}_\ell$ the physical electric field on that link, the top
    row being gauge-fixed to $\hat{\vartheta}=0$. The angles transform with
    $M=\mathcal{K}|_{\rm A}$ and the fields with $M^{\mathsf T}$ in the opposite direction, which preserves conjugacy. (c) Both frames expressed from the
    primitive data. Every link of the unreduced theory carries a pair
    $(\hat{\vartheta}_\ell,\hat{E}_\ell)$. The link frame keeps it as it stands,
    while the loop frame builds its angle as a local difference of primitive angles
    and its field as a partial sum of primitive fields up the column. The frames
    coincide at $N=1$, where all numerics of this work are performed.}
    \label{fig:two_frames}
\end{figure}
Both assignments are drawn in Fig.~\ref{fig:two_frames}. They describe the same theory and differ in which geometric object carries a mode, and therefore in which of the two non-Gaussian terms is elementary.
\paragraph{Loop basis:} The modes are the plaquette pairs: $\hat{\chi}_p=B_p$ is the
magnetic flux and $\hat{\eta}_p$ the conjugate loop electric field
(Fig.~\ref{fig:two_frames}a).
\paragraph{Link basis:} The modes are the retained links: $\hat{\chi}_\ell=\vartheta_\ell$
is the link angle, $\hat{u}_\ell=e^{i\hat{\vartheta}_\ell}$, and
$\hat{\eta}_\ell=\hat{e}_\ell\equiv\hat{E}_{h(\bm{n})}$ the physical link field
(Fig.~\ref{fig:two_frames}b). The retained links are the bottom edges of the
plaquettes, so the link modes sit half a cell below the loop modes of the first
frame.
\paragraph{Dictionary:} With $M\equiv\mathcal{K}|_{\rm A}$ the incidence matrix
restricted to the retained columns,
\begin{equation}
    \hat{\chi}^{\rm loop} = M\,\hat{\chi}^{\rm link},
    \qquad
    \hat{\eta}^{\rm link} = M^{\mathsf T}\hat{\eta}^{\rm loop},
    \qquad
    \det M = 1 ,
    \label{eq:basis_M}
\end{equation}
the second because family A carries no charge string. The angles are carried one way
by $M$ and the fields the other way by $M^{\mathsf T}$, which is the content of the
two opposing arrows in Fig.~\ref{fig:two_frames}. Component-wise,
\begin{equation}
\begin{aligned}
    B_{(p_x,p_y)} &= \vartheta_{(p_x,p_y)} - \vartheta_{(p_x,p_y+1)},
    &\qquad
    \vartheta_{(n_x,n_y)} &= \sum_{j\ge n_y}B_{(n_x,j)},\\
    \hat{e}_{(n_x,n_y)} &= \hat{\eta}_{(n_x,n_y)} - \hat{\eta}_{(n_x,n_y-1)},
    &\qquad
    \hat{\eta}_{(n_x,n_y)} &= \sum_{j\le n_y}\hat{e}_{(n_x,j)},
\end{aligned}
\label{eq:basis_dictionary}
\end{equation}
with $\vartheta_{(p_x,N)}\equiv0$ (gauge-fixed) and $\hat{\eta}_{(n_x,-1)}\equiv0$.
Equivalently $\hat{U}_p=\hat{u}_{(p_x,p_y)}\hat{u}^\dagger_{(p_x,p_y+1)}$: the
plaquette Wilson loop $\hat{u}_{\rm b}\hat{u}_{\rm r}\hat{u}_{\rm t}^\dagger\hat{u}_{\rm l}^\dagger$
with the vertical factors removed by the gauge fixing. All four relations act within
a single column, and each frame is reached from the primitive link pair in a
different way, as shown in Fig.~\ref{fig:two_frames}c: the link frame simply retains
$(\hat{\vartheta}_\ell,\hat{E}_\ell)$, while the loop frame obtains its angle by a
local difference and its field by a partial sum. Difference and partial sum are adjoint operations, and this adjointness preserves conjugacy,
$[\hat{\vartheta}_\ell,\hat{e}_{\ell'}]=i(M^{-1}M)_{\ell\ell'}=i\delta_{\ell\ell'}$.
For $N=1$ the frames coincide.
Consequently, in the notation of Eq.~\eqref{eq:U1_ham_dynam},
\begin{equation}
\begin{aligned}
    \text{loop:}\quad && \hat{B}_p &= \hat{\chi}_p, &
    \hat{\vartheta}_{\bm{n}} &= \sum_{j\ge n_y}\hat{\chi}_{(n_x,j)},\\
    \text{link:}\quad && \hat{B}_p &= \hat{\chi}_{(p_x,p_y)}-\hat{\chi}_{(p_x,p_y+1)}, &
    \hat{\vartheta}_{\bm{n}} &= \hat{\chi}_{\bm{n}} .
\end{aligned}
\label{eq:B_Theta_bases}
\end{equation}
Each frame makes one of the two non-Gaussian terms simple and the other composite.
The bilinear still needs exactly one link element in both, and what changes is only
whether that element is a mode or a product of modes.
\subsection{The transformed matrices}
\label{sec:transformed_matrices}
Since $\hat{E}_\ell=(\mathcal{K}^{\mathsf T}\hat{\eta}-\mathcal{C}\hat{Q})_\ell$, the
entire change of frame is the row transformation
\begin{equation}
    \tilde{\mathcal{K}} \equiv M^{-1}\mathcal{K},
    \qquad
    \mathcal{C}\ \text{unchanged},
    \label{eq:Ktilde}
\end{equation}
so the Gram structure of Appendix~\ref{app:GL} survives and
\begin{equation}
    \tilde{\mathcal{H}}^{(2)} = M^{-1}\mathcal{H}^{(2)}M^{-\mathsf T},
    \qquad
    \tilde{\mathcal{H}}^{(1)} = M^{-1}\mathcal{H}^{(1)},
    \qquad
    \tilde{\mathcal{H}}^{(0)} = \mathcal{H}^{(0)}.
    \label{eq:Htilde}
\end{equation}
The charge--charge block is exactly invariant, involving only $\mathcal{C}$. By
construction $\tilde{\mathcal{K}}|_{\rm A}=\Id$: the retained links contribute the
identity block, which is the statement that $\hat{e}_\ell$ is a mode.
In either frame the electric kernel is the Gram matrix of the incidence rows,
$\HH^{(2)}=\mathcal{K}_{\rm frame}\mathcal{K}_{\rm frame}^{\mathsf T}$, so it is
positive definite and transforms by congruence rather than by similarity, which is
what allows the sparsity pattern and the trace to be frame dependent. In the loop
frame $\mathcal{H}^{(2)}=4\Id-A$ with $A$ the adjacency matrix of the $N\times N$
plaquette graph, so $\mathrm{nnz}=5N^2-4N$ and
$\mathrm{tr}\,\mathcal{H}^{(2)}=4N^{2}$. In the link frame $M^{-1}=L^{\mathsf T}$,
with $L$ the column partial-sum matrix of Eq.~\eqref{eq:basis_dictionary}, and the
congruence $\tilde{\mathcal{H}}^{(2)}=L^{\mathsf T}(4\Id-A)L$ evaluates in closed
form, for $a=(a_x,a_y)$ and $b=(b_x,b_y)$, to
\begin{equation}
    \tilde{\mathcal{H}}^{(2)}_{ab}
    = \delta_{ab} + \delta_{a_xb_x}
    + \big(2\delta_{a_xb_x} - \delta_{|a_x-b_x|,1}\big)\big(N-\max(a_y,b_y)\big),
    \label{eq:H2tilde_closed}
\end{equation}
dense within each column and nearest neighbor between columns, the mirror image of
$\mathcal{H}^{(2)}$, which is nearest neighbor in both directions. Carrying out the
counts,
\begin{equation}
    \mathrm{nnz}\big(\tilde{\mathcal{H}}^{(2)}\big) = N^2(3N-2),
    \qquad
    \mathrm{nnz}\big(\tilde{\mathcal{H}}^{(1)}\big) = \tfrac12 N^2(N^2+4N+1),
    \qquad
    \mathrm{tr}\,\tilde{\mathcal{H}}^{(2)} = N^2(N+3),
    \label{eq:nnz_tilde}
\end{equation}
against $5N^2-4N$, $\tfrac12(3N^3+2N^2+N)$ and $4N^2$ in the loop basis.
\subsection{Frame invariants and non-invariants}
\label{sec:basis_invariants}
Because $M$ is integral with $\det M=1$, the frames are equivalent over
$\mathbb{Z}$, not merely over $\mathbb{R}$, which is what the compact theory
requires.
\paragraph{Flux quantization:} $\hat{e}=M^{\mathsf T}\hat{\eta}$ with
$M^{\mathsf T}$ unimodular, so $\hat{e}\in\mathbb{Z}^{N^2}\iff\hat{\eta}\in\mathbb{Z}^{N^2}$,
and $M\mathbb{Z}^{N^2}=\mathbb{Z}^{N^2}$ means $\hat\chi=M\hat\vartheta$ is a
re-coordinatization of the same torus ($T^{N^2}\cong (S^1)^{\times N^2}$). Hence the CV
map~\eqref{eq:CCR_rep_alpha}, the stabilizer~\eqref{eq:stab}, the
encoder~\eqref{eq:encoder} and $\braket{\hat{S}_\alpha}=e^{-\pi\Delta^2/2}$
hold in both, and the penalty costs the same.
\paragraph{Twist:} $\tilde{d}=M^{\mathsf T}d$, i.e.\
$\tilde{d}_{(n_x,n_y)}=d_{(n_x,n_y)}-d_{(n_x,n_y-1)}$. Unimodularity gives
$d\in\mathbb{Z}^{N^2}\iff\tilde{d}\in\mathbb{Z}^{N^2}$: the trivial-twist
classification of Sec.~\ref{sec:displacement} is frame-independent.
\paragraph{Energies:} $E_{\rm cl}$ and $\Delta_{\rm tw}$ are scalars and
unchanged. Indeed, $\HH^{(0)\rm eff}$ is invariant term by term, since
$\tilde{\mathcal{H}}^{(1)\mathsf T}(\tilde{\mathcal{H}}^{(2)})^{-1}\tilde{\mathcal{H}}^{(1)}
= \mathcal{H}^{(1)\mathsf T}(\mathcal{H}^{(2)})^{-1}\mathcal{H}^{(1)}$.
\paragraph{Gauss's law:} The reference corner, its vanishing columns and the
neutrality condition involve only $\mathcal{C}$, untouched by
Eq.~\eqref{eq:Ktilde}.
The exception is the encoding bias~\eqref{eq:dE_trace}. The GKP error is per-mode,
$M$ is unimodular but not orthogonal, and the trace is not preserved, which is the
frame comparison of Eq.~\eqref{eq:bias_ratio} and Sec.~\ref{sec:frames}.
\begin{table}[h]
\centering
\small
\begin{tabular}{lll}
\hline
Block & Loop basis & Link basis\\
\hline
$\hat{H}_E$ kernel & $4\Id-A$; nnz $=5N^2-4N$ & Eq.~\eqref{eq:H2tilde_closed}; nnz $=N^2(3N-2)$\\
\quad SUM gates    & $2N(N-1)$                & $\tfrac32N^2(N-1)$\\
$\hat{H}_E$ linear & $\tfrac12(3N^3{+}2N^2{+}N)$ D/CD$_z$ & $\tfrac12N^2(N^2{+}4N{+}1)$ D/CD$_z$\\
$\hat{H}_E$ charge & $\tfrac12N(N{+}1)(N^2{+}N{+}2)$ & identical\\
$\hat{H}_B$        & $N^2$ single-mode $\cos$ & $N(N{-}1)$ two-mode $+$ $N$ single-mode $\cos$\\
$\hat{H}_K$        & $\tfrac12N^2(N{+}1)$ CD & $N^2$ CD\\
penalty            & $N^2$ single-mode $\cos$ & identical\\
$\braket{\delta E}$ & $\propto 4N^2$ & $\propto N^2(N+3)$\\
\hline
\end{tabular}
\caption{Gate content and encoding bias of the two frames. The frames coincide at $N=1$. For $N=1,\dots,5$ the (SUM, D/CD$_z$,
$\hat{H}_K$) triples are $(0,3,1)$, $(4,17,6)$, $(12,51,18)$, $(24,114,40)$,
$(40,215,75)$ in the loop basis and $(0,3,1)$, $(6,26,4)$, $(27,99,9)$,
$(72,264,16)$, $(150,575,25)$ in the link basis.}
\label{tab:bases}
\end{table}

\subsection{Gate decomposition of the time evolution}
\label{sec:gates}
With the two frames in hand we can count what a simulation costs in each. We
summarize the gate content of one Trotter step $U_H(\delta t)=\exp(-i\,\delta t\,H)$,
frame-agnostically in terms of $\HH^{(2,1,0)}$, $\hat{B}_p$ and
$\hat{\vartheta}_{\bm{n}}$, with the circuits given in the two subsections that follow
and the counts collected in Table~\ref{tab:bases}.
The electric term splits into three pieces. The quadratic part is diagonalized mode by
mode with a quadratic phase gate $P(s)=e^{-is\hat{x}^2/2}$ between Fourier gates, and
each off-diagonal pair is synthesized with a two-mode SUM gate $e^{-is\hat{x}_i\hat{p}_j}$.
The SUM count is $\tfrac12[\mathrm{nnz}(\HH^{(2)})-N^2]$, which in the loop frame is
$2N(N-1)$. The linear part is a product of displacements for static charges, and for
dynamical ones splits into a bosonic displacement and a conditional displacement
$\text{CD}_z(\beta)=\exp[(\beta\hat{a}^\dagger-\beta^*\hat{a})\otimes Z]$, at a cost
$\mathrm{nnz}(\HH^{(1)})$. The charge--charge part is frame-invariant and needs $\tfrac12N(N+1)(N^2+N+2)$ rotations.

The magnetic term is built from conditional displacements along $\hat{B}_p$, $3N^2$ of
them in the loop frame where $\hat{B}_p$ is a single quadrature. In the link frame each
column is instead an open Josephson chain with its upper end pinned. The mass term is
$(N+1)^2$ single-qubit rotations. The kinetic term reduces, after a CNOT-conjugated
basis change, to controlled trigonometric gates
$\exp[-i\lambda f(\hat{\vartheta})\otimes Z]$~\cite{Rainaldi:2025ymn}. Because
$\hat{\vartheta}_{\bm{n}}$ is one quadrature in the link frame but a sum of $N-n_y$ in the
loop frame, this costs $N^2$ conditional displacements against $\tfrac12N^2(N+1)$.
\subsection{Electric term}
\paragraph{Quadratic part:} The diagonal contributions
\begin{equation}
    \exp\!\Big(-i\,\delta t\,\tfrac{g^2}{2}\HH_{ii}^{(2)}\tfrac{\hat{p}^2_i}{\alpha^2_i}\Big)
    = \bigotimes_{i}F_i^\dagger P\!\big(\delta t\,g^2\HH_{ii}^{(2)}/\alpha_i^2\big)F_i
\end{equation}
use a quadratic phase gate $P(s)=\exp(-is\hat{x}^2/2)$ and Fourier gates.
\begin{equation}
    \begin{quantikz}
    \lstick{$\ket{\psi_1}$} \setwiretype{b} & \gate{F} & \gate{P(\delta t\,g^2\HH^{(2)}_{11} /\alpha_1^2)} & \gate{F^\dagger} & \qw \\
    \setwiretype{b} & \push{\vdots} & \push{\vdots} & \push{\vdots} & \\
    \lstick{$\ket{\psi_{N^2}}$} \setwiretype{b} & \gate{F} & \gate{P(\delta t\,g^2\HH^{(2)}_{N^2N^2} /\alpha_{N^2}^2)} & \gate{F^\dagger} & \qw \\
\end{quantikz}
\end{equation}
Each off-diagonal pair $i\neq j$ gives
\begin{equation}
    \exp\!\Big(-i\,\delta t\,\tfrac{g^2}{\alpha_i\alpha_j}\HH^{(2)}_{ij}\hat{p}_i\hat{p}_j\Big)
    = F_i^\dagger\,\text{SUM}_{ij}\!\Big(\delta t\,\tfrac{g^2}{\alpha_i\alpha_j}\HH^{(2)}_{ij}\Big)F_i,
    \qquad
    \text{SUM}_{ij}(s)\equiv e^{-is\hat{x}_i\hat{p}_j},
    \label{eq:SUM_gate}
\end{equation}
the argument doubled because the sum over $i\neq j$ counts each unordered pair
twice. The number of SUM gates is
$\tfrac12[\mathrm{nnz}(\HH^{(2)})-N^2]$. In the loop basis $\HH^{(2)}=4\Id-A$ has
$\mathrm{nnz}=5N^2-4N$, giving $2N(N-1)$, whose $3\times3$ circuit geometry is
\begin{equation}
    \begin{quantikz}
    \lstick{$\ket{\psi_1}$} \setwiretype{b} & \gate[style={diamond, draw}]{} \vqw{1} &   &  \gate[style={diamond, draw}]{} \vqw{3} &  & & & & & \qw \\
    \lstick{$\ket{\psi_2}$} \setwiretype{b} & \gate[style={circle, draw}]{} &  \gate[style={diamond, draw}]{} \vqw{1} &  & \gate[style={diamond, draw}]{} \vqw{3}& & & & & \qw \\
    \lstick{$\ket{\psi_3}$} \setwiretype{b} & & \gate[style={circle, draw}]{}  &  &  & \gate[style={diamond, draw}]{} \vqw{3} & & & & \qw \\
    \lstick{$\ket{\psi_4}$} \setwiretype{b} & \gate[style={diamond, draw}]{} \vqw{1} &   & \gate[style={circle, draw}]{}  & & & \gate[style={diamond, draw}]{} \vqw{3}& & & \qw \\
    \lstick{$\ket{\psi_5}$} \setwiretype{b} & \gate[style={circle, draw}]{} &  \gate[style={diamond, draw}]{} \vqw{1} &  &\gate[style={circle, draw}]{} & & & \gate[style={diamond, draw}]{} \vqw{3}& & \qw \\
    \lstick{$\ket{\psi_6}$} \setwiretype{b} & & \gate[style={circle, draw}]{}  &  & & \gate[style={circle, draw}]{} & & & \gate[style={diamond, draw}]{} \vqw{3} & \qw \\
    \lstick{$\ket{\psi_7}$} \setwiretype{b} & \gate[style={diamond, draw}]{} \vqw{1} &   &  & & &\gate[style={circle, draw}]{} & &  & \qw \\
    \lstick{$\ket{\psi_8}$} \setwiretype{b} & \gate[style={circle, draw}]{} &  \gate[style={diamond, draw}]{} \vqw{1} &  & & & & \gate[style={circle, draw}]{} & & \qw \\
    \lstick{$\ket{\psi_9}$} \setwiretype{b} & & \gate[style={circle, draw}]{}  &  & & & & &\gate[style={circle, draw}]{}  & \qw \\
\end{quantikz}
\label{eq:Adjacency_SUM_gates_3x3}
\end{equation}
Diamonds and circles are the left and right indices of Eq.~\eqref{eq:SUM_gate}. Each
diamond is Fourier transformed.
\paragraph{Linear part:} For static charges this is a product of displacements,
\begin{equation}
    \exp\!\Big(-i\,\delta t\,\tfrac{g^2}{2}\tfrac{\hat{p}_i}{\alpha_i}\HH^{(1)}_{ij}Q_j\Big)
    = \bigotimes_{i}D\!\Big(\tfrac{g^2\,\delta t}{2\sqrt{2}\,\alpha_i}\HH^{(1)}_{ij}Q_j\Big),
    \label{eq:static_linear}
\end{equation}
using $e^{-is\hat{p}}=D(s/\sqrt{2})$. For dynamical charges,
Eq.~\eqref{eq:Q_pauli} splits it into a bosonic piece from
$\hat{Q}_j|_{\Id}=(-1)^{n_x+n_y}/2$ and a hybrid piece from $\hat{Q}_j|_{Z}=-Z_j/2$:
\begin{equation}
    \bigotimes_{i}D\!\Big(\tfrac{g^2\delta t}{4\sqrt{2}\alpha_i}\HH^{(1)}_{ij}(-1)^{n_x+n_y}\Big)
    \quad\text{and}\quad
    \bigotimes_{i,j}\text{CD}_z\!\Big({-}\tfrac{g^2\delta t}{4\sqrt{2}\alpha_i}\HH^{(1)}_{ij}\Big),
    \label{eq:dyn_linear}
\end{equation}
the minus sign following from the $Z$ convention. The count is
$\mathrm{nnz}(\HH^{(1)})$.
\paragraph{Charge--charge part:} Relevant only for dynamical charges. Being
quadratic, it needs only single-qubit rotations and CNOT-conjugated rotations. Since
$\HH^{(0)}$ is frame-invariant, so is the count,
\begin{equation}
    \#QQ = \tfrac12\big[\mathrm{nnz}(\mathcal{H}^{(0)})+\mathrm{nnz}(\mathrm{diag}\,\mathcal{H}^{(0)})\big]
         = \tfrac12 N(N+1)(N^2+N+2),
\end{equation}
using $\mathrm{nnz}(\mathcal{H}^{(0)})=N^4+2N^3+2N^2$ and
$\mathrm{nnz}(\mathrm{diag}\,\mathcal{H}^{(0)})=(N+1)^2-1=N^2+2N$, the deficit of one
being the reference corner.
\subsection{Magnetic, mass and kinetic terms}
\paragraph{Magnetic:} Up to a global phase,
$U_B(\delta t)=\bigotimes_{p}\exp(i\tfrac{\delta t}{g^2}\cos\hat{B}_p)$, each factor
approximated to $\mathcal{O}(\delta t^2/g^4)$ by
\begin{equation}
        \begin{quantikz}
    \lstick{$\ket{0_p}$} & \ctrl{1} & \gate{\text{R}^\dagger_z(\delta t/g^2)} & \ctrl{1} & \gate{\text{R}^\dagger_z(\delta t/g^2)} & \ctrl{1}  & \\
    \lstick{$\ket{\psi_p}$} \setwiretype{b} & \gate{\text{CD}_x} & \qw & \gate{(\text{CD}_x^\dagger)^2} & \qw & \gate{\text{CD}_x} & \qw
\end{quantikz},
\end{equation}
with $\text{CD}_x$ displacing along $\hat{B}_p$, the controlled-trigonometric
primitive of Ref.~\cite{Rainaldi:2025ymn}. In the loop basis $\hat{B}_p=\alpha_p\hat{x}_p$
is single-mode and the count is $3N^2$ conditional displacements plus $2N^2$
rotations. In the link basis $N(N-1)$ plaquettes give a two-mode
$\cos(\alpha\hat{x}_p-\alpha\hat{x}_{p+\bm{e}_y})$ and $N$ give a single-mode cosine
(the top-row link being gauge-fixed), so that per column $\hat{H}_B$ is an open
Josephson chain with its upper end pinned and the columns decouple. The ancilla
qubits here are not the fermions of the theory.
\paragraph{Mass:} From $\hat{H}_M=m_0\sum_{\bm{n}}(-)^{n_x+n_y}(\Id-Z_{s(\bm{n})})/2$,
the $Z$ piece gives $(N+1)^2$ rotations
$R_{z}\big(-\delta t\,m_0(-1)^{n_x+n_y}\big)$.
\paragraph{Kinetic:} With Eq.~\eqref{eq:JW} the horizontal hops are string-free and
the vertical ones carry a string. Keeping the prefactors of
Eq.~\eqref{eq:U1_ham_dynam},
\begin{equation}
\begin{split}
    \hat{H}_{K,\rm hor} &= \frac{1}{2}\sum_{\bm{n},\,n_x<N,\,n_y<N}
        \Big[\,i\,\hat\Psi^\dagger_{\bm{n}}\,\hat{u}^\dagger_{\bm{n},\bm{e}_x}\,\hat\Psi_{\bm{n}+\bm{e}_x} + \text{h.c.}\Big]\\
    &= \frac{1}{4}\sum_{\bm{n},\,n_x<N,\,n_y<N}\Big[
        \sin\hat{\vartheta}_{\bm{n}}\,\big(X_{\bm{n}}X_{\bm{n}+\bm{e}_x}+Y_{\bm{n}}Y_{\bm{n}+\bm{e}_x}\big)
      - \cos\hat{\vartheta}_{\bm{n}}\,\big(X_{\bm{n}}Y_{\bm{n}+\bm{e}_x}-Y_{\bm{n}}X_{\bm{n}+\bm{e}_x}\big)\Big],
\end{split}
\label{eq:HK_hor}
\end{equation}
the top row following at $\hat{\vartheta}\to0$, and
\begin{equation}
    \hat{H}_{K,\rm vert} = -\frac{1}{4}\sum_{\bm{n}}(-1)^{n_x+n_y}\,Z_{\rm str}(\bm{n})
        \big(X_{\bm{n}}X_{\bm{n}+\bm{e}_y}+Y_{\bm{n}}Y_{\bm{n}+\bm{e}_y}\big),
\label{eq:HK_vert}
\end{equation}
with the snake string
\begin{equation}
    Z_{\rm str}(\bm{n}) =
    \begin{cases}
        \prod_{x=n_x+1}^{N}Z_{(x,n_y)}Z_{(x,n_y+1)} & n_y \text{ even},\\[1ex]
        \prod_{x=0}^{n_x-1}Z_{(x,n_y)}Z_{(x,n_y+1)} & n_y \text{ odd},
    \end{cases}
\end{equation}
of $2(N-n_x)$ and $2n_x$ factors. Writing $\hat{\mathcal{S}}=X_{\bm{n}}X_{\bm{n}+\bm{e}_x}+Y_{\bm{n}}Y_{\bm{n}+\bm{e}_x}$ and
$\hat{\mathcal{A}}=X_{\bm{n}}Y_{\bm{n}+\bm{e}_x}-Y_{\bm{n}}X_{\bm{n}+\bm{e}_x}$, one has
$\hat{\mathcal{S}}^2=\hat{\mathcal{A}}^2=2(\Id-Z_{\bm{n}}Z_{\bm{n}+\bm{e}_x})$ and
$\{\hat{\mathcal{S}},\hat{\mathcal{A}}\}=0$: on the two-dimensional subspace where they act they
are a pair of anticommuting Pauli operators, so
$\sin\hat\vartheta\,\hat{\mathcal{S}}-\cos\hat\vartheta\,\hat{\mathcal{A}}$ is a single rotation of
fixed angle about a $\hat\vartheta$-dependent axis. After a CNOT-conjugated basis change
each term is therefore a controlled trigonometric gate
$\exp[-i\lambda f(\hat{\vartheta})\otimes Z]$~\cite{Rainaldi:2025ymn}. The vertical terms are
the same with $f\to1$ and $Z_{\rm str}$ absorbed into the control by a CNOT ladder
of depth $\mathcal{O}(N)$. By Eq.~\eqref{eq:B_Theta_bases}, $\hat{\vartheta}_{\bm{n}}$
is a single quadrature in the link basis, giving $N^2$ conditional displacements,
but a sum of $N-n_y$ quadratures in the loop basis, giving
\begin{equation}
    \sum_{n_x=0}^{N-1}\sum_{n_y=0}^{N-1}(N-n_y) = \tfrac12 N^2(N+1).
    \label{eq:HK_count_loop}
\end{equation}

\subsection{Choosing a frame}
\label{sec:basis_choice}
The displaced form of Eq.~\eqref{eq:displaced} is for static charges. Once the
charges are dynamical its charge--charge block densifies,
$\mathrm{nnz}(\HH^{(0)\rm eff})=[(N+1)^2-1]^2$, and the twist
$\hat{\theta}=\pi(\HH^{(2)})^{-1}\HH^{(1)}\hat{Q}$ becomes an operator supported on
all non-reference qubits, turning the twisted penalty into an operator-valued
cosine whose Pauli expansion carries $2^{k_i}$ terms per mode. This is the cost
comparison behind the static-versus-dynamical rule of Sec.~\ref{sec:frames}: one
prefers the undisplaced Eq.~\eqref{eq:undisplaced} whenever $\hat{H}_K$ is present.

The loop-basis advantage under uniform weighting, quoted in Sec.~\ref{sec:scaling_gates},
originates in the column partial-sum rule: what makes $\hat{u}_\ell$ single-mode also
spreads $\HH^{(1)}$ over the whole column.
However, uniform weighting is not the only metric to consider. The link basis buys a reduction of
the non-Gaussian kinetic cost from $\tfrac12N^2(N+1)$ to $N^2$ conditional
displacements (cubic to quadratic) paid for in Gaussian SUM gates and a
doubling of the $\hat{H}_B$ cosines. On hardware where hybrid qubit-qumode gates
dominate the error budget and Gaussian two-mode operations are nearly free, that
trade may be worth making. At presently reachable $N$ the two effects are of
comparable size. In the pure-gauge or static-charge sector, where $\hat{H}_K$ is
absent and the displacement removes the linear term, the loop basis wins outright,
its electric term needing only $2N(N-1)$ SUM gates.
This is a choice of computational frame, not of physics: by
Sec.~\ref{sec:basis_invariants} the spectrum, $E_{\rm cl}$, $\Delta_{\rm tw}$, flux
quantization and the trivial-twist classification are identical. 
\section{Measurement circuits and spectroscopy}
\label{app:measurement}
In this appendix we give the hardware protocols for the observables of
Secs.~\ref{sec:testing} and~\ref{sec:pair_add}, together with the demodulation
pipeline and the assumptions behind the shot budgets of
Sec.~\ref{sec:scaling_meas}. The measurements all reduce to a Hadamard test with a
single ancilla~\cite{Somma:2001kjh,Pedernales:2014izf}, whose bosonic ingredient is
the qubit-controlled displacement reading out the characteristic function
$\braket{e^{i\alpha\hat{x}}}$~\cite{Fluhmann:2019slh,Campagne-Ibarcq:2019nmy}. We
interleave controlled trigonometric operators, built from the gates of
Ref.~\cite{Rainaldi:2025ymn}, with the Hamiltonian evolution to extract the full
complex one- and two-point functions of Hermitian trigonometric observables of the
encoded rotors.

\subsection{Hadamard tests}
\label{app:two_point}
For Hermitian $\hat{A}$ the Hadamard test with one ancilla gives
$\braket{e^{i\hat{A}}}$: initializing in $\ket{+}$, applying
$e^{i\hat{A}\otimes Z/2}$ and a final Hadamard, a $Z$-basis measurement returns
$P_{\uparrow}-P_{\downarrow}=\braket{\cos\hat{A}}$, and inserting $S^{\dagger}$
before the final Hadamard returns $\braket{\sin\hat{A}}$ instead. Quadratic moments
follow from $\hat{A}\to\epsilon\hat{A}$ with $\epsilon\ll1$, through
$\braket{\hat{A}^2}=2\epsilon^{-2}(1-P_{\uparrow}+P_{\downarrow})+\mathcal{O}(\epsilon^2)$,
improved by fitting a parabola over several $\epsilon$.

For the complex correlator $\braket{\hat{O}(t)\hat{O}(0)}$ of a Hermitian $\hat{O}$,
define
\begin{equation}
  C_{\sigma\sigma'}(t;\epsilon)
  \equiv \bra{\psi}e^{i\sigma\epsilon\hat{O}(t)}e^{i\sigma'\epsilon\hat{O}(0)}\ket{\psi},
  \qquad \sigma,\sigma'\in\{+,-\},
  \label{eq:Css}
\end{equation}
each accessible by a Hadamard test. Expanding to $\mathcal{O}(\epsilon^2)$, the
cross term carries a factor $\sigma\sigma'$, the $\mathcal{O}(\epsilon)$ one-point
terms are purely imaginary, and combining circuits at $\pm\epsilon$ cancels both
the one-point and the single-time contamination:
\begin{equation}
\begin{split}
  \operatorname{Re}\big[\braket{\hat{O}(t)\hat{O}(0)}\big]
  &= \frac{\operatorname{Re}C_{+-} - \operatorname{Re}C_{++}}{2\epsilon^2} + \mathcal{O}(\epsilon^2),\\
  \operatorname{Im}\big[\braket{\hat{O}(t)\hat{O}(0)}\big]
  &= -\frac{\operatorname{Im}C_{++}(\epsilon) + \operatorname{Im}C_{++}(-\epsilon)}{2\epsilon^2} + \mathcal{O}(\epsilon),
\end{split}
  \label{eq:two_point_real}
\end{equation}
the standard variant giving the real parts and the $S^\dagger H$ variant the
imaginary, so four circuits per time $t$ determine the full complex correlator. The
differing error orders arise because the next correction is
$\mathcal{O}(\epsilon^4)$ in the first case and $\mathcal{O}(\epsilon^3)$ in the
second. As a by-product the same circuits give the one-point functions from the
odd-in-$\epsilon$ parts of $\operatorname{Im}C_{\sigma\sigma'}$.

\subsection{Correlator, spectral function and demodulation}
\label{app:demod}
The time-domain correlator and its spectral function are
\begin{equation}
    C_b(t) = e^{iE_0 t}\bra{\rm GS}\hat{O}_b^\dagger e^{-iHt}\hat{O}_b\ket{\rm GS}
           = \sum_s |Z_s|^2 e^{-i\Omega_s t},
    \qquad
    A(\omega) = \int_0^{T_{\rm max}}\!\! C_b(t)e^{i\omega t}\mathrm{d}t,
\end{equation}
with $\Omega_s=E_s-E_0$ and $Z_s=\braket{s|\hat{O}_b|{\rm GS}}$. A finite
window gives peaks of width $\delta\omega=2\pi/T_{\rm max}$. An envelope
$e^{-\Gamma t}$ suppresses ringing at the cost of a Lorentzian of half-width
$\Gamma$. We use $\Gamma=0.02$ for a two-level toy model carrying the pair-addition
channel at coupling $\kappa$, on which we validate the pipeline below, and
$\Gamma=0.05$ for the full theory of Sec.~\ref{subsec:qed3_plaquette}. Neither is
physical, since the simulated systems are closed, and both bias the extracted line,
because the Lorentzian tails of the neighboring lines drag the peak. That bias is
quadratic in $\Gamma$, the ratio $(\text{bias})/\Gamma^{2}$ holding constant to
better than $1\%$ over a decade, so it is removed exactly as the finite-squeezing bias
of Sec.~\ref{sec:err_active} is: run at several $\Gamma$ and read the
$\Gamma^{2}\to0$ intercept. The construction needs every line in its Lorentzian limit,
$\Gamma T_{\rm max}\gtrsim10$, so the extrapolation is run at $T_{\rm max}=600$ with
$\Gamma\in\{0.05,0.04,0.03,0.02\}$, while the $T_{\rm max}=150$ spectra sit below this criterion and serve to locate the lines, every number entering
Eq.~\eqref{eq:double_extrap} coming from the $T_{\rm max}=600$ runs. In the
pure-gauge limit of the toy model, $\kappa\to0$ and $m_0\to\infty$, with the
correlator computed by direct time evolution and no shortcut through the
eigendecomposition, it sharpens the twist energy from $+1.18\%$ at the best single
$\Gamma$ to $-0.04\%$ at the intercept, $0.0126316$ against the exact $0.0126361$,
the numerical verification of Eq.~\eqref{eq:Omega0} in the electric and the compact
encoded basis alike.
The demodulated correlator of Sec.~\ref{sec:pair_add} is
$\tilde C_b(t) = e^{i(2m_0+E_{\rm cl})t}C_b(t)$, the reference needing only the classical
Gauss-law matrices and Eq.~\eqref{eq:Q_bottom}.
\subsection{Resolution and shot-budget assumptions}
\label{app:meas_scaling}
The estimates of Sec.~\ref{sec:scaling_meas} rest on two structural facts and a set of assumptions, which we state separately. The first fact is that the twist line never has
to be separated from a line an exponentially small distance away, the nearest
large-weight feature after demodulation being the first gauge or spectator
excitation at
\begin{equation}
    \Delta_{\rm gap} = \mathcal{O}(1) \quad\text{in lattice units}
    \label{eq:gap_def}
\end{equation}
away at fixed volume, so that isolating the line from it requires only
\begin{equation}
    T_{\max}\ \gtrsim\ \frac{2\pi}{\Delta_{\rm gap}} \;=\; \mathcal{O}(1),
    \label{eq:Tmax_req}
\end{equation}
The finite-$m_0$ satellites of Sec.~\ref{subsec:qed3_plaquette} set the finer,
$1/m_0$-collapsing scale discussed there. The second fact is that for a single line
of Fourier width $\delta\omega=2\pi/T_{\max}$
measured with signal-to-noise $S$, the centroid is determined to $\delta\omega/S$, so a
relative precision $\epsilon$ needs
\begin{equation}
    T_{\max}\,S \;\gtrsim\; \frac{2\pi}{\epsilon\,\Delta_{\rm tw}} ,
    \label{eq:Tmax_precision}
\end{equation}
which with $S\sim\sqrt{N_{\rm shot}}$ per time point on a Nyquist grid of
$T_{\max}/\delta t$ points gives the Fourier-route budget
\begin{equation}
    N_{\rm shot}^{\rm tot} \;\sim\; \frac{1}{\delta t}
    \left(\frac{2\pi}{\epsilon\,\Delta_{\rm tw}}\right)^{2}\frac{1}{T_{\max}} ,
    \label{eq:shot_ft}
\end{equation}
quoted against the single-time phase estimate in Sec.~\ref{sec:scaling_meas}: at
$\epsilon=3\%$, $T_{\max}=150$ and $\delta t=0.1$ it evaluates to $2\times10^{7}$
shots against the $3\times10^{2}$ of Eq.~\eqref{eq:shot_phase}. The
exponentially small scale has not disappeared, but it constrains the product of window length and
signal-to-noise not just the window length alone. Longer times than
Eq.~\eqref{eq:wrap_ceiling} allows are reachable with a ladder of interrogation times
in the manner of robust phase estimation, at logarithmic overhead.

Equation~\eqref{eq:shot_phase} treats the twist line as sharp, whereas the fiber distribution of Sec.~\ref{sec:fiber_dist} gives it an inhomogeneous width
$\sqrt{\pi}\,\Delta\cot(\theta/2)$ relative to $\Delta_{\rm tw}$, which at ten decibels
and $\theta=\pi/2$ is half the quantity being measured and which enters the budget as
a further reduction of the usable contrast. For static charges the band edge
$\theta=\pi$ removes most of it at no cost, the width there being second order in
$\Delta$. With dynamical matter the sector is not ours to choose, and recovering the
same factor becomes part of designing a probe that populates it, alongside raising
$w$. The equation also assumes the twist line has been separated from the
satellites, which costs a handful of time points or a spectral filter and is what
keeps $w$ from being free, and it carries neither the circuit depth needed to reach
$T$ nor the accumulated gate and decoherence errors over that depth. A defensible
number requires a Cram\'er--Rao analysis carrying all of these, which we do not
attempt.

The argument has two caveats. It assumes the demodulation reference is exact, which
holds for static charges but only up to the $\delta_{\rm dyn}$ term of
Eq.~\eqref{eq:Omega0} once matter is dynamical, and it assumes the twist line is
resolvable from the satellites, which fails if $\Delta_{\rm gap}$ itself closes. Note that the gap is not set by the coupling of the theory but by the lattice volume of the system. Expanding $\hat{H}_{\rm gauge}$ of
Eq.~\eqref{eq:Hgauge} to quadratic order gives $\tfrac{g^2}{2}\hat\eta\HH^{(2)}\hat\eta
+\tfrac{1}{2g^2}\hat\chi^{\mathsf T}\hat\chi$ in the loop frame, whose normal
frequencies are $\omega_k=\sqrt{\lambda_k(\HH^{(2)})}$: the factors of $g$ cancel
between the electric and magnetic terms, and the harmonic photon spectrum is
$g$-independent at fixed volume. With $\HH^{(2)}=4\Id-A$ the eigenvalues are
$4-2\cos\tfrac{\pi j}{N+1}-2\cos\tfrac{\pi k}{N+1}$ for $j,k=1,\dots,N$, so the
smallest is $\omega_{\min}=2\sqrt{2}\,\sin\tfrac{\pi}{2(N+1)}$, set by the lattice
volume and shrinking as $N$ grows. At weak coupling it is $\Delta_{\rm tw}$ itself that becomes exponentially small, Eq.~\eqref{eq:Dtw_instanton}, so the line to be resolved
moves down while the satellites around it do not. The obstruction at weak coupling is
thus the shot budget of Eq.~\eqref{eq:Tmax_precision} together with the reach in
charge separation $R$, not the coherent window alone.

\end{document}